\documentclass[graybox, envcountchap, vecphys]{svmult}
\usepackage[margin=31mm]{geometry}

\usepackage{graphicx}        
\usepackage{multicol}        
\usepackage[bottom]{footmisc}

\usepackage{newtxtext}       %
\usepackage{newtxmath}       

\usepackage{epsfig}
\usepackage{epstopdf}
\usepackage{scalerel}

\usepackage{hyperref}
\hypersetup{colorlinks=true,linkcolor=blue,citecolor=red,urlcolor=blue}
\usepackage[numbers,compress]{natbib}
\usepackage{chapterbib}
\usepackage{comment}
\usepackage{siunitx}
\usepackage{slashed}
\usepackage[normalem]{ulem}

\DeclareSIUnit \pc {pc}
\DeclareSIUnit \gauss {G}

\usepackage{calligra}
\DeclareMathAlphabet{\mathcalligra}{T1}{calligra}{m}{n}
\DeclareFontShape{T1}{calligra}{m}{n}{<->s*[2.2]callig15}{}

\usepackage{euscript}

\def\bra#1{{\langle#1|}}

\def\cg(#1,#2)(#3,#4)(#5,#6){\bra{#1,#2,#3,#4}#5,#6\rangle}

\def\threej(#1,#2)(#3,#4)(#5,#6){\begin{pmatrix}#1&#3&#5\\#2&#4&#6\end{pmatrix}}
\def\sixj(#1,#2,#3)(#4,#5,#6){\begin{Bmatrix}#1&#2&#3\\#4&#5&#6\end{Bmatrix}}
\def\ninej(#1,#2,#3)(#4,#5,#6)(#7,#8,#9){\begin{Bmatrix}#1&#2&#3\\#4&#5&#6\\#7&#8&#9\end{Bmatrix}}

\usepackage[arrows]{ezedits}

\begin{document}

\mainmatter

%
%
%

\title{Astrophysical Searches and Constraints on Ultralight Bosonic Dark Matter}


\author{David J. E. Marsh and Sebastian Hoof}

\authorrunning{Marsh and Hoof}


\institute{David J. E. Marsh \at Department of Physics, King's College London, London, United Kingdom, \\ \email{david.j.marsh@kcl.ac.uk}
\and
Sebastian Hoof \at Institut f\"ur Astrophysik, Georg-August-Universit\"at G\"ottingen, Germany, \\ \email{hoof@uni-goettingen.de}}

\maketitle

\abstract{Starting from the evidence that dark matter (DM) indeed exists and permeates the entire cosmos, various bounds on its properties can be estimated. Beginning with the cosmic microwave background and large scale structure, we summarize bounds on the ultralight bosonic dark matter (UBDM) mass and cosmic density. These bounds are extended to larger masses by considering galaxy formation and evolution, and the phenomenon of black hole superradiance. We then discuss the formation of different classes of UBDM compact objects including solitons/axion stars and miniclusters. Next, we consider astrophysical constraints on the couplings of UBDM to Standard Model particles, from stellar cooling (production of UBDM) and indirect searches (decays or conversion of UBDM). Throughout, there are short discussions of ``hints and opportunities'' in searching for UBDM in each area.}
\renewcommand{\thesection}{\arabic{section}}
\renewcommand{\thefigure}{\arabic{figure}}
\renewcommand{\thetable}{\Roman{table}}
\renewcommand{\theequation}{\arabic{equation}}
\section{Astrophysical search channels}
\label{sec:astro-review}

Astrophysics and cosmology, as outlined in Chapter~1, give convincing evidence that dark matter~(DM) exists in the form of new particles beyond the Standard Model of particle physics. The space of possible theories in Chapter~2, even for the subclass of ultralight bosonic DM (UBDM) models considered in this book, is vast. Beyond the basic fact of the existence of DM, astrophysics can be used to reign in this vast theoretical parameter space, with a view to direct detection and measurement of model parameters.

The most basic astrophysical route to constrain UBDM is via the relic density. There are three channels for UBDM production:\index{relic density}
\begin{enumerate}
\item Coherent field oscillations.
\begin{enumerate}
\item Vacuum realigment.\index{vacuum realignment}
\item Topological defect decay.\index{topological defect}
\end{enumerate}
\item Thermal production.
\item Non-thermal production by direct decay.
\end{enumerate}
Without going into the specifics (see Ref.~\cite{Marsh:2015xka}), it suffices to say that only channel~1 produces UBDM with the required properties as outlined in the chapter ``Introduction to Dark Matter.'' Production channels~2 and 3 produce hot DM, or indeed dark radiation, each of which are strongly constrained by the CMB anisotropies~\cite{2013JCAP...10..020A,Aghanim:2018eyx}. 

In channel~1a (vacuum realignment) the UBDM relic density is a function of two parameters, $(m,\phi_i)$, where $\phi_i$ is the initial field displacement, i.e.\ the location of the field in its potential relative to the minimum at ``the initial time'' (in practice, at the end of inflation). In this scenario, the initial field displacement is taken to be completely uniform throughout space, this state of affairs having been arranged by the same mechanism that causes the large scale observed homogeneity of the cosmic microwave background (CMB), inflation, or otherwise. The correct relic abundance can be achieved across many orders of magnitude, covering all the masses of interest $(\SI{e-33}{\eV}, \SI{e-1}{\eV})$ for $\phi_i\leq M_{pl}$.\footnote{$M_{pl}=1/\sqrt{8\pi G_N}$ is the \emph{reduced} Planck mass,\index{Planck mass} related to the mass scale given in the ``Units and Conversions'' section by the factor of $\sqrt{8\pi}$ coming from Einstein's equation in general relativity.} For an axion-like particle~(ALP) the fundamental parameter from theory is $f_a$: the scale of spontaneous symmetry breaking, also called the axion decay constant.\index{axion!decay constant} The parameter $\theta_i$, defined via $\phi_i \equiv \theta_i f_a$, is the initial angle that the axion field takes (recall that the axion is the phase of a complex field). At early times the axion possesses a shift symmetry,\index{shift symmetry} $\phi\rightarrow \phi+\text{constant}$, and thus $\theta_i$ has no preferred value and can be considered a free random variable (although very small values, or values very close to $\pi$ are considered fine tuned). Because $\theta_i$ is undetermined there is a wide range of allowed values for the fundamental parameters $(m,f_a)$ consistent with the required relic density. In particular, in this channel large values of the decay constant at the grand unified scale ($\sim \SI{e16}{\GeV}$) or the reduced Planck scale ($\sim \SI{e18}{\GeV}$) are allowed.

Production via channel 1b (topological defect decay)\index{topological defect!decay} is possible only for UBDM that is a Goldstone boson of a spontaneously broken global symmetry (the ``Kibble-Zurek mechanism''~\cite{1976JPhA....9.1387K,1985Natur.317..505Z}\index{Kibble-Zurek mechanism} described in Section~\ref{sec:miniclusters}). In particular it applies to the QCD axion and other ALPs, where topological strings\index{cosmic string} and domain walls\index{domain walls} are formed when the global $\mathbb{U}(1)$ symmetry\index{U(1) symmetry} is spontaneously broken.\index{spontaneous symmetry breaking} If symmetry breaking occurs after inflation,\index{inflation} then the defects cannot be smoothed out and inflated away, and the axion field takes on a very inhomogenous distribution (in contrast to the case of vacuum realignment). The defects later decay when non-perturbative effects give the ALP a mass. This process must be simulated using classical lattice field theory, and has only been studied in detail for the QCD axion~\cite{Hiramatsu:2012gg,Klaer:2017ond,Gorghetto:2018myk}. Large numerical uncertainties related to extrapolation to physical couplings prevent an agreed estimation of the relic density.\index{relic density} The correct relic abundance can be achieved within numerical and model uncertainty (extrapolation, domain wall number, explicit symmetry breaking) for all values of $f_a\lesssim \SI{e12}{\GeV}$~\cite{Armengaud:2019uso}. 

The production mechanism channel 1b works for $f_a < T_\text{max}$, where $T_\text{max}$ is the maximum thermalization temperature of the Universe, and the bound arises since defects only form if symmetry breaking occurs during the ordinary thermal history of the Universe. $T_\text{max}$ is bounded from above due to observational constraints on the theory of inflation. In particular $H_I$, the inflationary Hubble rate,\index{Hubble parameter} is bounded from above by the fact that tensor-type CMB anisotropies have relative amplitude $r\lesssim 0.1$ compared to scalar-type perturbations leading to the constraint $H_I\lesssim \SI{e14}{\GeV}$. $H_I$~sets the temperature of the Universe during inflation to be the Gibbons-Hawking temperature,\index{Gibbons-Hawking temperature} $T_\text{GH}=H_I/2\pi$. The maximum thermalization temperature could actually be larger than this, which can easily be seen from the Friedmann equation\index{Friedmann equation} during radiation domination, $3H^2M_{pl}^2=\pi^2 g_\star T^4/30$, where the quantity $g_\star$ counts the effective number of relativistic degrees of freedom~\cite{1990eaun.book.....K}:
\begin{equation}
g_\star = \frac{7}{8}\sum_{i\in \text{fermions}}g_i\left(\frac{T_i}{T}\right)^4+\sum_{i\in \text{bosons}}g_i\left(\frac{T_i}{T}\right)^4\,,
\label{eqn:gstar_def}
\end{equation}
where $g_i$ is the degrees of freedom of species $i$ (e.g.\ two polarizations for the photon) and $T_i$ is the temperature of species $i$, and $T$ is the photon bath temperature. The value of $g_\star$ at very high temperatures is bounded from below by the Standard Model contribution, $g_{\star,\text{SM}}=106.75$. $H$ monotonically decreases, and so $H_\text{max}=H_I$. If reheating after inflation is instantaneous and 100\% efficient, we find an upper bound for $T_\text{max}\lesssim \SI{8e15}{\GeV}$. ALPs with values of $f_a$ larger than this upper bound on $T_\text{max}$ cannot be produced by mechanism 1b, and must be produced by mechanism 1a. The observational lower bound on $T_\text{max}$ arises from demanding successful Big Bang nucleosythesis,\index{Big Bang nucleosynthesis (BBN)} $T_\text{max}\gtrsim \SI{1}{\MeV}$. For values of $f_a$ in this very large range of allowed $T_\text{max}$ values, it is not determined whether ALPs are produced by mechanism 1a or 1b, either being possible depending on the model of inflation and reheating. 

There are various astrophysical search channels we can use to constrain UBDM:
\begin{enumerate}
\item Gravitational probes.
\item ``Indirect detection.''
\begin{enumerate}
\item Production of UBDM (e.g.\ in stars or from background radiation).
\item Decay/conversion of existing UBDM.
\end{enumerate}
\end{enumerate}
Gravitational probes are the most general form of constraints on UBDM and give us powerful bounds on the key parameters of mass and density (both cosmic and local), which are important for the design of direct DM searches. Indirect detection depends on the UBDM interactions with ordinary matter: null results provide baseline constraints on couplings to which laboratory searches are compared, and anomalous results give hints for promising regions of parameter space to search.

In this Chapter, unless stated otherwise, we use natural units where $\hbar=c=k_B=1$ and express all quantities in electronvolts (eV). We use the Einstein summation convention for repeated indices. Roman indices $i$, $j$, etc.\ run from~1 to~3, while greek indices $\mu$, $\nu$, etc.\ run from~0 to~3, with zero labelling the time-like direction. In relativity, we distinguish covariant (lower)\index{covariant} and contravariant (upper)\index{contravariant} indices, with the metric being responsible for raising and lowering: $x_\mu = g_{\mu\nu}x^\nu$.

\section{Gravitational probes of UBDM}
\label{sec:grav-constraints}

The goal of this section is to assess the validity of UBDM as a model of DM. Since all current observations are consistent with cold dark matter\index{cold dark matter (CDM)} (CDM, defined as a pressureless fluid), the bounds we estimate on the UBDM mass~$m$ can be thought of as answering the question: ``is UBDM observationally equivalent to CDM?'' The answer to this question depends on the observable and leads to lower bounds on $m$ (and upper bounds on the UBDM density if we allow for multi-component DM). In order to derive our bounds we must specify the ways in which UBDM is not equivalent to CDM. These differences further suggest astrophysical phenomena that could distinguish between UBDM and CDM in the future, possibly providing evidence for one model over the other. 

\subsection{The CMB and linear structure formation\index{cosmic microwave background radiation (CMB)}\index{CMB}}
\label{ch3:sec:CMB-structure-formation}

Considering how the gravitational effects of DM dominate the formation of structure in the Universe, one can derive bounds on the UBDM properties from the theory of cosmological structure formation in general relativity~\cite{Dodelson:2003ft}. Consider a flat, homogeneous, and isotropic spacetime described by the Friedmann-Robertson-Walker metric:\index{Friedmann-Robertson-Walker metric}
\begin{eqnarray}
g = {\rm diag}[-1,a(t)^2,a(t)^2,a(t)^2]\, . \label{eqn:FRW}
\end{eqnarray}
The scale factor is $a(t)$,\index{scale factor} which obeys Friedmann's equation for the Hubble rate $H(t)=\dot{a}/a$:
\begin{eqnarray}
H(t)^2 = \frac{8\pi G_N}{3}\bar{\rho} \, , \label{eqn:friedmann}
\end{eqnarray}
where $\bar{\rho}$ is the total, spatially averaged, energy density. $\rho$ is composed of photons, ``baryons'' (by convention in cosmology we do not separately consider the small mass density of electrons), neutrinos, DM, and the cosmological constant or dark energy. Objects ``on the Hubble flow'',\index{Hubble flow} i.e., feeling negligible local gravitational potentials, appear to recede from an observer at the origin with a velocity $\vec{v}_H=H r \vec{\hat{r}}$, where $r$ and $\vec{\hat{r}}$ are the distance and direction from the observer to the object, respectively. We begin with a Newtonian approximation\index{Newtonian approximation} to cosmology (see e.g.\ Ref.~\cite{mukhanov}). Consider an observer at the origin, and a single particle of UBDM on the Hubble flow. The UBDM de Broglie wavelength\index{de Broglie wavelength} is $\lambda_H=1/(mv)=1/(mHr)$, which gives the radial position uncertainty, $\Delta r$. A net gravitational force in the positive direction along the line of centres between the observer and the UBDM requires $\Delta r\lesssim r \Rightarrow r\gtrsim (mH)^{-1/2}$, which defines a critical separation $r_\text{crit}=(mH)^{-1/2}$. On average, UBDM separations larger than $r_\text{crit}$ undergo gravitational clustering,\index{gravitational clustering} and those smaller than it do not.

\begin{figure}[t]
\center
\includegraphics[width=4.6in]{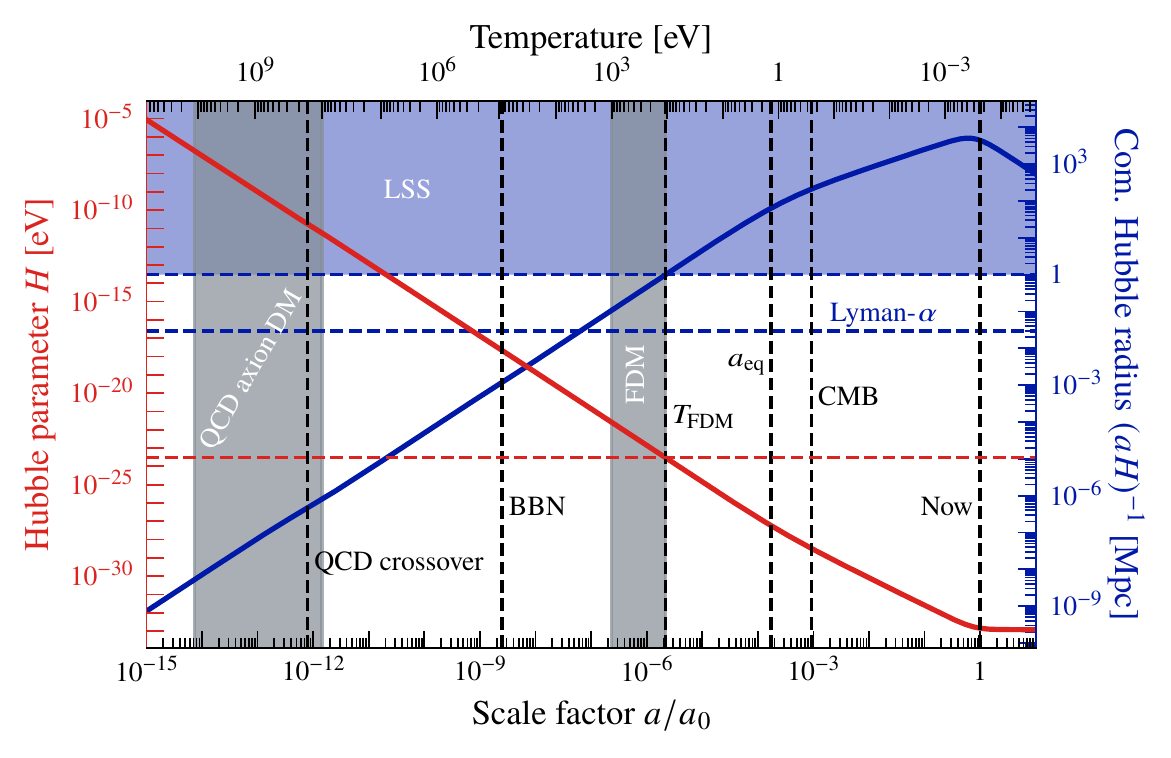}
\caption{The evolution of cosmic quantities as a function of scale factor or temperature. We show the evolution of the Hubble parameter~(red line, left axis) and the comoving Hubble radius~(blue line, right axis) together with various relevant cosmological events. The blue shaded area approximately encompasses the large~scale structure~(LSS) of the Universe, while grey shaded areas indicate where QCD~axion~(with $f_a \in [\SI{e6}{\GeV},\SI{e18}{\GeV}]$) and fuzzy dark matter~(FDM)\index{fuzzy dark matter} start to become dynamical. Note that the temperature scale on the top is not exactly regular due to the scaling with the number of relativistic degrees of freedom for entropy, $g_{\star,S}$. The quantity $g_{\star,S}$ gives the number of effective relativistic degrees of freedom contributing to the entropy density; $g_{\star,S}$ takes the same form as Eq.~\eqref{eqn:gstar_def} with the fourth powers replaced by cubes (see, e.g., Ref.~\cite{1990eaun.book.....K}, Chapter 3).}
\label{fig:ch3:cosmic_history}
\end{figure}

The cosmological horizon\index{horizon!cosmological} size is approximated by the Hubble length scale\index{Hubble length scale} $R_H=H^{-1}$. In order for UBDM to have any inhomogeneous gravitational effect within this radius requires $r_\text{crit} < R_H$. We show the cosmological evolution of~$H = R_H^{-1}$, and the related comoving Hubble radius~$(aH)^{-1}$, as functions of temperature and scale factor in Fig.~\ref{fig:ch3:cosmic_history}. The bounds and other cosmological events mentioned in what follows can often be read off directly from that figure, and we will occasionally highlight this fact going forward.

Evaluating the Hubble length scale today, and using that $H_0 = 100\,h\,\si{\km\per \second\per\mega\pc} = \SI{2.13e-33}{\eV} \times h$ (where $h$ is the dimensionless Hubble parameter with approximate observed value $h \sim 0.7$), we arrive at our first bound on the UBDM mass:
\begin{eqnarray}
m > \SI{1.5e-33}{\eV} \times \left(\frac{h}{0.7} \right) \, \quad\text{(size of the observable universe)}. \label{eqn:h0_bound}
\end{eqnarray}
UBDM violating this bound does not cluster within our cosmological horizon,\index{horizon!cosmological} is thus indistinguishable from the cosmological constant, and will not concern us in this book.\footnote{Very light scalar fields which are homogeneous on the scale of the cosmological horizon provide models for dark energy.\index{dark energy} The simplest such models are described by a canonical kinetic term in the Lagrangian, and a scalar potential $V(\phi)$, and are known as ``quintessence''~\cite{Copeland:2006wr}.\index{quintessence} An ultralight bosonic field with a mass less than the bound Eq.~\eqref{eqn:h0_bound} is one such very simple model, with $V(\phi)=m^2\phi^2/2$. More complex models invoke different potentials, or more fields, or even generalize the kinetic term, at which point they cross over into theories of ``modified gravity'' and ``beyond Horndeski'' scalar-tensor\index{scalar-tensor gravity} theory~\cite{Clifton:2011jh}.}

Assuming that UBDM constitutes the entirety of the DM, we can extend the bound to any redshift of interest where we know that DM exerted a discernible gravitational effect by simply substituting the Hubble parameter at that redshift. For temperatures below about 1 MeV we can use the expression for the Hubble parameter~\cite{Dodelson:2003ft}:
\begin{eqnarray}
H(z) = H_0 E(z) = H_0 \, \sqrt{\Omega_\Lambda +\Omega_m\left[ (1+z)^3+\frac{(1+z)^4}{1+z_\text{eq}}\right]}\, , \label{eqn:hubble}
\end{eqnarray}
where the second equality defines the energy function $E(z)$. The quantities $\Omega_m$ and $\Omega_\Lambda$ are the density parameters of matter and the cosmological constant,\index{cosmological constant} defined as the density divided by the critical density,\index{critical density} i.e.\ $\Omega_i=\bar{\rho}_i/\rho_\text{crit}$ and $\rho_\text{crit}=3M_{pl}^2H_0^2$. The last term in the brackets arises from the radiation energy density, which is defined relative to the matter density via the redshift\index{redshift} of matter-radiation equality,\index{matter-radiation equality} $z_\text{eq}$. The epoch of matter-radiation equality can be found via the relative redshifting of matter and radiation components: $\rho_m(1+z_\text{eq})^3=\rho_r(1+z_\text{eq})^4$, with the density parameters defined today. CMB observations fix $z_\text{eq}\approx 3390$, and it is thus slightly earlier in cosmic history than decoupling, $z_{\rm dec}\approx 1100$.

The \emph{baryon acoustic oscillations} (BAO, see Sec.~1.1)\index{baryon acoustic oscillations (BAO)} observed in the CMB and galaxy surveys like the \emph{Sloan Digital Sky Survey}~\cite{2019ApJS..240...23A}\index{Sloan Digital Sky Survey} require that DM was gravitationally relevant at and before matter-radiation equality: if it was not, because baryons are coupled to the photons at early times and perturbations in them cannot grow in the radiation era, the amplitude of galactic fluctuations on scales of order 1 Mpc would not not be consistent with the amplitude and scale dependence of the CMB anisotropies. Again assuming that UBDM is all the DM and substituting $H(z_\text{eq})$ we arrive at the tighter bound~(cf.\ Fig.~\ref{fig:ch3:cosmic_history}):\index{matter-radiation equality}
\begin{align}
m > \SI{1.6e-28}{\eV} \times \left(\frac{h}{0.676}\right)\left(\frac{\Omega_m}{0.311}\right)^{1/2}\left(\frac{1+z_\text{eq}}{3390}\right)^{3/2}\nonumber \\
\text{(matter-radiation equality)} \, , \label{eqn:equality_bound}
\end{align}
where we have neglected the small contribution of $\Omega_\Lambda$ at equality, and taken reference parameters from the CMB+BAO combination\index{LambdaCDM@$\Lambda$CDM model} in Ref.~\cite{Aghanim:2018eyx}.\footnote{Using these reference parameters further assumes that UBDM is sufficiently CDM-like that we can use the standard CMB parameters (which are derived under the assumption of $\Lambda$CDM).}

The matter-radiation equality bound, Eq.~\eqref{eqn:equality_bound}, is the UBDM equivalent of saying that DM is not ``hot''~\cite{Primack:2001ib}: gravitational clustering is required before matter-radiation equality in order for bottom-up hierarchical structure formation\index{structure formation} (rather than top-down fragmentation) of galaxies, consistent with observations of extremely high redshift galaxies. We could progress further with such estimates (and we will in due course), but now we must make our model more precise.

\begin{example}{Tutorial: The Growth of Cosmic Structure \index{cosmological perturbation theory}}
The challenge in cosmological perturbation theory~\cite{bertschinger1995}\index{perturbation theory!cosmological} is to compute the \emph{transfer function}\index{transfer function}, $T_X(t,k)$ for the mode evolution of each cosmological species $X$ (baryons, photons, neutrinos, dark matter) with Fourier wavenumber\index{wavenumber} $k$, which fully specifies linear evolution of cosmological fields from Gaussian initial conditions. That is: 
\begin{eqnarray}
\zeta_X(k,t) = \zeta_{X,i}(k)T_X(t,k)\xi_X \, ,
\end{eqnarray}
where $\zeta_{X,i}(k)$ is the initial condition of the field and $\xi_X$ is a Gaussian random field defining the initial correlation functions of the field $\zeta_X$.

The codes \textsc{camb}~\cite{camb} and \textsc{class}~\cite{class} are the standards for numerical computation for CDM (and many other things), while \textsc{axionCAMB}~\cite{Hlozek:2014lca}\footnote{Available at \url{https://github.com/dgrin1/axionCAMB}.} can be used for UBDM that is a real scalar field with the self-interaction potential approximated by $V(\phi)=m^2\phi^2/2$. This tutorial gives a brief overview of the most relevant aspects of cosmological perturbation theory for UBDM constraints. 

Cosmological perturbation theory deals with the evolution of fluctuations relative to a homogeneous and isotropic background. Background quantities are labeled with an overbar, since they represent the spatial average, and thus depend only on cosmic time~$t$. The perturbation modes have spatial dependence captured by their wavenumber, and perturbations at the initial time all have relative amplitude much less than one with the respect to the background quantities. The fields $\zeta$ of interest are the components of the energy momentum tensor,\index{energy momentum tensor} written as $T^0_{\,\,0}=-(\bar{\rho}+\delta\rho)$, $T^i_{\,\,j}=(\bar{P}+\delta P)\delta^i_j+\Sigma^i_j$, $ik^iT^0_{\,\,i}=(\bar{\rho}+\bar{P})\theta$, which defines the energy density, $\rho$, pressure, $P$, and heat flux, $\theta=\nabla\cdot \vec{v}$, and we assume anisotropic stresses $\Sigma^i_j$ vanish\index{energy momentum tensor}. This gives the fields $\delta_X=\delta\rho_X/\bar{\rho}_X$, and $\theta_X$, while pressure is typically described it terms of a sound speed, $c_s^2=\delta P/\delta \rho$.

Next, perturb the metric\index{metric} Eq.~\eqref{eqn:FRW}, and switch to conformal time,\index{conformal time} $\tau$, via ${\rm d}t=a{\rm d}\tau$. The Newtonian gauge\index{Newtonian gauge} considers only scalar metric perturbations: 
\begin{eqnarray}
g = a^2 {\rm diag}[-(1+2\Psi), \, 1-2\Phi, \, 1-2\Phi, \, 1-2\Phi] \, .
\label{eqn:newton_line_element}
\end{eqnarray}
The potential $\Phi$ is the usual Newtonian potential, and $\Psi$ is the curvature perturbation: they are equal in the non-relativistic limit. The energy momentum tensor is coupled to the metric degrees of freedom by the Einstein equation:\index{Einstein equation}
\begin{eqnarray}
G_{\mu\nu} = 8\pi G_N T_{\mu\nu}\, ,
\label{eqn:einstein}
\end{eqnarray}
where $G_{\mu\nu}$ is the Einstein tensor,\index{Einstein tensor} and depends on the metric potentials and their derivatives. This is the dynamical equation determining the evolution of the metric.

The equation of motion for the UBDM field with self-interaction potential $V(\phi)$ is:
\begin{eqnarray}
\Box\phi - \partial_\phi V = 0\, , \label{eqn:KG_eq}
\end{eqnarray}
where the d'Alembertian~($\Box$)\index{d'Alembertian} is:
\begin{eqnarray}
\Box = \frac{1}{\sqrt{-g}}\partial_\mu\sqrt{-g}g^{\mu\nu}\partial_\nu\, ,
\label{eqn:box}
\end{eqnarray}
and where $g$ and $g^{\mu\nu}$ are the metric determinant and the inverse of the metric, respectively. Setting $V=\frac{1}{2}m^2\phi^2$ for simplicity, this leads to the equations of motion for the UBDM background field, $\bar{\phi}$ and fluctuation mode $\delta\phi_k$:
\begin{align}
\bar{\phi}''+2\mathcal{H}\bar{\phi}'+a^2m^2\bar{\phi} &= 0 \label{eqn:background_conformal} \, , \\
\delta\phi_k''+2\mathcal{H}\delta\phi_k'+(m^2a^2\delta\phi_k+k^2)\delta\phi_k &= (\Psi'+3\Phi')\bar{\phi}'-2m^2a^2\Psi\bar{\phi} \, , \label{eqn:perturbations_conformal}
\end{align}
where primes denote derivatives with respect to conformal time, and $\mathcal{H}=a'/a=aH$. For the UBDM field, we find $T^{\mu\nu}=\delta S/(\delta g_{\mu\nu})$ by variation of the action with respect to the metric tensor, giving\index{energy momentum tensor}\index{equation of state}:
\begin{equation}
T^{\mu\nu} = g^{\mu\alpha}\partial_\alpha\phi\partial^\nu\phi-g^{\mu\nu}\left[\frac{1}{2}g^{\alpha\beta}\partial_\alpha\phi\partial_\beta\phi+V(\phi)\right] \, .
\label{eqn:energy-momentum}
\end{equation}
Working to first order in the metric perturbations and $\delta\phi$, and with potential $V=m^2\phi^2/2$ the components are:
\begin{align}
\bar{\rho} &= \frac{1}{2}a^{-2}(\bar{\phi}')^2+\frac{1}{2}m^2\bar{\phi}^2\, , \\
\bar{P} &= \frac{1}{2}a^{-2}(\bar{\phi}')^2-\frac{1}{2}m^2\bar{\phi}^2\, , \\
\delta\rho &= a^{-2}[\bar{\phi}'\delta\phi_k'-\Psi(\bar{\phi}')^2]+m^2\bar{\phi}\delta\phi_k\, , \\
\delta P &= a^{-2}[\bar{\phi}'\delta\phi_k'-\Psi(\bar{\phi}')^2]-m^2\bar{\phi}\delta\phi_k\, , \\
(\bar{\rho}+\bar{P})\,\theta&=a^{-2}ik^2\bar{\phi}'\delta\phi_k\, .
\end{align}

\begin{question}{Problem 1: Background evolution of UBDM}
Assuming a single-fluid universe with constant equation of state $w$ satisfying $\dot{\rho}=-3H(1+w)\rho$, first solve Friedmann's equation Eq.~\eqref{eqn:friedmann} for $a(t)$, and thus $H(t)$. Then change variables in Eq.~\eqref{eqn:background_conformal} to physical time ${\rm d}t=a {\rm d}\tau$. Substituting your solution for $H(t)$, derive the solution for $\bar{\phi}(t)$ (you may use exact functions or asymptotic methods). Given that the energy density and pressure of UBDM are $\bar{\rho}=\frac{1}{2}\dot{\phi}^2+V(\phi)$ and $\bar{P}=\frac{1}{2}\dot{\phi}^2-V(\phi)$, derive the behaviour of the equation of state for UBDM, $w_\text{UBDM}=\bar{P}/\bar{\rho}$. What is the asymptotic value of $w_\text{UBDM}$ for $m\ll H$, and $\langle w\rangle$ for $m\gg H$ (brackets denote period average)? Repeat this exercise for a $\lambda\phi^4$ potential. Comment on the results for $w_\text{UBDM}$ in relation to the approximate UBDM mass bounds above.

~\

{\it{Solution on page~\pageref{ch11:prob-sol:3-1}.}}

\end{question}

CDM is defined as a collisionless and uncoupled fluid, $w_c=c_c^2=0$. Baryons have a sound speed, $c_b^2 \neq 0$ (computed from the evolution of the baryon temperature), equation of state $w_b=0$ (on average the baryons have negligible pressure), and are coupled to photons via Thomson scattering.\index{Thomson scattering} The photon equation of motion is derived from the Boltzmann equation,\index{Boltzmann equation} which is expanded in Legendre polynomials\index{Legendre polynomials} to capture the dependence on the angle between the momentum coordinate on phase space and the wavevector. The hierarchy of moment equations are labeled by the order~($l$) of the Legendre polynomial: the zeroth moment gives the equation of motion for the density, the first, for the velocity, the second, for the anisotropic stress, and so on (a recursion relation can be used to approximately close the hierarchy above some $l_\text{max}$). Truncating this Boltzmann hierarchy at the velocity moment, the photons resemble a fluid with $w=c_s^2=1/3$, collisionally coupled to the baryons. We consider perturbations to the energy density $\delta_X = \delta\rho_X/\bar{\rho}_X$ and heat flux $\theta_X$, defined via $\bar{\rho}_X(1+w_X)\theta_X=ik^j\delta (T^0{}_j)_X$, where $(T^\mu{}_\nu)_X$ is the $X$ energy momentum tensor\index{energy momentum tensor}.

Let us now consider a number of limits of the full equations of motion, which can be found in Ref.~\cite{bertschinger1995}. At early times, photons have enough energy to keep hydrogen and other atoms ionized, giving rise to a large free electron density. Thus, the photons and baryons are tightly coupled by Thomson scattering and can be treated as a single fluid with $\theta_\gamma=\theta_b$. Considering only sub-horizon modes ($k\gg aH$), and using the Poisson equation\index{Poisson equation} and the $ii$~pressure component of the Einstein equation Eq.~\eqref{eqn:einstein}, the photon fluid at early times obeys the equation of motion:
\begin{eqnarray}
\delta_{\gamma}''+\left(c_{s,\gamma}^2k^2-\frac{16 \pi}{3}G a^2\rho_\gamma\right)\delta_\gamma
 = 4\pi G a^2\sum_i(1+c_i^2)\rho_i\delta_i \, , 
\label{eqn:photon_eom}
\end{eqnarray}
where the photon sound speed is $c_{s,\gamma}=1/\sqrt{3}$ (speed of pressure perturbations in a gas of photons in thermodynamic equilibrium). At very early times all $\rho_i$ in the driving term on the right hand side can be neglected. Then this equation has sound wave solutions for $k>(16\pi G a^2\rho_\gamma)^{1/2}=\sqrt{6}aH$. This defines the Jeans scale\index{Jeans scale} of the photon-baryon fluid, which is of order the comoving horizon size. Perturbations with wavelength shorter than the Jeans scale undergo coherent, pressure supported oscillations. Perturbations with wavelength longer than the Jeans scale grow due to gravitational instability. The sound waves prevent the formation of gravitationally bound structures in the photon-baryon fluid\index{photon-baryon fluid} and lead to BAO\index{baryon acoustic oscillations (BAO)}. At \emph{recombination}\index{recombination} temperatures of around 0.2 eV (redshift $z\approx 1100$)~\cite{1990eaun.book.....K}, the energy of the ambient photon fluid is no longer sufficient to keep neutral hydrogen from forming. At this time, the free electron density drops to zero, the photon-baryon fluid decouples, and the sound wave stalls. This \emph{sound horizon}\index{sound horizon}\index{horizon!sound} for the BAO is given by:
\begin{eqnarray}
r_s = \int_0^{t_0} \frac{{\rm d}t}{a} c_{s,b}\approx \frac{1}{\sqrt{3}}\int^{t_{\rm dec}}_0 \frac{{\rm d}t}{a}\, ,
\label{eqn:sound_horizon_bao}
\end{eqnarray}
where $c_{s,b}$ is the baryon sound speed in the plasma, $t_0$ is the time today, and $t_{\rm rec}$ is the time at recombination when $c_{s,b}$ drops rapidly from $c_{s,\gamma}$ to zero. The BAO scale leads to oscillations in the CMB angular power spectrum, which we have seen already in Chapter 1. The gauge invariant temperature anisotropy of the CMB is given by:\footnote{This equation ignores the effect of gravitational lensing\index{gravitational lensing} along the line of sight. This second-order effect is important at high multipoles and is sensitive to the UBDM sound speed and structure growth. See Refs.~\cite{Lewis:2006fu,Hlozek:2017zzf}.}
\begin{eqnarray}
\frac{\delta T}{T} = \int_0^{\tau_0}\left[\dot{\mu}\left(\Phi+\frac{\delta_\gamma}{4}+\vec{\hat{n}}\cdot\vec{v}_b+2\dot{\Phi}\right)\right]e^{-\mu}{\rm d}\tau\, , \label{eqn:cmb_temp}
\end{eqnarray}
where $\mu$ is the Thomson scattering opacity, $\vec{v}_b$ is the baryon velocity, $\vec{\hat{n}}$ is a unit vector giving the sky position, and the integral is along the line of sight. The four terms in Eq.~\eqref{eqn:cmb_temp} correspond, respectively, to: the gravitational redshift,\index{gravitational redshift}\index{redshift} the photon anisotropy, the Doppler effect,\index{Doppler effect} and the final term gives rise to the integrated Sachs-Wolfe effect,\index{Sachs-Wolfe effect} which is an additional form of gravitational redshift. 

Decoupling occurs at a redshift $z_{\rm dec}\approx 1100$\index{CMB decoupling}, which gives the angular scale of the first CMB acoustic peak. The driving term on the right hand side of Eq.~\eqref{eqn:photon_eom} is dominant for $z<z_{\rm eq}\approx 3400$, corresponding to angular scales slightly smaller than the first peak, including the second and third peak. Thus, the relative heights of these peaks can be used to measure the matter content and its behaviour near matter-radiation equality. 

How do UBDM perturbations evolve? The first transition in behaviour is in the equation of state, which becomes zero (i.e.\ pressureless) shortly after $H(a_\text{osc})=m$ (this defines the value of the scale factor $a_\text{osc}$ when the background UBDM field, $\bar{\phi}$, begins to undergo coherent oscillations, see Problem~1). Prior to this time the UBDM is relativistic and perturbations cannot grow.\footnote{For an axion-like potential, the equation of state is $w=-1$ prior to $a_\text{osc}$. For a scalar field with potential $V = m^2\phi^2/2 + \lambda\phi^4$, the equation of state is $w=1/3$ at early times for large $\phi$ initial conditions. For a complex scalar, the early time equation of state is $w=1$ due to the conserved charge and Goldstone mode~\cite{2014PhRvD..89h3536L}. In each case, perturbations are suppressed relative to pressureless CDM.} For $H\ll m$ the UBDM perturbations Eq.~\eqref{eqn:perturbations_conformal} can be approximated as a fluid with sound speed~\cite{2009PhLB..680....1H}:\footnote{This expression is exact in the UBDM comoving gauge. Additional terms due to the gauge transformation to a standard gauge, e.g., Newtonian or synchronous, decay on sub-horizon scales as all gauge artifacts do in cosmological perturbation theory~\cite{Hlozek:2014lca}.}
\begin{eqnarray}
c_\text{UBDM}^2 = \frac{k^2/4m^2a^2}{1+k^2/4m^2a^2}\, .
\end{eqnarray}
The non-relativistic limit of this expression is derived later on in this Chapter from the Schr\"{o}dinger-Poisson equation, see Section~\ref{sec:schrodinger-poisson} and Problem~2.

Now compare the behaviour of UBDM and CDM+baryons for subhorizon modes in the matter-dominated era. The baryon sound speed can be neglected after decoupling, so CDM and baryons can be combined into a single pressureless fluid. In the sub-horizon $k \gg aH$, super-Compton $k \ll m$ limit, the CDM+baryon and UBDM fluids obey the coupled equations of motion:
\begin{eqnarray}
&\ddot{\delta}_{c+b}+2H\dot{\delta}_{c+b} = -\frac{k^2}{a^2}\Phi \, , \label{eqn:cdm_growth}\\
&\ddot{\delta}_\text{UBDM}+2H\dot{\delta}_\text{UBDM}+\frac{k^4}{4m^2a^4}\delta_\text{UBDM} = -\frac{k^2}{a^2}\Phi \, ,\label{eqn:ubdm_growth} \\
&\frac{k^2}{a^2}\Phi = 4\pi G_N(\bar{\rho}_{c+b}\delta_{c+b}+\bar{\rho}_\text{UBDM} \, \delta_\text{UBDM})\, . \label{eqn:poisson}
\end{eqnarray}
Setting $\bar{\rho}_\text{UBDM}=0$ and substituting the Poisson equation\index{Poisson equation} Eq.~\eqref{eqn:poisson} into Eq.~\eqref{eqn:cdm_growth} gives the solution $\delta_{c+b}=A_+(k)a+A_-(k)a^{-3/2}$. The growing mode\index{cosmic growing mode} initial conditions sets $A_-(k)=0$, and the inflationary initial conditions and matter transfer function fix $A_+(k)$. Due to the zero pressure and sound speed of CDM, all the $k$-dependence in the solution is fixed by the initial conditions, and the dynamics are scale invariant. 

Now consider a UBDM-dominated universe by taking $\bar{\rho}_{c+b}=\bar{\rho}_b\ll\bar{\rho}_\text{UBDM}$ (i.e., no CDM and treating the baryons as sub-dominant) in Eq.~\eqref{eqn:ubdm_growth}, and again substituting the Poisson equation. The substitution of the Poisson equation\index{Poisson equation} gives rise to a negative contribution on the left hand side proportional to $\delta_\text{UBDM}$, which drives growth of $\delta_\text{UBDM}$, while the positive contribution from the sound speed term leads to acoustic oscillations. The sign of the term proportional to $\delta_\text{UBDM}$ depends on $k$ and as such different modes evolve differently. That is, we find Eq.~\eqref{eqn:ubdm_growth} exhibits a \emph{Jeans scale}\index{Jeans scale}, $k_J$, separating growing/decaying and oscillating modes. The exact solution for pure UBDM is $\delta_\text{UBDM}=A_+(k)D_+(k,a)+A_-(k)D_-(k,a)$, where the growth functions are:
\begin{eqnarray}
&D_+(k,a)= \frac{3\sqrt{a}}{\tilde{k}^2}\sin \left( \frac{\tilde{k}^2}{\sqrt{a}} \right) +\left[ \frac{3a}{\tilde{k}^4}-1 \right]\cos \left( \frac{\tilde{k}^2}{\sqrt{a}} \right)		\, , \label{eqn:delta_exact2}\\
&D_-(k,a) = \left[ \frac{3a}{\tilde{k}^4}-1\right] \sin \left( \frac{\tilde{k}^2}{\sqrt{a}} \right) -\frac{3\sqrt{a}}{\tilde{k}^2}\cos \left( \frac{\tilde{k}^2}{\sqrt{a}} \right) \, .
\label{eqn:delta_exact3} \\
&\tilde{k}=k/\sqrt{m H_0}
\end{eqnarray}

Consider the evolution of three wavenumbers in the pure UBDM case: the horizon size, $k_\star=aH$; the Jeans scale, $k_J=a\sqrt{Hm}$; and the Compton scale, $k_c=ma$. The Compton scale\index{Compton scale} defines relativistic modes where $c_\text{UBDM}^2=1$; $k_c$ increases with time, and more modes become non-relativistic. If $k_\star<k_c$, then a mode is non-relativistic when it enters the horizon and behaves as CDM (``long modes''). If a mode is relativistic when it enters the horizon (``short modes'') then the sound speed cannot be neglected, and modes will not grow until the later time when the Jeans wavenumber\index{wavenumber!Jeans} enters the horizon. The evolution of these three modes is illustrated in Fig.~\ref{fig:scales_linear_perts}. All modes intersect at the time $a_\text{osc}$, which defines the special mode $k_m$, the horizon size when the UBDM background becomes non-relativistic. All $k<k_m$ evolve similarly to CDM. All $k>k_m$ have suppressed growth. 
\begin{figure}[t]
\center
\includegraphics[width=0.85\textwidth]{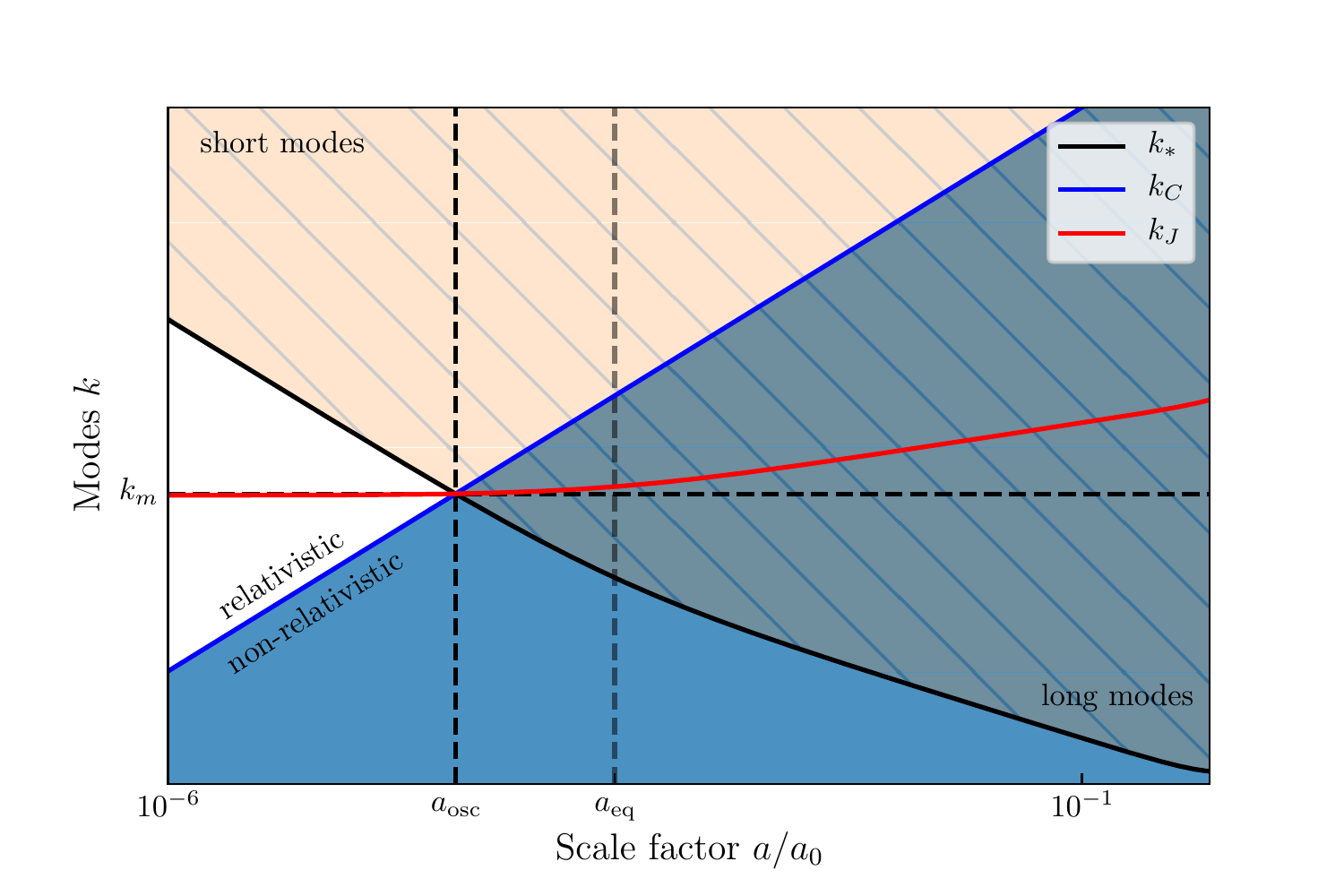}
\caption{Evolution of scales for linear perturbations with $m=10^{-26}\text{ eV}$. The Jeans scale, Compton scale, and horizon scale, all intersect at $a_\text{osc}$ when the field begins to oscillate. This defines the scale of power suppression as the comoving horizon size at this time, $k_m=a_\text{osc}H_\text{osc} = R_H(a_{\rm osc.})^{-1}$. Due to the slow evolution of $k_J$ with $a$, and the logarithmic growth of density perturbations during the radiation epoch, the suppression scale is also approximated by the Jeans scale at matter-radiation equality. Adapted from Ref.~\cite{Bauer:2020zsj}.}
\label{fig:scales_linear_perts}
\end{figure}

The scale that determines suppression of growth compared to CDM is the Jeans scale at matter-radiation equality. Using Eq.~\eqref{eqn:ubdm_growth} in the pure UBDM limit with $c_\text{UBDM}\approx k^2/4m^2a^2$, substituting the Poisson equation, and solving for $k_J$ where the effective mass term in the oscillator equation for the overdensity vanishes, we find:
\begin{eqnarray}
k_{J,\text{eq}} = 9.0 \left( \frac{3390}{1+z_\text{eq}}\right)^{1/4}\left( \frac{\Omega_\text{UBDM}}{0.12}\right)^{1/4}\left( \frac{m}{10^{-22}\text{ eV}}\right)^{1/2}\text{ Mpc}^{-1}\, .
\label{eqn:kJ_eq}
\end{eqnarray}
Recall that by definition CDM has zero sound speed. Thus CDM possesses no Jeans scale (the growing mode solution above is scale invariant), and we see that UBDM is only equivalent to CDM exactly in the limit $m \rightarrow \infty$. In practice, they are equivalent as long as $k_J$ does not play a role in any observation.

An observable related to the matter clustering is the \emph{matter power spectrum}\index{matter power spectrum}\index{power spectrum!matter} defined by $\langle\delta_m(\vec{k}_1)\delta_m(\vec{k}_2)\rangle=(2\pi)^3\delta_D(\vec{k}_1-\vec{k}_2)P(k)$, where $\delta_m$ is the total matter (baryon+CDM+UBDM+neutrino) overdensity, and $\delta_D$ is the Dirac delta distribution. The presence of the sound speed and consequent Jeans scale\index{Jeans scale} for UBDM leads to a suppression of $P(k)$ relative to CDM at large wavenumbers. A fit for the relative suppression in $P(k)$ for UBDM with $V(\phi)=m^2\phi^2/2$ versus CDM is~\cite{hu2000}: 
\begin{eqnarray}
P_\text{UBDM}(k) = T_\text{UBDM}(k)^2P_{\rm CDM}\, , \\
T_\text{UBDM}(k) = \frac{\cos x_J^3(k)}{1+x_J^8(k)} \, , \label{eqn:fdm_transfer} \\
x_J(k)=1.61 \left(\frac{m_a}{10^{-22}\text{ eV}} \right)^{1/18}\frac{k}{k_{J,\text{eq}}} \, .
\end{eqnarray}
For the mixed CDM-UBDM system, the behaviour of $P(k)$ can also be derived~\cite{amendola2005,2010PhRvD..82j3528M}: perturbations with $k>k_m$ experience a finite amplitude suppression which increases with the ratio $\Omega_\text{UBDM}/\Omega_m$.

~\

\noindent {\bf End of Tutorial}
\end{example}

As we have just seen in the above tutorial, two effects distinguish UBDM from other ingredients in the $\Lambda$CDM model:\index{LambdaCDM@$\Lambda$CDM model} (1)~the background expansion rate, $H(z)$, driven by the transition in the equation of state\index{equation of state} $w_\text{UBDM}$ at the epoch $a_\text{osc}$, and (2)~the growth of perturbations, driven by the gradient energy in the Klein-Gordon equation\index{Klein-Gordon equation} and manifested as an effective sound speed, $c^2_\text{UBDM}$.

Depending on the value of $a_\text{osc}$, the change in $H(z)$ affects different CMB multipoles. This can be understood by considering Eqs.~\eqref{eqn:photon_eom} and~\eqref{eqn:sound_horizon_bao} in the tutorial. First consider UBDM \emph{violating} the bound Eq.~\eqref{eqn:equality_bound}. We know such UBDM must be a sub-dominant component of the DM. How does the CMB tell us this? Such UBDM changes the expansion rate \emph{after} matter-radiation equality.\index{matter-radiation equality} This changes the distance to the surface of last scattering,\index{surface of last scattering} and the angular size of the BAO\index{baryon acoustic oscillations (BAO)} in the CMB: it moves the first acoustic peak from its observed position $\ell\approx 200$. This can be compensated by a change in the value of the Hubble constant, $H_0$.\index{Hubble constant} After such a compensation there is a residual \emph{integrated Sachs-Wolfe effect}\index{Sachs-Wolfe effect} which differs from $\Lambda$CDM. If $w\neq 0$ in the post-recombination Universe, then the gravitational potential $\dot{\Phi}\neq 0$ into Eq.~\eqref{eqn:cmb_temp}. Due to the fact that the equation of state $w_\text{UBDM}\neq 0,-1$ (the two available equations of state in $\Lambda$CDM), the evolution of $\Phi$ is different in the presence of a small contribution of UBDM, and the shape of the $\ell<200$ CMB multipoles is very sensitive to the value of $\Omega_\text{UBDM}$.\footnote{This is one of the ways the CMB is used to constrain the equation of state of dark energy.\index{dark energy}} 

Now consider UBDM satisfying the bound Eq.~\eqref{eqn:equality_bound}. The change in the expansion rate compared to $\Lambda$CDM now occurs during the radiation dominated epoch. The horizon size at the time $a_\text{osc}$ was smaller than one degree on the sky, corresponding to multipoles $\ell>200$, i.e.\ the higher acoustic peaks. UBDM changes the distance scales for sound waves in the photon-baryon plasma, and alters the radiation driving term by changing the relative densities of matter (including UBDM) and radiation. These effects change the relative heights of the CMB acoustic peaks. An additional effect occurs in the \emph{diffusion damping} (Silk damping)\index{diffusion damping}\index{Silk damping} at larger multipoles, since the diffusion scale depends on the expansion rate during the radiation era. 

\begin{figure}[t]
\center
\includegraphics[width=1\textwidth]{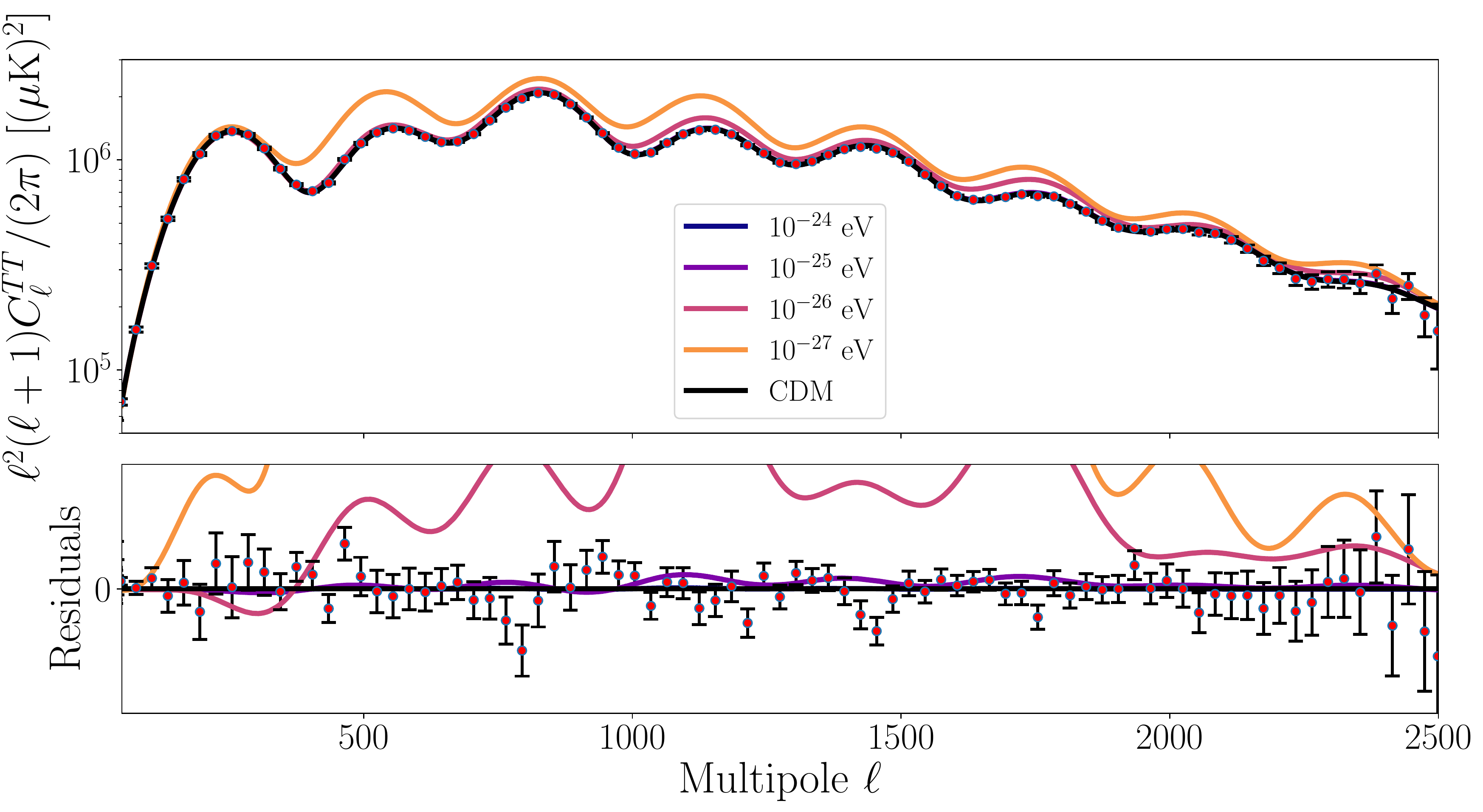}
\caption{UBDM effects on the CMB temperature power spectrum. UBDM changes the expansion rate compared to CDM in the early radiation dominated epoch, $z\gtrsim 3000$, which affects the damping of the BAO, visible through the heights of the power spectrum peaks at large multipoles. By eye, it is clear that the \emph{Planck} data strongly exclude UBDM with $m\leq 10^{-26}\text{ eV}$. Combining the temperature data with polarization, lensing, and cross-correlations~\cite{Hlozek:2017zzf}, tightens the bound to be roughly consistent with our estimate, Eq.~\eqref{eqn:cmb_bound}. On the other hand, UBDM with $m\geq 10^{-24}\text{ eV}$ is indistinguishable from the black best-fit CDM curve. Note that this plot rescales the $y$-axis in the upper panel by one power of $\ell$ compared to the usual convention, to enhance the visibility of high-$\ell$ features, and that the $x$-axis begins at $\ell=50$, since the large scales are not sensitive to this particular physics.}
\label{fig:CMB_TT_UBDM}
\end{figure}

Due to the above mentioned effects, the CMB is sensitive to the relative contribution of $\Omega_\text{UBDM}(a_\text{osc})$. However, any fluid component with $w<1/3$ becomes increasingly sub-dominant to the radiation at early times (as is the case for axion-like UBDM) and so $\Omega_\text{UBDM}$ decreases moving deeper into the radiation era.\footnote{A complex scalar with $w=1,1/3$ prior to $a_\text{osc}$ \emph{increases} its energy density relative to radiation at early times. The effect in the expansion rate is similar to adding additional neutrino species, which are also strongly constrained by the CMB~\cite{2014PhRvD..89h3536L}.} Because of this decrease in $\rho_\text{UBDM}/\rho_\gamma$, the CMB is unable to distinguish between axion-like UBDM and CDM for $a_\text{osc}\lesssim 10^{-5}$~\cite{Poulin:2018dzj}. Plugging $z=10^5$ in Eq.~\eqref{eqn:hubble} and requiring $m>H(z_\text{osc})$ gives the bound (see Fig.~\ref{fig:ch3:cosmic_history}
\begin{equation}
m > \SI{2.6e-25}{\eV} \quad \text{(primary CMB anisotropies)} \, , \label{eqn:cmb_bound}
\end{equation}
using the same reference parameters as Eq.~\eqref{eqn:equality_bound}. UBDM effects on the CMB are illustrated in Fig.~\ref{fig:CMB_TT_UBDM}. A detailed study of these effects on the \emph{Planck} CMB anisotropies constrains axion-like UBDM violating Eq.~\eqref{eqn:cmb_bound} (but satisfying Eq.~\ref{eqn:h0_bound}) to be at most a few percent of the total DM density~\cite{Hlozek:2014lca,Hlozek:2017zzf,Poulin:2018dzj}. We have spent a considerable time deriving what will turn out to be a rather weak lower bound on $m$. However, this bound is extremely rigorous in practice, in a way that our later bounds are not. The bound Eq.~\eqref{eqn:cmb_bound} relies only on linear physics, and on the extremely well understood statistics of the CMB that give us our most rigorous evidence for the existence of DM in the first place.

\begin{example}{UBDM Hints: Precision Cosmology and ALPs from the GUT Scale}
The realignment production mechanism of ALPs gives the relic density\index{relic density} $\Omega_a$ as a function of mass and initial field value, $\phi_i$. Taking $\phi_i$ to be near the GUT scale,\index{grand unification theory (GUT)} $\phi_i\in [10^{15},10^{17}]\text{ GeV}$ gives a DM relic density compatible with the observed value $\Omega_d h^2\leq 0.12$ for all masses $m \lesssim 10^{-18}\text{ eV}$. At lower masses, a sub-dominant population is predicted, with the fraction of ALP DM saturating at around 0.1\%. Upcoming cosmological surveys, including lensing tomography and intensity mapping, will greatly increase the sensitivity to sub-dominant components of the DM. The CMB is a 2D probe, and the number of modes measured with a cosmic variance precision is $\ell_\text{max}^2$. An intensity mapping survey is 3D, measuring in the line-of-sight redshift direction, and thus has many more modes. The combination of a next generation CMB survey like the Simons Observatory or CMB-S4 with an intensity mapping survey by the \emph{Square Kilometer Array}~\cite{SKA_RedBook} or HIRAX~\cite{HIRAX} could make significant inroads into the GUT scale predictions~\cite{Bauer:2020zsj}, as will next generation Lyman-$\alpha$\index{Lyman alpha forest@Lyman-$\alpha$ forest} forest surveys (see below) and ``pulsar timing arrays''~\cite{Khmelnitsky:2013lxt,Porayko:2018sfa}.\index{pulsar timing arrays} These forecasted opportunities are shown as open regions in Fig.~\ref{fig:decadal}.
\end{example}

\subsection{Schr\"{o}dinger-Poisson equations}
\label{sec:schrodinger-poisson}

The UBDM condensate\footnote{In the sense that all classical fields can be thought of as condensates.} coupled to general relativity\index{general relativity} obeys the Einstein-Klein-Gordon equations,\index{Einstein-Klein-Gordon equations} derived from variation of the relevant fundamental action. In the non-relativistic limit (Newtonian approximation)\index{Newtonian approximation}, for all forms of UBDM (be they ALPs, real, or complex scalars) these equations reduce to the Schr\"{o}dinger-Poisson equations (SPEs):\index{Schrodinger-Poisson equations@Schr\"{o}dinger-Poisson equations}
\begin{eqnarray}
i\dot{\psi}+\frac{\nabla^2}{2m}\psi-m\Phi\psi +\frac{\lambda_{\rm GP}}{m}|\psi|^2\psi=0 \, ,\label{eqn:SP_1}\\
\nabla^2\Phi = 4\pi G m^2\left(|\psi|^2-\int {\rm d}^3 x|\psi|^2 \right)\, ,\label{eqn:SP_2}
\end{eqnarray}
where we are using the convention that the Newtonian potential is dimensionless, and the field $\psi$ has canonical mass dimension one such that the average number density is:
\begin{eqnarray}
\bar{n} = m \int {\rm d}^3 x|\psi|^2\, .
\label{eqn:def_nbar}
\end{eqnarray}
The subtraction of the background density in the Poisson equation follows from the background-perturbation split of the Einstein equations on the Friedmann background.

Equations~\eqref{eqn:SP_1}--\eqref{eqn:SP_2} are a nonlinear Schr\"{o}dinger equation\index{Schrodinger equation@Schr\"{o}dinger equation} for the UBDM condensate, with Gross-Pitaevski\index{Gross-Pitaevski equation} self-coupling $\lambda_{\rm GP}$, which can be computed from the relativistic self interaction potential, $V$. The SPEs fully describe the nonlinear, non-relativistic, structure formation in most astrophysical environments at low redshifts ($a\gg a_\text{osc}$, $L\gg 1/m$, $v\ll 1$, $\Phi\ll 1$), i.e.\ the gravitational structure of UBDM at the coherence scale. One should avoid letting the name ``Schr\"{o}dinger'' cause confusion; these equations have nothing quantum about them: $\psi$ is not a probability density, and there is no measurement problem or wavefunction collapse. The SPEs are simply the non-relativistic limit of the classical field equations, valid whenever the particle number is large: they are the UBDM equivalent of Maxwell's equations.
\begin{question}{Problem 2: Derivation of the Schr\"{o}dinger-Poisson equations for UBDM}
Take the metric Eq.~\eqref{eqn:newton_line_element} in the non-relativistic limit ($\Phi=\Psi$) on a non-expanding background ($a=1$). Evaluate the d'Alembertian, Eq.~\eqref{eqn:box}, to first order in $\Psi$. Substitute the ansatz:\index{Schrodinger-Poisson equations@Schr\"{o}dinger-Poisson equations}
\begin{equation}
\phi = \frac{1}{m\sqrt{2}}\left(\psi e^{imt}+\psi^*e^{-imt} \right)\, ,
\end{equation}
into the Klein-Gordon equation\index{Klein-Gordon equation} with potential $V(\phi)=m^2\phi^2/2+\lambda\phi^4$. In the Wentzel-Kramers-Brillouin (WKB) limit,\index{WKB approximation} $\dot{\psi}/(m\psi)\ll 1$, and making the non-relativistic approximation $k/m\ll 1$, $\dot{\Psi}/m\ll 1$, show that the complex field amplitude $\psi$ obeys the Schr\"{o}dinger equation Eq.~\eqref{eqn:SP_1}. Now take the general form of the stress energy tensor, Eq.~\eqref{eqn:energy-momentum}, and show that in the same limits $\rho = |\psi|^2$ at leading order, and hence that the Poisson equation Eq.~\eqref{eqn:SP_2} is obeyed for the overdensity $\delta\rho$.

~\

{\it{Solution on page~\pageref{ch11:prob-sol:3-2}.}}

\end{question}

An instructive change of variables on the SPEs makes use of the \emph{Madelung transformation}\index{Madelung transformation}, $\psi=\sqrt{\rho}e^{i\theta}/m$ to write the wave function as a fluid with density $\rho$ and velocity $\vec{v}=\nabla\theta$. Substitution into the SPEs yields the continuity and Euler equations:
\begin{eqnarray}
\dot{\delta}_\text{UBDM} +a^{-1}\vec{v}_\text{UBDM}\cdot\nabla\delta_\text{UBDM}=-a^{-1}(1+\delta_\text{UBDM})\nabla\cdot\vec{v}_\text{UBDM} \, , \label{eqn:ubdm_conservation}\\
\dot{\vec{v}}_\text{UBDM}+a^{-1}\left( \vec{v}_\text{UBDM}\cdot \nabla\right)\vec{v}_\text{UBDM}= -a^{-1}\nabla (\Phi+Q)-H\vec{v}_\text{UBDM}\, , \label{eqn:ubdm_euler}\\
\text{where } Q \equiv -\frac{1}{2m^2 a^2}\frac{\nabla^2\sqrt{1+\delta_\text{UBDM}}}{\sqrt{1+\delta_\text{UBDM}}} \, .\label{eqn:q_def_for_eoms}
\end{eqnarray}
The continuity and Euler equations differ from those of CDM by the presence of the so-called ``quantum pressure''\index{quantum pressure}~$Q$ -- a misleading term, as it is neither quantum, nor a pressure. Expanding these equations to first order in $\delta_\text{UBDM}$ and going to Fourier space, one can to verify that they are equivalent to the fluid equation Eq.~\eqref{eqn:ubdm_growth} for pure UBDM: in the non-relativistic and linearised limit, the quantum pressure and sound speed are equivalent.

For UBDM, the SPEs replace the normal Newtonian dynamics of particle DM. 
Solving gravitational collapse and dynamics in generality requires methods of solution of nonlinear partial differential equations. The challenge in this system is the non-local interaction from the Newtonian potential, the wide range of scales in gravitational collapse, and the need to accurately resolve the phase of the field $\psi$ in low density and large cosmic voids. Common numerical methods include lattice field theory (discretizing derivatives in real space), spectral methods (numerical Fourier analysis), or finite elements (alternative real space discretizations). A public code is \textsc{pyultralight}~\cite{Edwards:2018ccc}. Particle-based hydrodynamics using Eqs.~\eqref{eqn:ubdm_conservation}--\eqref{eqn:q_def_for_eoms} is also useful on some scales, but it fails to resolve interference fringes (as can be seen from the co-ordinate singularity in $Q$ when $\rho\rightarrow 0$) and vortex lines, which appear generically in complex fields (the fluid has $\nabla\times \vec{v}=0$). On scales larger than the UBDM de~Broglie wavelength,\index{de Broglie wavelength} standard Newtonian particle mechanics is accurate e.g.\ the public code \textsc{gadget}~\cite{Springel:2005mi}. The convergence of the SPEs to the ordinary collisionless limit of CDM on super-de Broglie scales can be shown rigorously via the \emph{Schr\"{o}dinger-Vlasov} correspondence~\cite{Widrow&Kaiser1993,Uhlemann:2014npa,Mocz:2018ium},\index{Schrodinger-Vlasov correspondence@Schr\"{o}dinger-Vlasov correspondence} and is well known in the field of quantum hydrodynamics~\cite{wyatt_trajectories}.

A kinetic description of the SPEs begins by writing the field $\psi$ using the Wigner distribution\index{Wigner distribution} (see e.g.\ Ref.~\cite{ballentine_book}), which describes the occupation probability of modes~$k$. This distribution function obeys a collisional Boltzmann equation,\index{Boltzmann equation} with scattering time scale~\cite{Levkov:2018kau}: 
\begin{eqnarray}
\tau_{\rm gr} \approx \frac{\sqrt{2}}{12\pi^3}\frac{mv^6}{G^2\bar{n}^2\log\Lambda}\, ,
\label{eqn:tau_grav}
\end{eqnarray}
where $v$ is the typical speed in the system (i.e., the virial velocity)\index{virial velocity} and $\log\Lambda=\log(r_\text{max}/r_\text{min})$ is the Coulomb scattering logarithm for $r_\text{min}$ and $r_\text{max}$ the minimum and maximum length scales in the problem, respectively. This gravitational scattering time scale governs the time over which wavelike effects cause UBDM to depart dynamically from CDM.

In addition to the scattering timescale, solution of the SPEs leads to UBDM having distinctive effects on scales of order the de~Broglie wavelength.\index{de Broglie wavelength} There are three important consequences:
\begin{enumerate}
\item Transient ``quasi-particle'' fluctuations.\index{quasi-particles}
\item Formation of long-lived self-bound objects.
\item Interference fringes.
\end{enumerate} 
We discuss the first in Section~\ref{sec:galaxies}, and the second in Section~\ref{sec:axion_stars}. Interference fringes are observed prominently in numerical simulations of galactic filaments\index{galactic filaments} composed of UBDM with $m\approx 10^{-22}\text{ eV}$~\cite{Schive:2014dra,Mocz:2019emo}, though the observational consequences are at present unclear.
\subsection{Galaxies and nonlinear structure}\label{sec:galaxies}
%
%
The scale of suppression Eq.~\eqref{eqn:kJ_eq} can be converted into a DM halo\index{halo!dark matter} mass by considering the average DM density in a sphere with radius of one half wavelength, $R_J=\pi/k_{J,\text{eq}}$:
\begin{equation}
\frac{M_0}{M_\odot} = \num{5.9e9} \left(\frac{\Omega_\text{UBDM}h^2}{0.12}\right)^{1/4}\left(\frac{h}{0.676}\right)\left( \frac{1+z_\text{eq}}{3390}\right)^{3/4}\left( \frac{m}{\SI{e-22}{\eV}}\right)^{-3/2} .
\label{eqn:m0_halo_mass}
\end{equation}
Halos that are significantly more massive than $M_0$ will have the same abundance as in a CDM universe, while halos much lighter than $M_0$ are largely absent. Our estimate for $M_0$ from inspection of the linear equations of motion is within a factor of two of the suppression scale found in $N$-body simulations of nonlinear cosmological structure formation:\index{structure formation} Ref.~\cite{Schive:2015kza} finds $M_0 = \SI{1.9e10} \, M_\odot \, (m/\SI{e-22}{\eV})^{-4/3}$, where the different scaling with $m$ results from using the half-mode of the transfer function Eq.~\eqref{eqn:fdm_transfer}, $T(k_{1/2})=1/2$, instead of the Jeans scale.\index{Jeans scale} The half-mode is always at $k<k_J$ since $T(k)$ decreases below $k_{1/2}$, and $T(k_J)=0$. It is possible for structures to form at the half-mode, though they will have suppressed number with respect to CDM. The Jeans scale represents the absolute limit below which no structures form, and corresponds to lower mass halos. Thus using the Jeans scale gives more conservative limits on $m$.

\begin{figure}[t]
\center
\includegraphics[width=0.75\textwidth]{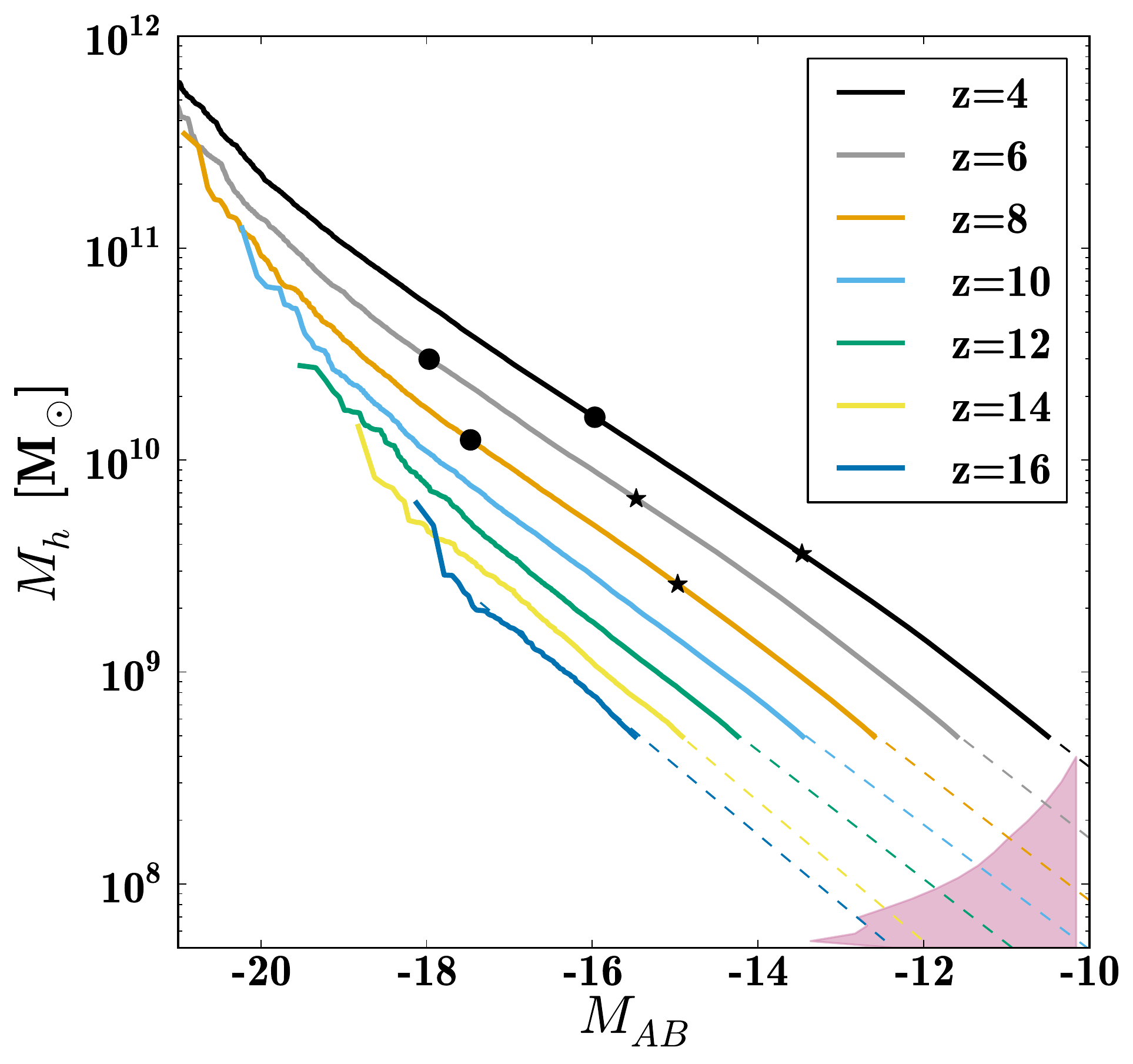}
\caption{``Abundance matching'' between halo mass, $M_h$ measured in solar masses ($M_\odot$), and UV magnitude, $M_{AB}$, assuming CDM, evaluated at different redshifts, $z$. Taken from Ref.~\cite{Schultz:2014eia}. Filled circles show the limiting magnitudes for the \emph{Hubble Ultra Deep Field} observation~\cite{Bouwens:2014fua}, while stars are for the future \emph{James Webb Space Telescope}~\cite{Windhorst:2005as}. The dotted lines represent power law extrapolation from the simulations, while the shaded region denotes the cooling limit below which galaxies cannot form efficiently.}
\label{fig:abundance}
\end{figure}
How can we constrain UBDM using our estimate for $M_0$? In hierarchical structure formation, low mass halos form first, i.e., at high redshift. Halos with low masses can be identified at high redshift from the light emitted by the galaxies that they host, which is in the form of UV flux from stars, which in turn ionizes hot gas. An approximate relationship between UV flux and halo mass can be derived by so-called abundance matching\index{halo!abundance matching}. One assumes that there is a one-to-one mapping between UV magnitude, $M_\text{AB}$ (the ``AB system'' for defining magnitude), and halo mass. This can be found assuming that the number of UV sources at some redshift $z$, $n_\text{UV}(z)$, statistically matches the number of DM halos, $n_h(z)$. The matching depends on the observations used to calibrate it, and monotonicity of each function. Current observations (e.g.\ Ref.~\cite{Bouwens:2014fua}) are largely consistent with monotonicity (however, see Ref.~\cite{Leung:2018evj}), which is consistent with all sources being in halos with mass above $M_0$. In this case, all the halos we observe are formed on scales far from the Jeans scale, and so the relationship between UV magnitude, $M_{AB}$, and halo mass, $M_h(M_{AB})$ is as in Fig.~\ref{fig:abundance} (computed from simulations of CDM with no Jeans scale). The limiting magnitude of the \emph{Hubble Ultra Deep Field} UV luminosity function at $z=8$ is $M_\text{AB,lim}\approx-18$ which we read off from Fig.~\ref{fig:abundance} as giving a limiting halo mass of $M_h\approx 10^{10}M_\odot$. Demanding $M_0>M_h(M_\text{AB,lim})$ we find the bound:
\begin{align}
m > \num{0.7e-22}\left(\frac{\Omega_\text{UBDM}h^2}{0.12}\right)^{1/6} \left(\frac{h}{0.676}\right)^{2/3}\left( \frac{1+z_\text{eq}}{3390}\right)^{1/2}\si{\eV}\nonumber\\ \hfill \text{(high redshift galaxies)}\, . \label{eqn:high-z_bound}
\end{align}
The estimate Eq.~\eqref{eqn:high-z_bound} agrees favourably with complete analyses of similar data~\cite{Bozek:2014uqa,Schive:2015kza,Corasaniti:2016epp}. 

Another important bound to consider is from the Lyman-$\alpha$ forest flux power spectrum\index{Lyman alpha forest@Lyman-$\alpha$ forest}. This observable traces the matter power spectrum,\index{matter power spectrum} $P(k)$, on quasi-linear scales at high redshifts. It can be used to infer the existence of otherwise of the UBDM Jeans scale. Current observations see no evidence for the Jeans scale, and can be used to infer that the UBDM de Broglie wavelength must be correspondingly small.

The light from distant quasars is absorbed by neutral hydrogen (HI) along the line of sight. The differing optical thickness of dense clouds of HI leads to a ``forest'' of absorption features: the optical depth for the absorption traces the HI density, and (since HI clouds lie in gravitational potential wells) the total matter density including DM. A survey of cosmological quasars can then be used to estimate the matter power spectrum\index{matter power spectrum} by correlation of the absorption feature. For example, \emph{HIRES/MIKE} covers $k$ as large as $k_\text{max}\approx 50\, h\,\si{\mega\pc^{-1}}$~\cite{Viel:2013apy}.\footnote{We convert from Lyman-$\alpha$ units for $k$ in s km$^{-1}$ to the more standard Mpc$^{-1}$ by multiplication with $H_0 (1+z)$. For reviews and discussions of the Lyman-$\alpha$ forest as a probe of the matter power spectrum see Refs.~\cite{Gnedin:2001wg,Tegmark:2002cy,Hui:2016ltb,Chabanier:2019eai}.} The data are well described by CDM with no evidence for a suppression of power, and so we can derive an approximate bound on the UBDM mass. Using Eq.~\eqref{eqn:kJ_eq} with the quoted $k_\text{max}$ gives the bound~(cf.\ Fig.~\ref{fig:ch3:cosmic_history}):
\begin{eqnarray}
m > \SI{1.5e-21}{\eV} \quad \text{(Lyman-}\alpha\text{ forest)}\, .
\label{eqn:ly-a_bound}
\end{eqnarray}
which again agrees well with the result derived from more careful analysis~\cite{Armengaud:2017nkf,Irsic:2017yje}.

Caution is advised with all our estimates on UBDM mass bounds in this section, since they assume that the observations agree perfectly with CDM, and thus that on scales observed UBDM can be treated as such. Strong self-interactions of UBDM also change these bounds, and any other bound based on the suppression of structure formation\index{structure formation} relative to CDM. A particular example of this is an ALP with large initial field displacement. The ALP potential is $V(\phi) = f_a^2 m^2 [1-\cos(\phi/f_a)]$. An initial displacement $\theta=\phi/f_a=\pi-\delta\theta$ with small $\delta\theta$ leads to large self-interactions at early times, and the field is near an unstable local maximum of the potential. This \index{tachyonic instability}{\emph{tachyonic instability}}\footnote{A potential is said to have a tachyonic region if $V''(\phi)<0$, i.e.\ a local maximum, and negative effective mass squared.} in the evolution of $\delta$ leads to an \emph{increase} in the UBDM power spectrum relative to CDM on scales close to the Jeans scale~\cite{Zhang:2017dpp}. A displacement $\delta\theta\approx 0.02$ is sufficient to evade the bound Eq.~\eqref{eqn:ly-a_bound} and allow $m\approx10^{-22}\text{ eV}$ to fit the Lyman-$\alpha$ power spectrum as well as CDM, while a value $\delta\theta\approx 0.003$ leads to a \emph{better} fit than CDM~\cite{Leong:2018opi}. The tuned values of $\delta\theta$ require smaller $f_a$ to get the correct relic abundance than in the harmonic approximation, which could make direct detection of this type of tuned UBDM easier by increasing the matter couplings.

UBDM displays dynamics distinct from CDM on scales of order the de Broglie wavelength.\index{de Broglie wavelength} A complete description of the effects of sub-de~Broglie physics requires numerical simulation. However, analytical understanding is possible in varying degrees of complexity, which has largely been developed in recent years (see e.g.\ Refs.~\cite{Hui:2016ltb,2019ApJ...871...28B,El-Zant:2019ios,Lancaster:2019mde,Marsh:2018zyw,2019MNRAS.485.2861C}). We will give only the simplest description useful for estimates.

UBDM in a gravitational potential well has a coherence length,\index{coherence length} $L\sim\lambdabar_\text{dB}=1/mv$ ($\hbar/mv$ in physical units), and coherence time\index{coherence time} $\tau \sim 1/mv^2$, where $v$ is the characteristic velocity. The heuristic picture of a wave distribution with these properties is one of quasi-particles\index{quasi-particles} of size $L$ and lifetime $\tau$. The quasiparticle mass is :
\begin{eqnarray}
M_{\rm qp}\sim \lambdabar_\text{dB}^3\bar{\rho}\, ,
\end{eqnarray}
where $\bar{\rho}$ is the average local density in a volume encompassing a large number of quasi-particles (i.e., in the solar neighbourhood, 0.4 GeV cm $^{-3}$). 
Two body relaxation between quasi-particles leads to the relaxation time (see Problem~3)~\cite{Hui:2016ltb}:\index{relaxation timescale} 
\begin{eqnarray}
t_{\rm relax} \sim \frac{10^{10}}{\log\Lambda}\left(\frac{m}{10^{-22}\text{ eV}}\right)^3\left(\frac{v}{100 \text{ km s}^{-1}}\right)^2\left(\frac{R}{5\text{ kpc}}\right)^4\text{ yr} \, ,
\label{eqn:t_relax_hui}
\end{eqnarray}
where the Coulomb logarithm in the quasi-particle picture is $\log\Lambda=\log(R/\lambdabar_\text{dB})$. On timescales longer than $t_{\rm relax}$ UBDM departs from the SHM (in the sense that the density distribution is not time-independent) due to heating and cooling. Note the similarity of the relaxation time Eq.~\eqref{eqn:t_relax_hui} to the gravitational scattering timescale Eq.~\eqref{eqn:tau_grav} in the kinetic picture if we substitute $v^2 = Gm\bar{n}R^2$.

Heating and cooling on the timescale $t_{\rm relax}$ can be observed if a tracer population of stars with mass $m_t$ is present in the UBDM halo (when the gravitational potential due to DM is dominant, stars are tracer particles). For $m_t\ll M_{\rm qp}$, heating dominates, while for $m_t\gg M_{\rm qp}$, cooling dominates. Let's estimate $M_{\rm qp}$ for some systems of interest. In the solar neighbourhood, $\bar{\rho}\approx 0.4\text{ GeV cm}^{-3}=10^7\,M_\odot\text{ kpc}^{-3}$ and $v\approx 100\text{ km s}^{-1}\Rightarrow \lambdabar=0.2 (10^{-22}\text{ eV}/m)\text{ kpc}$, which gives $M_{\rm qp}\approx 7\times 10^4 (10^{-22}\text{ eV}/m)^3 M_\odot $. In the solar neighbourhood tracers are stars with $m_t\sim 1M_\odot$, and the transition from heating to cooling occurs for UBDM mass $m\approx 4\times 10^{-21}\text{ eV}$, with lighter masses giving rise to heating. The Milky Way in fact possesses a ``thick disk'' of old stars~\cite{BinneyTremaine2008}, and this has been argued to provide evidence that in fact DM is composed of UBDM in this so-called \textit{fuzzy DM}\index{fuzzy dark matter} regime~\cite{Hui:2016ltb,2019MNRAS.485.2861C} (for more information, see the ``Fuzzy Dark Matter Hints'' box below). On the other hand, if heating is too efficient then the disk will be destroyed completely. Demanding that the relaxation time is shorter than the age of the Universe, i.e.\ $10^{10}\text{ years}$, and applying Eq.~\eqref{eqn:t_relax_hui} we find:
\begin{eqnarray}
m\gtrsim 10^{-22}\text{ eV}\, \quad \text{(Milky Way disk heating)}\, ,
\label{eqn:mw_disk_heating}
\end{eqnarray}
which agrees with more accurate modelling~\cite{2019MNRAS.485.2861C}.

A very strong bound from UBDM heating can be derived by considering the existence of the old, centrally located star cluster in the ultrafaint dwarf galaxy Eridanus~II\index{Eridanus II}. Observations~\cite{Li:2016utv,2016ApJ...824L..14C} indicate that the DM density is $\bar{\rho}=0.15 M_\odot\text{ pc}^{-3}$, and the velocity dispersion is $\sigma_v=6.9^{+1.2}_{-0.9}\text{ km s}^{-1}$. For UBDM this gives $M_{\rm qp}=3 (10^{-19}\text{ eV}/m)^3 M_\odot$, implying that heating dominates for masses $m\lesssim 10^{-19}\text{ eV}$. The star cluster has a half-light radius of $r_h= 13 \text{ pc}$, an estimated age $t\sim 10^{10}\text{ years}$, and is close to the centre of Eridanus~II. Using Eq.~\eqref{eqn:t_relax_hui}, replacing $R$ with the half-light radius (since the star cluster is approximately centrally located), $v$ with $\sigma_v$, taking $\log\Lambda\sim \mathcal{O}(1)$, and demanding that the star cluster is stable on the time scale of its age, gives the bound:
\begin{eqnarray}
m\gtrsim 10^{-19}\text{ eV} \quad \text{(Eridanus~II)}\, ,
\end{eqnarray}
which again agrees very favourably with a more rigorous treatment~\cite{Marsh:2018zyw}. 

Based on the present analysis, the bound from Eridanus~II does \emph{not}, however, apply for $m\lesssim 10^{-21}\text{ eV}$, where the fluctuation time scale becomes longer than the star cluster orbital period, and potential fluctuations become adiabatic. Another time dependent feature of UBDM halos\index{halo!dark matter} becomes important at $m\lesssim 10^{-21}\text{ eV}$: the central soliton (see Section~\ref{sec:axion_stars}) undergoes a random walk on scales of order its own radius (which is much larger than the star cluster radius in this case) due to collisions with the quasi-particles in the halo. This again leads to star cluster disruption and could exclude $m\approx 10^{-22}\text{ eV}$ from the Eridanus~II star cluster stability. However, the Milky Way tidal potential may lead to sufficient tidal stripping of the quasi-particle atmosphere to quell this random walk, and leave $m\approx 10^{-22}\text{ eV}$ safe from this bound~\cite{Schive:2019rrw}.

\begin{question}{Problem 3: Relaxation of UBDM}
The timescale for gravitational two-body relaxation (diffusion of a body's velocity caused by gravitational interaction in two-body close encounters) of particles with mass $m$ moving with velocity $v$ in a host of mass $M$ with radius $R$ can be written as~\cite{BinneyTremaine2008}:
\begin{eqnarray}
t_\text{relax} = 0.1 \, \frac{R}{v} \, \frac{M}{m\log\Lambda} \, ,
\end{eqnarray}
Use this to derive the relaxation timescale, $t_{\rm relax}$, in Eq.~\eqref{eqn:t_relax_hui}.

~\

{\it{Solution on page~\pageref{ch11:prob-sol:3-3}.}}

\end{question}

\begin{example}{UBDM Hints: Fuzzy Dark Matter}\index{fuzzy dark matter}
We have seen a large variety of constraints on UBDM with mass $m \lesssim \SI{e-22}{\eV}$ from cosmic large scale structure. We have also seen how heating in Eridanus~II excludes the range $\SI{e-21}{\eV} \lesssim m \lesssim \SI{e-19}{\eV}$, and we will see shortly that Black Hole superradiance excludes $\SI{e-19}{\eV} \lesssim m \lesssim \SI{e-16}{\eV}$. There is only one strong bound in the range just above $\SI{e-22}{\eV}$ coming from the Lyman-$\alpha$ forest flux power spectrum. This bound is sensitive to aspects of astrophysical modelling and, in particular, can be relaxed if the baryon temperature evolves non-monotonically or if significant ionizing photons are produced outside of galactic halos, e.g., in filaments (however, see the recent Ref.~\cite{Rogers:2020ltq}). Another possible window is afforded by the Eridanus~II bounds around $\SI{e-21}{\eV}$, where the statistical modelling is uncertain, and Eridanus~II can survive sandwiched between orbital resonances. If either of these bounds (Ly-$\alpha$ or Eridanus~II) can be relaxed, then there are some hints that DM may in fact be UBDM with masses between about~\SI{e-22}{\eV} and~\SI{e-21}{\eV}, the so-called \emph{fuzzy dark matter} (FDM) model~(cf.\ Fig.~\ref{fig:ch3:cosmic_history}).\index{fuzzy dark matter} These hints include:
\begin{itemize}
\item {\textbf{The Milky Way ``thick disk''}}: FDM just outside the bound Eq.~\eqref{eqn:mw_disk_heating} can help explain the old thick disk in our galaxy~\cite{2019MNRAS.485.2861C}.
\item {\textbf{Suppressed high-$z$ galaxy formation}}: The redshift of reionization is known to be around $z_{\rm reion}\approx 8$. This relatively low value naturally explained by FDM, which suppresses formation of galaxies at $z\gtrsim 8$. 
\item {\textbf{Solitons and galactic cores}}: Solitons in FDM halos (see Section~\ref{sec:axion_stars}) may help explain cored density profiles in dwarf galaxies without baryonic feedback~\cite{Schive:2014dra,Marsh:2015wka}. 
\item {\textbf{Relic density}}: The relic density\index{relic density} is naturally explained by an FDM ALP with $f_a$ close to the GUT scale, as expected in certain string compactifications.\index{string theory}
\end{itemize} 
Each hint provides a method to search for FDM. Furthermore, The FDM mass range corresponds to field oscillation frequencies of order one inverse month, making it challenging, but not impossible, to search for via direct detection.

\end{example}

\subsection{Black hole superradiance}
\label{sec:BHSR}

In the following we adopt different units: so-called geometric units\index{geometric units} where $G_N=c=1$.

Spinning black holes (BHs) are described by the Kerr metric\index{black hole}\index{Kerr metric}, which has two parameters: mass, $M$, and dimensionless spin $a_J = J/M\in [0,1]$. In ``Boyer-Linquist'' coordinates\index{Boyer-Linquist coordinates} the line element is:\footnote{An accessible introduction to general relativity\index{general relativity} can be found in Ref.~\cite{2004sgig.book.....C}.}
\begin{eqnarray}
{\rm d}s^2_{\rm Kerr} &=& - \left( 1 - \frac{2Mr}{\Sigma}\right){\rm d}t^2 - \frac{4M a_Jr \sin^2 \theta} {\Sigma}{\rm d}t {\rm d}\phi + \frac{\Sigma}{\Delta}{\rm d}r^2 + \Sigma {\rm d}\theta^2 \nonumber \\
&+& \frac{(r^2+a_J^2)^2 - a_J^2 \Delta \sin^2\theta}{\Sigma}\sin^2\theta {\rm d}\phi^2 \,, \label{eqn:kerr_metric}\\
\Sigma &\equiv& r^2 + a_J^2\cos^2 \theta\,,	\\
\Delta &\equiv& r^2 + a_J^2 - 2Mr\,, \label{eqn:Kerr_delta} \\
r_{\pm} &\equiv& M \pm \sqrt{M^2 - a_J^2}\,. \label{eqn:Kerr_rpm} \\
r_\text{ergo}&\equiv&M+\sqrt{M^2-a_J^2\cos^2\theta}\, ,
\end{eqnarray}
where we use spherical polar coordinates. The zero solutions of Eq.~(\ref{eqn:Kerr_delta}) define the two horizons $r_\pm$: an inner Cauchy (causal) horizon\index{horizon!Cauchy}\index{Cauchy horizon} at $r_{-}$, and the outer physical event horizon\index{horizon!event}\index{event horizon} at $r_+$\index{event horizon}. The ``ergoregion''\index{ergoregion} is defined as radii smaller than $r_\text{ergo}$, where $g_{00}=0$ (the co-efficient of $dt^2$ in the line element). If an object enters the ergoregion between $r_+<r<r_\text{ergo}$, and ejects some mass which falls into the event horizon, then the object will emerge from the ergoregion with a larger energy than it went in with, and the BH will lose a small amount of energy in the form of mass and spin. This is known as the Penrose process.\index{Penrose process} 

A wavepacket has a finite extent, and can ``eject'' part of itself into the BH if it passes through the ergoregion and overlaps with the event horizon. If the wave is trapped near the BH, then this process continually extracts energy from the BH, growing the wavepacket amplitude and becoming ``superradiant.''\index{supperadiance!black hole}\index{black hole superradiance} The process only ends when the ergoregion has shrunk small enough to remove the overlap (ultimately, the process must stop if $a_J=0$, i.e., a Schwarzschild BH with no ergoregion). Such a situation is in fact realized naturally for a massive bosonic field. Gravitational bound states trap the field near the BH, and the hydrogen-like wavefunctions overlap with the superradiant region between the ergosphere and the event horizon. The field in question must be bosonic in order that the wavepacket energy levels can continue to be filled as energy is extracted. ``Black hole superradiance''\index{black hole superradiance} (BHSR) for bosonic fields is discussed in detail in Refs.~\cite{2011PhRvD..83d4026A,2015LNP...906.....B}.

Consider a scalar field near a Kerr BH. Just like in the tutorial on cosmic structure above, the field obeys the Klein-Gordon equation,\index{Klein-Gordon equation} Eq.~\eqref{eqn:KG_eq}, except that now the d'Alembertian~($\Box$)\index{d'Alembertian} should be evaluated with the metric Eq.~\eqref{eqn:kerr_metric}. Let us write the field as
\begin{eqnarray}
\phi = \sum_{\ell,\alpha}e^{-i\omega t+i\mu\varphi}S_{\ell \mu}(\theta)\psi_{\ell \mu}(r)+\text{h.c.}\, ,
\end{eqnarray}
where $S_{\ell\mu}(\theta)$ are the spheroidal harmonics\index{spheroidal harmonics} (eigenfunctions of the Laplacian on the surface of a spheroid, respecting the axial symmetry of the Kerr spacetime). To avoid confusion, we have labeled the magnetic quantum number $\mu$ and the azimuthal angle $\varphi$. The Klein-Gordon equation can then be reduced to a time-independent Schr\"{o}dinger equation\index{Schrodinger equation@Schr\"{o}dinger equation} for the radial eigenfunctions $\psi_{\ell\mu}$, with eigenvalue $\omega$. The BH provides a background potential $V(r,\omega)$, which possesses a barrier separating the bound states from the horizon, and a potential well with size of order the boson Compton wavelength, $1/m$. The system resembles a hydrogen atom with effective fine structure constant $\alpha_\text{eff} \equiv G_N M m$, where we temporarily re-instated~$G_N$.

The existence of superradiant solutions is determined by the imaginary part of the eigenvalue $\omega$, which leads to growth of the occupation number of the mode $\psi_{\ell\mu}$. The superradiant rate is $\Gamma_{\rm SR}\propto \alpha_\text{eff}^{4\ell+4}m$, and numerically it is found to be maximised around $\alpha_\text{eff} \sim 1$. This gives an approximate criterion for BHSR:\index{black hole superradiance}
\begin{eqnarray}
m \sim \frac{8\pi M_{pl}^2}{M} = 1.33\times 10^{-10}\text{ eV} \left(\frac{1M_\odot}{M}\right)\, . \label{eqn:sr_condition}
\end{eqnarray}
For BHSR to be effective, the superradiant time scale should be longer than any timescale of relevance for the BH, e.g.\ accretion. If BHSR is effective, then the BH will lose spin. Thus large observed values of $a_J$ will be disfavoured if a boson exists satisfying Eq.~\eqref{eqn:sr_condition}. 

Astrophysical observations indicate the existence of BHs across a wide range of masses, from those formed by collapse of stars at the Chandrasekhar limit\index{Chandrasekhar limit} $M\approx 1.4 M_\odot$, to the supermassive BHs (SMBHs)\index{black hole!supermassive} at the centres of galaxies. The spins of BHs can also be estimated, using X-ray spectroscopy of the accretion disk, or by measurement of the gravitational waveform in the inspiral phase of binary systems. Detectable spins are generally large, $a_J\gtrsim 0.5$. Assuming that these large values would be disfavoured by a boson satisfying Eq.~\eqref{eqn:sr_condition} we can estimate exclusions on UBDM. First consider the stellar BHs,\index{black hole!stellar} and assume a full spectrum of observations from the Chandrsekhar mass to the LIGO inspiral masses $M\approx 30 M_\odot$\index{Laser Interferometer Gravitational-Wave Observatory (LIGO)}\index{LIGO}~\cite{Abbott:2016blz}. This excludes UBDM for:
\begin{eqnarray}
\SI{4e-12}{\eV} < m < \SI{8e-11}{\eV} \quad \text{(stellar BHs)} \, .
\label{eqn:bhsr_stellar}
\end{eqnarray}
Next, consider SMBHs. The lightest currently known SMBH is in NGC4051, with mass $M\approx 1.9\times 10^6 M_\odot$, while the \emph{Event Horizon Telescope}\index{Event Horizon Telescope} has imaged the BH at the centre of M87 and determined the mass $M\approx 6.5\times 10^9 M_\odot$. Again, assuming a continuous spectrum in between we can exclude the range of UBDM masses:
\begin{eqnarray}
\SI{2e-20}{\eV} < m < \SI{7e-17}{\eV} \quad \text{(supermassive BHs)} \, .
\label{eqn:bhsr_smbh}
\end{eqnarray}
These estimates agree somewhat favourably with more accurate treatments of BHSR modelling and BH population statistics~(e.g.\ Ref.~\cite{Stott:2018opm}).

To obtain the more accurate picture, the bosonic field equations on the Kerr background should be solved numerically. The oscillation time scale of the field is $\tau\sim 1/m$. For real scalar fields the gravitational pressure oscillates with a frequency $2m$, sourcing oscillations of the metric potentials on a time scale faster than the superradiant timescale. This makes brute force numerical solution challenging, but many approximation methods are available. 

BHSR also works for massive spin-one and spin-two fields (which are also UBDM candidates). The superradiant timescales can be vastly different, and specific treatments are necessary. Reference~\cite{Dolan:2018dqv} considers spin-one vectors which have much smaller instability rates, and thus weaker constraints. Reference~\cite{Brito:2013wya} considers spin two fields, which possess a particular mode mimicking the spin zero case, and thus have similar constraints. A significant difference occurs for complex fields. Due to the underlying $\mathbb{U}(1)$ symmetry and conserved particle number, the complex vector $A_\mu$ field does not source oscillations in the metric potentials with frequency $m$. This greatly simplifies the numerical task, and has allowed direct simulation of superradiance with these so-called Proca fields~\cite{East:2017ovw}\index{Proca field}. The simulations are important because they include nonlinear back-reaction of the superradiant cloud on the Kerr spacetime and demonstrate that BHSR occurs in this more realistic setting.

One known ``showstopper'' for BHSR is the so-called ``Bosenova''\index{Bosenova} caused by attractive quartic self-interactions, which shut off the instability and prevent growth of the scalar cloud. The self-interaction term in the potential is $V_\text{int}=\lambda\phi^4/4!$, for some coupling constant $\lambda$. As the cloud grows, this term can become as large as the other terms in the energy budget. At this time, the scalar cloud collapses and superradiance is shut off. This introduces a new timescale into the problem and practically gives rise to a maximum $\lambda$ above which the superradiance rate is subdominant to the Bosenova rate and no spin extraction can occur. Numerical simulations~\cite{Yoshino:2012kn} determine the maximum cloud occupation number before Bosenova occurs~\cite{Arvanitaki:2014wva}: 
\begin{eqnarray}
N_\text{Bose} \sim 150\frac{n^4}{\alpha_\text{eff}\lambda} = \num{5e78}\frac{n^4}{\alpha_\text{eff}^3}\left(\frac{M}{10M_\odot}\right)^2\left(\frac{f_a}{M_{pl}}\right)^2 \, ,
\label{eqn:BHSR_Nbose}
\end{eqnarray}
where $n$ is the energy level of the occupied cloud and $M_{pl}$ is the reduced Planck mass. In the last equality we assumed that the scalar potential is of the ALP form $V(\phi)\propto -\cos(\phi/f_a)$, giving $\lambda=m^2/f_a^2$. Using this formula for stellar mass BHs, Ref.~\cite{Arvanitaki:2014wva} finds that BHSR is shut off for $f_a \lesssim \SI{e13}{\GeV}$; for SMBHs this turns out to be $f_a \lesssim \SI{e16}{\GeV}$. 

Any UBDM interactions can compete with superradiance, and possibly shut it off. Examples include interactions between the cloud and the Standard Model particles in the BH environment or the ALP interaction $g_{a\gamma\gamma}$, which leads to stimulated decay of the cloud~\cite{Rosa:2017ury,Ikeda:2018nhb}. Of course, both ``showstoppers'' (Bosenova and axion-photon interactions) also predict new observables in the form of emission from BH regions for UBDMs of particular masses. Finally, we note that the superradiance phenomenon need not be limited strictly to BHs, and can occur also near stars and neutron stars~\cite{Day:2019bbh} -- even though the astrophysical uncertainties are far greater.

\begin{question}{Problem 4: Estimating superradiance properties of UBDM}
A simple way to estimate the relevance of BHSR is to inspect terms in the action,\index{black hole superradiance}
\begin{eqnarray}
	S = \int {\rm d}^4 x \, \sqrt{-g} \left[ \frac{M_{pl}^2}{2} R - \frac{1}{2}(\partial\phi)^2-\frac{1}{2}m^2\phi^2+\frac{\lambda}{4!}\phi^4\right]\, ,
\end{eqnarray}
where $R$ is the Ricci~scalar of the Kerr background~metric~$g$ and $m$, $\lambda$, and $\phi$ are the UBDM mass, (dimensionless) self-coupling, and field value, respectively. Note that it is useful to re-instat $M_{pl}$~(or $G_N$) for this exercise. Assuming a suitable setup in which superradiance indeed occurs, estimate both the superradiance and Bosenova conditions i.e.\ Eqs~\eqref{eqn:sr_condition} and~\eqref{eqn:BHSR_Nbose}. Note the similarity between your estimate of $N_\text{Bose}$ and Eq.~\eqref{eqn:BHSR_Nbose} when $\lambda = m^2/f_a^2$.

~\

{\it{Solution on page~\pageref{ch11:prob-sol:3-4}.}}

\end{question}

\begin{example}{UBDM Hints: LIGO and the QCD Axion}
The exclusion estimates, Eqs.~\eqref{eqn:bhsr_stellar}--\eqref{eqn:bhsr_smbh}, assumed continuous BH distributions between the minimum and maximum values. In reality, the distributions are of course incomplete. In fact, this can serve as a discovery tool for UBDM. If light bosons with particular masses exist, then the observed BH mass and spin distribution should contain forbidden regions, and astrophysical BHs should cluster along superradiant ``trajectories'' in the $(m,a_J)$ plane. Gravitational wave observations will, over time, provide a very complete survey of this plane. Furthermore, superradiant clouds emit their own gravitational waves\index{gravitational waves} due to level transitions and annihiliation. From these effects, the LIGO observatory provides a discovery channel for UBDM with $10^{-13}\text{ eV}\lesssim m\lesssim 10^{-12}\text{ eV}$~\cite{Arvanitaki:2016qwi}. This region is disfavoured by current measurements of BH spins, but the excluded region is determined by the uncertainty on BH masses with a small number of measurements. Thus there is the possibility to make discoveries with precise measurements and greater statistics. The accessible mass region for LIGO corresponds to the QCD axion with $f_a\sim M_{pl}$. For proposed GW detectors in lower frequency bands corresponding to higher mass BHs (e.g., Laser Interferometer Space Antenna, LISA),\index{Laser Interferometer Space Antenna (LISA)}\index{LISA} discovery potential moves to lower UBDM masses.\index{gravitational waves}
\end{example}

\subsection{Summary of gravitational constraints}
\begin{figure}[t]
\centering
\includegraphics[width=12cm]{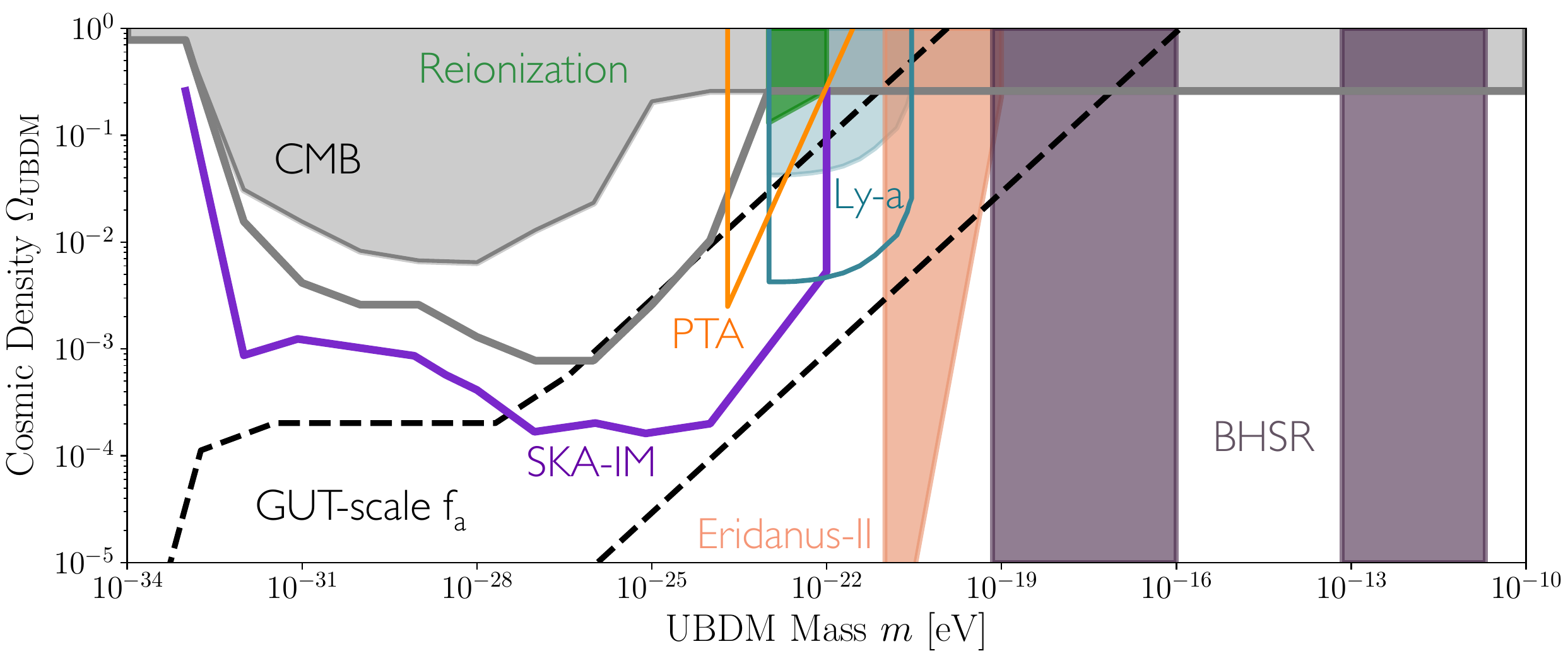}
\caption{Summary of gravitational constraints (shaded) on UBDM, and forecasts (open) for upcoming surveys. Constraints assume a real scalar with potential $V(\phi)=m^2\phi^2/2$, see text for clarification on generalizing the bounds. CMB: cosmic microwave background~\cite{Hlozek:2017zzf,Hlozek:2016lzm}, PTA: pulsar timing array~\cite{Khmelnitsky:2013lxt}, BHSR: black hole superradiance~\cite{Stott:2018opm}, Ly-a: lyman alpha forest~\cite{Kobayashi:2017jcf,Aghamousa:2016zmz}. SKA-IM: Square Kilometer Array intensity mapping~\cite{Bauer:2020zsj}. Adapted from Ref.~\cite{Grin:2019mub}.}
\label{fig:decadal}
\end{figure}

Current constraints on UBDM mass and cosmic density from the CMB, galaxy formation, relaxation, and black hole superradiance, are combined in Fig.~\ref{fig:decadal}, along with a selection of forecasts for upcoming surveys. They cover an astonishing 24~orders of magnitude in mass and place sub-percent constraints on the density parameter. We caution that the limits apply strictly only to scalar UBDM with $w_\text{UBDM} = -1$ in the early Universe and negligible self-interactions, e.g., ALPs and similar cases. However, the limits apply by order-of-magnitude to all UBDM, particularly if they come from non-relativistic effects where model dependence is less important. In addition to the effects discussed in detail, we also show projections for the measurement of pulsar timing arrays\index{pulsar timing arrays} (PTA) with the \emph{Square Kilometer Array}~\cite{Khmelnitsky:2013lxt}. Current bounds from this technique~\cite{Porayko:2018sfa} are not yet at the $\mathcal{O}(1)$ level for $\Omega_\text{UBDM}$, and so do not appear.
%
\section{Axion compact objects}

ALP UBDM can form two different types of gravitationally bound objects which are distinct from ordinary DM galactic halos. These objects, miniclusters\index{miniclusters} and axion stars,\index{axion!star}\index{star!axion} are interesting phenomenologically since they are far denser than galactic halos. They can thus have observational effects as sources of enhanced DM decay and conversion, gravitational lensing,\index{gravitational lensing} or in direct detection if they happen to pass through the Earth.

\subsection{Axion stars}\label{sec:axion_stars}

There exist several classes of (pseudo-)solitonic solutions to the Einstein-Klein-Gordon equations.\index{Einstein-Klein-Gordon equations} These solutions go by many names, and have been discovered and re-discovered many times. They date back to Wheeler's idea of a ``geon'':\index{geon} a wave confined to a finite region by gravity, thus mimicking a lump of matter. Ruffinni and Bonnazola~\cite{1969PhRv..187.1767R} found explicit ``boson stars''\index{boson star}\index{star!boson} as time-independent fixed particle number state solutions for a complex scalar field coupled to general relativity: these are true solitons,\index{soliton} stabilised by the existence of the conserved $\mathbb{U}(1)$ scalar field charge. Solutions also exist for a real scalar field. However, in this case there is no conserved charge and instead the solutions have a time-dependent metric, and are known as ``oscillatons''~\cite{1991PhRvL..66.1659S}.\index{oscillatons} We could continue with the soliton bestiary for some time, but instead we will focus on the most well-motivated class of these objects: axion stars\index{axion!star}.\footnote{To continue the bestiary just a little further, solutions are named for all scalar fields: inflaton stars, moduli stars, Higgs stars, etc.}

First, consider the fully relativistic case. We are interested in time-dependent solutions for a scalar field coupled to general relativity.\index{general relativity} A public code is \textsc{GRChombo}~\cite{Clough:2015sqa}.\footnote{\url{http://www.grchombo.org/}} Like all stars, axion stars are stabilised by a balance between attraction (gravity, and axion quartic self-interactions) and repulsion (gradient pressure, and higher order interactions).\footnote{The axion potential is $V(\phi)=m^2f_a^2[1-\cos\phi/f_a)]$. Taylor expanding this we find that the $\phi^4$ term is attractive, while higher order terms alternate in sign.} Initial conditions are found solving the boundary value problem on the initial spacetime volume (hypersurface), and evolved forward in time to investigate their stability. The solutions are a two parameter family in mass, $M$, and axion decay constant, $f_a$, giving a ``phase diagram'' that can be explored numerically~\cite{Helfer:2016ljl}. 

The structure of the axion star phase diagram is easy to understand. As the mass of the star increases, the central value of the field $\phi_0$ also increases. There are two possible instabilities, and which wins depends on $f_a$. For large $f_a$, the self-interactions can be neglected. Now the ordinary GR lore applies: collapse to a BH\index{black hole} at large mass. At low $f_a$, the axion has strong self-interactions, and these also drive collapse. Collapse increases $\phi_0$ further until higher order repulsive interactions take over and expel relativistic axions from the collapsing core in an ``axion nova''~\cite{Levkov:2016rkk},\index{axion!nova} which occurs at critical mass $M_\text{nova} = 10.4 \, M_{pl} f_a/m g_4$, where $g_4$ is the coefficient of quartic interactions equal to unity for a cosine potential. For small $f_a$ the restoring interactions become important earlier during collapse, and bring the star back to a stable configuration with only slightly lower mass than before the nova. As $f_a$ increases, it takes more and more of the mass of the star to contract and reach the repulsive core, thus expelling a larger mass in the nova, and reducing the mass of the stable remnant. The two types of instability are divided by a particular value of $f_a$. As $f_a \rightarrow \infty$, oscillatons and boson stars are found to be unstable when $\phi_0\sim M_{pl}$ (this defines the ``Kaup mass'', $M\sim M_{pl}^2/m$), while self-interactions become important when $\phi_0\sim f_a$, and so the boundary between the two unstable regions occurs for $f_a\sim M_{pl}$. A third phase boundary exists between the nova and BH regions, which simulations have found to be fractal in structure~\cite{Michel:2018nzt}. It is not clear this boundary could be reached by any astrophysical process, and so it is likely only a mathematical curiosity. The ``triple point'' between all three phases is found numerically to be near $(M,f_a)=(2.4 M_{pl}^2/m,0.3 M_{pl})$, where $M$ is the ``Arnowitt-Deser-Misner''\index{Arnowitt-Deser-Misner mass} mass~\cite{1959PhRv..116.1322A}.\footnote{Due to co-ordinate transformations, mass is not a straightforward quantity to define in general relativity (indeed, sometimes it is not defined). The Arnowitt-Deser-Misner mass is defined in the Hamiltonian formulation of general relativity, and is essentially the conserved mass measured in the infinite future.}

Non-relativistic axion stars are far simpler to study: in the non-relativistic limit the real scalar field possesses an effective conserved particle number. In this case the solutions are simply referred to as solitons\index{soliton} and the results apply generically to UBDM in the non-relativistic limit. Solitons are stationary waves of the form $\psi(r,t) = M_{pl}\chi(mr)e^{-i\gamma m t}$, where $\chi$ is a dimensionless function giving the radial profile, and $\gamma$ is the energy eigenvalue. An important property of the SPEs (see Sec.~\ref{sec:schrodinger-poisson}) is their \emph{scaling symmetry}: 
\begin{eqnarray}
(t,x,\psi,\Phi)\rightarrow (\lambda^{-2}t,\lambda^{-1}x,\lambda^2\psi,\lambda^2\Phi)\, ,
\label{eqn:scale_symm}
\end{eqnarray}
where $\lambda$ is the scale parameter (not to be confused with the quartic interaction strength in Section~\ref{sec:BHSR}). The boundary value problem normalised to $\chi(0)=1$, $\lambda=1$, can be solved numerically and the results are fit by eigenvalue $\gamma=-0.692\lambda^2$ and radial density profile:
\begin{eqnarray}
\frac{\rho_{\rm sol} (r)}{m^2M_{pl}^2} = \chi^2(mr) = \frac{1}{[1+(0.230mr)^2]^8}\, .
\label{eqn:soliton_fit}
\end{eqnarray}
These solutions are the ground state of the SPEs. They are a balance of the nonlinear and non-local gravitational force in the Poisson equation, and the dispersive effect of the gradient energy term in the Schr\"{o}dinger equation. Soliton dynamics can be studied using the numerical methods already discussed. In the limit of vanishing self-interactions, the soliton solutions are a one-parameter family given by the mass, $M$. Thanks to the scaling symmetry, we only need to find the solution once, and then scale it using $\lambda$ (see Problem~5).

How might axion stars form in astrophysical environments? Two mechanisms are seen in simulation of the SPEs. Which occurs depends on the scale, $R$, of the gravitational fluctuations compared to the de Broglie wavelength: 
\begin{itemize}
\item Direct collapse: $R\sim \lambda_\text{dB}$ (e.g.\ Ref.~\cite{Schive:2014dra}).
\item Kinetic condensation: $R\gg \lambda_\text{dB}$ (e.g.\ Ref.~\cite{Levkov:2018kau}).
\end{itemize}

Direct collapse leads to rapid formation of axion stars on the gravitational free-fall time, and by definition occurs in the smallest objects near to the cut-off scale of gravitational fluctuations, i.e., $M\sim M_0$, Eq.~\eqref{eqn:m0_halo_mass}. This mechanism leads to an axion star in the centre of all DM halos close to the cut-off scale. If all mergers of this first generation are complete up to the largest scale of halos observed, then the numerically determined relationship between the star mass, $M_\star$, and the halo mass, $M$, is:
\begin{equation}
M_\star \propto \left(\frac{M}{M_0}\right)^{1/3}M_0\, ,
\label{eqn:core_halo}
\end{equation}
where the constant of proportionality can be found in Ref.~\cite{Schive:2014hza}, and depends on the definition of $M_0$. This relationship is believed to derive from a combination of the virial theorem, equilibrium between the soliton and its gravitationally bound ``atmosphere,'' and universal mass growth in the merger history of solitons~\cite{Du:2016aik}. Slow growth of solitons by accretion leads to significant scatter in the relation.\footnote{Very recently some authors have even found a different best-fit exponent~\cite{Mina:2020eik,Nori:2020jzx}, and numerical convergence may also play a part. The issue is not yet resolved at the time of writing.}

The direct collapse mechanism is particularly relevant to the formation of solitonic cores in dwarf galaxies in the FDM regime (see hint box above), and the formation of axion stars in miniclusters (discussed below). Axion stars formed by this mechanism are in virial equilibrium with their environment for $t<t_{\rm relax}$, and do not change appreciably in mass over such scales. The surrounding halo is a hot ``atmosphere'' for the star. The constant interaction with the halo causes the star to undergo radial oscillations at the normal mode frequencies~\cite{Veltmaat:2018dfz}.

Kinetic condensation gives rise to axion star formation in regions much larger than the de Broglie wavelength, for example in the solar neighbourhood for the QCD axion. The scattering timescale thermalizes the distribution function on time scales of order $\tau_{\rm gr}$, Eq.~\eqref{eqn:tau_grav}, and at this time the local ground state is found in the form of an axion star\index{axion!star}\index{star!axion} which condenses spontaneously. Axion stars formed in this way continue to grow over time as they swallow up matter from the environment, with $M\propto (t/\tau_{\rm gr})^p$. The index $p$ is to be determined numerically, and will evolve slowly in time with the wave distribution function. The growth process will eventually slow down when the star grows a gravitationally bound ``atmosphere,'' at which point it should enter a local virial equilibrium solution close to Eq.~\eqref{eqn:core_halo}. 

Despite progress in our understanding of the formation and growth of axion stars, at the time of writing their abundance and galactic distribution is not fully understood even in benchmark models. The problem is partly one of scale: we do not know the mass above which the relation Eq.~\eqref{eqn:core_halo} breaks down and halos have no central soliton, but instead grow many small solitons in the kinetic regime.

Axion stars have a host of possible phenomenological consequences:
\begin{itemize}
    \item {\textbf{Galactic cores}}: Solitons composed of Fuzzy DM with $m\sim 10^{-22}\text{ eV}$ may help explain flat central densities in Milky Way dwarf satellites (tracer stars reside within the soliton)~\cite{Schive:2014dra,Marsh:2015wka}, or central mass excesses (tracer stars outside the soliton). See Hint box above for more details.
    \item {\textbf{Direct detection}}: The passage of axion stars through Earth, though rare, will greatly enhance the signal in a direct search, and could be identified using a co-ordinated network of detectors like the Global Network of Optical Magnetometers to search for Exotic physics (GNOME)\index{Global Network of Optical Magnetometers to search for Exotic physics (GNOME)}\index{GNOME} and GPS.DM~\cite{JacksonKimball:2017qgk,Derevianko:2013oaa}.\index{GPS.DM}
    \item {\textbf{Indirect detection}}: The high axion density creates a larger radio signal from decay and conversion of axions into photons~(see Section~\ref{sec:axion-photon-astro}). Cataclysmic signals could arise if the stars can reach the critical mass for an axion nova or stimulated decay due to interactions.
    \item {\textbf{Relativistic axion stars}}: If dense enough axion stars can be formed, they may show up as ``Exotic Compact Objects'' in gravitational wave detectors~\cite{Giudice:2016zpa} and multi-messenger astronomy~\cite{Dietrich:2018jov}.\index{multi-messenger astronomy}
\end{itemize}
\subsection{Miniclusters}
\label{sec:miniclusters}

A second special class of UBDM compact objects is formed by the process of spontaneous symmetry breaking,\index{spontaneous symmetry breaking} if this occurs during the normal course of thermal evolution of the Universe (as opposed to during the initial conditions epoch, inflation or otherwise). The Peccei-Quinn (PQ) phase transition\index{Peccei-Quinn phase transition} (see Chapter~2) occurs when the temperature of the Universe drops below approximately $f_a$. Recall that we write the complex PQ field as $\varphi = Re^{i\theta}$, and spontaneous symmetry breaking occurs when the field $R$ takes on a vacuum expectation value. The following scenario applies specifically to ALPs where the field $R$ is heavy and unstable (such that it decays at late times), while the field $\theta$ is initially massless, but acquires a mass hierarchically smaller than the mass of $R$ and at some time much later than the time of PQ symmetry breaking.

When PQ symmetry breaking occurs, $R$ takes on a non-zero vacuum expectation value, and thus $\theta$ must also be specified. Since the axion field is massless at symmetry breaking, the only terms in the Lagrangian are proportional to $\partial\theta$, meaning there can be no preferred value for $\theta$. The axion thus takes on a random value on essentially all scales. Imagine a pencil falling over from its point: in the absence of an external preference, the pencil falls in a random direction specified by an angle, $\theta$, with the $\theta=0$ axis arbitrary.

\begin{figure}[t]
\center
\includegraphics[width=0.9\textwidth]{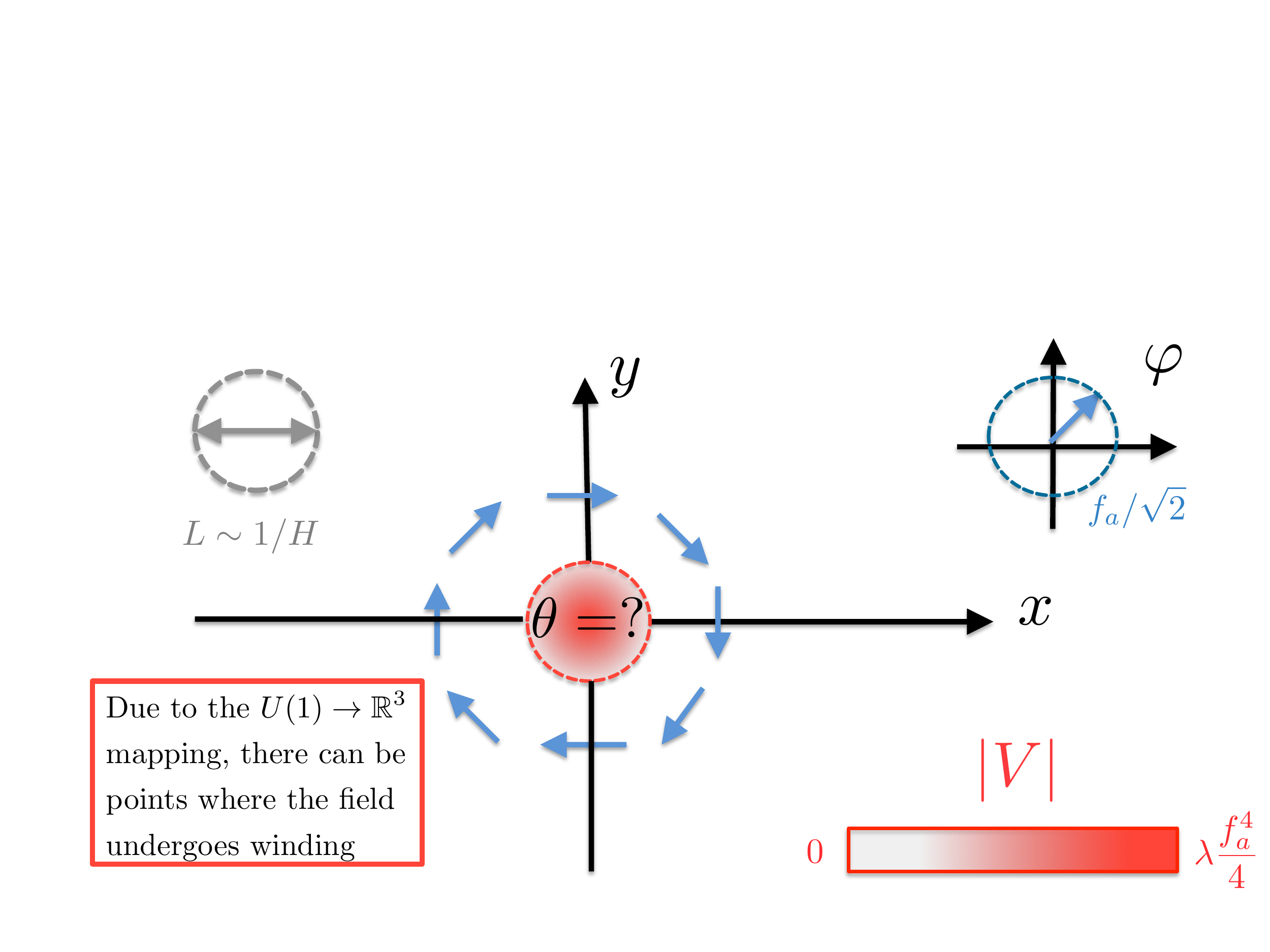}
\caption{Formation of axion strings from spontaneous symmetry breaking. The complex plane of the PQ field $\varphi$ is illustrated in the upper right corner with the radius marked where $R$ takes the vacuum expectation value, and the circle along which the axion field is defined. When spontaneous symmetry breaking occurs, this circle is mapped onto the co-ordinates in the real plane, $(x,y)$. The complex phase of the field (i.e.\ the value of the axion field) is shown by the orientation of the arrow. Complete wrappings of the field lead to defects where the phase is not defined. In this two-dimensional case, the defects are points, in thee dimensions they are lines, i.e.\ strings. In the centre of the string, the PQ field has $\langle\varphi\rangle=0$ and the potential takes the value indicated by the colour bar. Damping of sub-horizon modes $k>aH$ in Eq.~\eqref{eqn:Kibble_mode} smooths fluctuations on length scales $L\sim H^{-1}$. }
\label{fig:string_sketch}
\end{figure}

First, consider the simpler two-dimensional case, illustrated in Fig.~\ref{fig:string_sketch}. Because the PQ field is a continuous function (as all fields must be), for any random configuration of a complex field there will be points in space around which $\theta$ makes a complete wrapping. At the wrapped point, the axion field $\theta$ is undefined (imagine shrinking the circle to a point: at the point the circle must have zero size, and $\theta$ takes every value at once). The only way that this can be possible is if $R=0$ at the wrapped point. The point in the complex field space where the radial co-ordinate is zero indeed has undefined phase. As long as the complete windings of $\theta$ persist, then at the centre of these windings $R$ must remain at the origin, and thus symmetry breaking cannot happen. When the potential is $V(\varphi)=\lambda (|\varphi|^2-f_a^2/2)^2$ this implies that the potential at the origin is $V(0)=\lambda f_a^4/4$, and this is the value of the potential at the centre of a point around which $\theta$ wraps. 

In fact, in three dimensions, $\theta$ cannot wrap just a single point or else the field would be discontinuous. The field must wrap continuous one-dimensional lines (either infinitely long or in closed loops) known as cosmic strings,\index{cosmic strings} and in this particular case as axion strings, or global strings (since the symmetry breaking is of a global $\mathbb{U}(1)$)\index{cosmic string}.\footnote{This is generic for complex fields in three spatial dimensions. It is a topological property. Complex fields have symmetry group $\mathbb{U}(1)$ of rotations in the complex plane, i.e.\ loops. The mapping of $\mathbb{U}(1)$ onto $\mathbb{R}^3$ (Euclidean 3-space) is expressed by the first homotopy group\index{homotopy group} $\pi_1(\mathbb{R}^3)$. This group is not the empty set, i.e.\ it is non-trivial, which can be seen by noting that $\mathbb{R}^3$ is the universal cover of $T^3$, the 3-torus, and $\pi_1(T^n)=\mathbb{Z}^n$. This last can be seen since one cannot shrink circles on tori to points continuously, and there are $n$ distinct circles wrapping $T^n$.} This leads to the existence of PQ strings: continuous one dimensional structures around which $\theta$ makes complete windings and where the radial field is pinned at $R=0$. The strings are a class of topological soliton:\index{soliton!topological}\index{topological defect} localised field configurations stabilised by the topology of the field space. The formation mechanism is known as the \emph{Kibble-Zurek mechanism},\index{Kibble-Zurek mechanism} and is observed experimentally in condensed matter phase transitions with the same symmetries, for example the transition from normal fluid to superfluid helium~\cite{1976JPhA....9.1387K,1985Natur.317..505Z}.

What happens to the axion field? The equation of motion for the axion field in Fourier space is:
\begin{eqnarray}
\ddot{\tilde{\theta}}_k+3H\dot{\tilde{\theta}}_k+(k^2/a^2+m(T)^2)\tilde{\theta}_k=0\, ,\label{eqn:Kibble_mode}
\end{eqnarray}
where we are careful to distinguish between the field $\theta$ and the modefunction $\tilde{\theta}_k$: $\tilde{\theta}_k=0$ does not imply $\theta=0$ as a preferred value, only that the mode $k$ is absent from its spectrum and thus gradients of $\theta$ on the spatial scale $1/k$ are small. At early times, the QCD axion mass, $m(T)$\index{axion!mass} is vanishingly small comapred to $H$ and can be neglected.

Imagine initially that all modes are populated with some amplitude (for example, the inflationary fluctuations of the PQ field), and then the field configuration far from any string is allowed to evolve. Any mode ``inside the horizon'' has $k^2\gg a^2H^2$. These modes will undergo damped oscillation, and decay in amplitude. Modes larger than the horizon, $k^2\ll a^2H^2$ remain pinned to their initial value by the friction term (co-efficient of $\dot{\tilde{\theta}}_k$) in Eq.~\eqref{eqn:Kibble_mode} given by $3H$ (Hubble friction).\index{Hubble friction} Thus high frequency modes decay and low frequency modes remain static, smoothing the field on scales of order the horizon size.\footnote{Note that the mode function $\tilde{\theta}_k$ decaying to zero does not imply a preference for the axion field $\theta$ to move to zero: in the massless limit the shift symmetry\index{shift symmetry} prevents any such preference.} Around any string, the axion field is wound $\theta\in (-\pi,\pi]$, and so we have $\mathcal{O}(1)$ variation of the field on horizon size patches around the string. String formation is sketched in Fig.~\ref{fig:string_sketch}. Furthermore, numerical simulations indicate that string dynamics enter into a scaling solution with $\mathcal{O}(1)$ strings per horizon volume.

Strings decay when the axion mass becomes relevant to the mode evolution. Recall first that during cosmological expansion temperature, $T$, always decreases as time, $t$, increases. Second, recall that the axion mass and the Hubble rate $H$ are both decreasing functions of $T$. The mass term (co-efficient of $\tilde{\theta}_k$) in Eq.~\eqref{eqn:Kibble_mode} is comparable to the friction term when $m(T)\approx H(T)$. The axion mass term defines a preferred value for the field, $\theta=0$, which is exactly why the PQ mechanism solves the strong-CP problem. Eq.~\eqref{eqn:Kibble_mode} is just a damped harmonic oscillator, and so the mode functions for all $k<aH$ (not just the short wavelength modes inside the horizon) will begin to oscillate when $m(T)\approx H(T)$, defining the special temperature $T_\text{osc}$. Now, everywhere, the axion field is making harmonic oscillations about zero. There are thus no longer regions around which it makes a complete and continuous winding. The axion field everywhere has average value zero (but importantly of course, nonzero variance and energy density). This means that the radial mode is no longer required to take the value $R=0$ along the strings. The axion field is everywhere defined, the radial field is not pinned, and it can undergo symmetry breaking at the string locations, i.e.\ the strings decay (or ``unwind''). 

\begin{figure}[t]
\center
\includegraphics[width=0.75\textwidth]{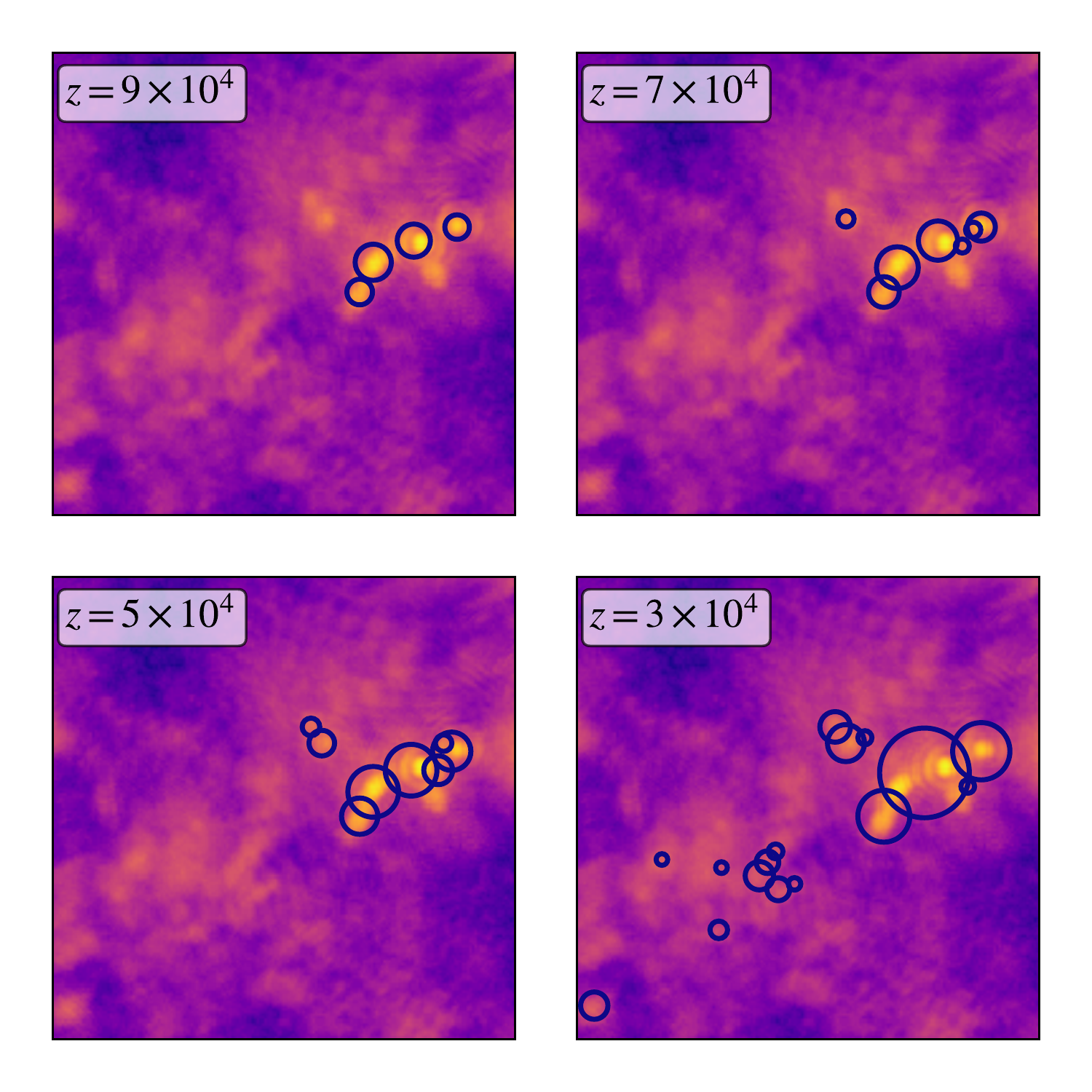}
\caption{Initial conditions for minicluster formation. After string decay, the axion field has large density perturbations, which subsequently collapse into the objects known as miniclusters. The figure shows a small patch of results from the numerical simulations of Ref.~\cite{Vaquero:2018tib}, which used lattice field theory methods to solve the equations of motion for the complex Peccei-Quinn field in the absence of gravity. The large hierarchies involved necessitate further approximations, and different simulation methods are not currently in precise agreement for the spectrum of perturbations extrapolated to physical values of the particle masses and couplings. Miniclusters are identified at different redshifts using a threshold derived from spherical collapse under gravity~\cite{Ellis:2020gtq}.}
\label{fig:PeakPatchEvo}
\end{figure}

At this time the axion field has a well specified distribution: in every horizon-size patch, it has $\mathcal{O}(1)$ fluctuations, while on larger scales it is uncorrelated. The power spectrum, $P(k)$, is flat (white noise) for $k\ll a(T_\text{osc})H(T_\text{osc})=k_J(T_\text{osc})$ and cut off by the Jeans scale\index{Jeans scale} for $k\gg a(T_\text{osc})H(T_\text{osc})$. The normalisation of the power spectrum is fixed by the variance, which should match the variance of the uniform distribution for $\theta$, $\langle \theta^2\rangle=\pi^2/3$. Just prior to $a(T_\text{osc})$ the axion equation of state is $w\approx-1$, and so the fluctuations in $\theta$ do not source any curvature perturbations in the metric:\footnote{Intuitively this can be understood because if $w=-1$ exactly then this is equivalent to a cosmological constant,\index{cosmological constant} which is constant in space and time, and thus cannot source a spatially varying curvature.} this is what is meant by the term \emph{isocurvature}.\index{isocurvature} It is this particular power spectrum (white noise isocurvature, with $\mathcal{O}(1)$ variance, truncated at the horizon size at $T_\text{osc}$), which gives rise to the structures known as \emph{axion miniclusters} as a remnant of string decay~\cite{1988PhLB..205..228H}. Figure~\ref{fig:PeakPatchEvo} shows a snapshot from numerical simulation of the axion field after string decay~\cite{Vaquero:2018tib}, and miniclusters are located using a threshold based on spherical collapse under gravity~\cite{Ellis:2020gtq}.

The mass scale of miniclusters is the same as the mass scale of ordinary axion halos: it is fixed by the Jeans scale when field oscillations begin, Eq.~\eqref{eqn:m0_halo_mass}, but now for very different values of the reference parameters. However, because of the large amplitude of these isocurvature fluctuations (variance of order unity), axion miniclusters begin to collapse earlier than ordinary DM halos and there is significant nonlinear structure formation before matter-radiation equality.\footnote{Contrast this to the evolution of large scale inhomogeneities seen in the CMB, and the inflationary ``adiabatic'' mode. These perturbations have small amplitude, and a red slope in the power spectrum leading to smaller amplitude fluctuations on small scales. These fluctuations undergo logarithmic growth at early times in the radiation era, which is important to seed galaxy formation and is one of the pieces of evidence for DM discussed in Chapter~1. However, due to the existence of a free-streaming scale/Jeans scale for particle DM models, there is, in general, no nonlinear structure formation during the radiation era. For WIMPs with $m\lesssim 1\text{ TeV}$, structure formation in the adiabatic mode begins at $z\approx 500$ with the formation of Earth mass, $10^{-6}M_\odot$, halos~\cite{Green:2003un}.} The density of any DM halo is related to the background density at the time when it first collapses and leaves the Hubble flow.\index{Hubble flow} Thus, miniclusters are denser than ordinary halos, and may survive repeated mergers up to the present day. Let us consider the phenomenology of these low mass, dense objects.

First, we need the minicluster mass which we estimate from the number density of axions within the comoving cosmological horizon at the time when field oscillations begin. We find $T_\text{osc}$ in the usual way, setting $H(T_\text{osc})=A_1m_a(T_\text{osc})$.\footnote{$A_1$ is simply a constant of proportionality to account for ambiguity defining $T_\text{osc}$, and its later use in analytical formulae for the evolution of the energy density. Our earlier choices, e.g.\ Eq.~\eqref{eqn:h0_bound}, set $A_1=1$, physically assuming structure formation begs when the de Broglie wavelength is equal to the Hubble length, $H^{-1}$. Many authors choose $A_1=3$ when estimating the relic density. This ambiguity in the use of $T_\text{osc}$ leads to significant uncertainty in analytical minicluster mass estimates, which can only be resolved by fitting results of numerical simulations.} Now we need to calculate the horizon volume. Take the volume of rotation of the spherical wave with comoving wavenumber\index{wavenumber} $k_\text{osc}=a(T_{\rm osc})H(T_\text{osc})$ over one-half wavelength:
\begin{eqnarray}
V(k) = \int {\rm d}^3x j_0(kr)=\frac{4\pi}{(ak)^3}\int_0^{\pi}{\rm d}y\, y\sin(y) = \frac{4\pi^2}{(ak)^3} \equiv \frac{V_H}{(ak)^3}\, .
\end{eqnarray}
This defines the Hubble volume $V_H$.\index{Hubble volume} Alternative definitions are the cubic volume, $V(k)=(ak)^{-3}$, and the spherical volume of of one half wavelength $V(k)=\frac{4}{3}\pi (\pi/ak)^3$. The expected minicluster mass is simply $M_\text{MC} = V(a_\text{osc}k_\text{osc}) \Omega_a\rho_\text{crit}$. We compute $M_\text{MC}(T_\text{osc})$ using the Friedmann equation to fix $k_\text{osc}$, $3H^2M_{pl}^2=(\pi^2/30)g_{\star}T^4$, and conservation of entropy to write $a(T)\propto T^{-1}g_{\star S}$ (normalised using the CMB measurement of $z_\text{eq}$). The result is well fit by:
\begin{eqnarray}
M_\text{MC}(T_\text{osc}) = \num{9.2e-13} \, V_H \, \left(\frac{T_\text{osc}}{2\text{ GeV}}\right)^{-3} \mathcal{S}\left(\log_{10}\left(\frac{T_\text{osc}}{\si{\GeV}}\right)\right) \, M_\odot\, , \\
\mathcal{S}(x) \equiv 0.5[1+\tanh 4(x- 8.2)]+1.3[1+\tanh 4(8.2-x)]\, ,
\label{eqn:minicluster_mass_tosc}
\end{eqnarray}
where $\mathcal{S}(x)$ is an activation function that accounts for the behaviour of $g_{\star}$ and $g_{\star,S}$ in the Standard Model (dominantly the quark-hadron phase transition) and has only been roughly fit here using the results for $g_\star$ from Ref.~\cite{Wantz:2009it}. Note that this expression is valid for any ALP with minicluster-like initial conditions.

From the QCD axion $m_{a,\text{QCD}}(T)$ dependence based on lattice QCD in Ref.~\cite{Borsanyi:2016ksw}, $T_\text{osc}(m_{a,\text{QCD}})$ is well fit by:\footnote{For the present purposes a simple power law $m\propto T^{-4}$, matched to the zero-temperature result $m_{a,\text{QCD}} \equiv m_{a,\text{QCD}}(0) = \SI{5.72}{\micro\eV}(\SI{e12}{\GeV}/f_a)$ at $T_{\rm QCD}=\SI{140}{\MeV}$ is accurate enough.}\index{axion!mass}
\begin{eqnarray}
T_\text{osc}(m_{a,\text{QCD}}) = 2 \, \left(\frac{m_{a,\text{QCD}}}{100A_1\,\si{\micro\eV}}\right)^{0.165}\text{ GeV}\, ,
\label{eqn:t_osc_qcd_axion}
\end{eqnarray}
over the range of interest (broadly $\SI{e-5}{\eV}\leq m_{a,\text{QCD}} \leq \SI{e-3}{\eV}$) for the relic density in this scenario~(see also Fig.~\ref{fig:ch3:cosmic_history}). For smaller axion masses there there is an $\mathcal{O}(1)$ change in the constant at the front of Eq.~\eqref{eqn:t_osc_qcd_axion}, while the power law remains approximately the same.

Now we would like to know the minicluster density profile. Kolb and Tkachev~\cite{Kolb:1994fi} wrote down the equation of motion for spherical collapse of an isolated top-hat density profile, with initial overdensity $\delta$, in an expanding Universe dominated by radiation. The perturbation first grows in size as the Universe expands. It then turns around, and collapses. Spherical collapse formally leads to a density singularity. However, in real collapses below the threshold for BH formation, aspherical perturbations lead to virialization\index{virialization} (equilibrium between average kinetic and gravitational potential energy expressed by the virial theorem, see, e.g., Ref.~\cite{sigl}) and the collapsed object becomes self-supported with a finite average density. virialization occurs when the radius of the perturbation is half of the turn-around radius. Using this information, one can compute the overdensity of the spherical system at the time of virialization. A numerical solution of the ordinary differential equation for spherical collapse gives the final average overdensity:\footnote{For the more standard case during matter domination, which applies to ordinary DM halos, see, e.g., Ref.~\cite{peacock}. In the standard case, the equations can be solved analytically, leading to the well-known result that the virial overdensity $\langle\rho_f\rangle/\bar{\rho}\approx 200$, independent of $\delta$.}
\begin{eqnarray}
	\langle \rho_f \rangle = 140 \bar{\rho}_\text{eq} \delta^3(1+\delta).
\end{eqnarray}
Assuming the minicluster has a constant density, the radius is then calculated to be
\begin{eqnarray}
	R_\text{MC} = \left(\frac{3M_\text{MC}}{4 \langle \rho_f \rangle} \right)^{1/3} \, .
\end{eqnarray}
Using $ \bar{\rho}_\text{eq} = 2\times10^{14} M_{\odot} \text{kpc}^{-3}$
\begin{eqnarray}
	R_\text{MC} = 3 \times 10^{-10}\,\text{kpc} \, \frac{1}{\delta (1+\delta)^{1/3}} \, \left( \frac{T_{\mathrm{osc}}}{2 \text{GeV}} \right)^{-1} \, .
\label{eqn:minicluster_radius}
\end{eqnarray}

To make more use of this result, we need to know the minicluster radial profile, $\rho(r)$. Collapse of isolated density perturbations is self-similar, and leads to a power law density profile with no preferred scale. Miniclusters are not isolated, and the formation proceeds much like ordinary DM halos, leading to Navarro-Frenk-White~(NFW) profiles~\cite{Zurek:2006sy,Gosenca:2017ybi,Eggemeier:2019khm}.\index{Navarro-Frenk-White profiles} It is then natural to associate the radius $R_{\mathrm{MC}}$ with the NFW scale radius. If the initial distribution of $\delta$ could be measured, one would know the mass and size distribution of miniclusters.

In reality, the problem of miniclusters is far more complex than the simple story given here. Firstly, miniclusters do not have one fixed mass. Structure formation always proceeds hierarchically, and there is a mass function of miniclusters. This can be computed numerically via N-body simulation, or semi-analytically from the initial power spectrum~\cite{Enander:2017ogx,Fairbairn:2017sil,Eggemeier:2019khm,Ellis:2020gtq}. The mass function takes on a power law spreading over many orders of magnitude around $M_\text{MC}$. Secondly, the distribution of $\delta$ in initial conditions is not uniquely defined. Numerical thresholding using the spherical collapse results allows some progress to be made~\cite{Ellis:2020gtq} but the results still require calibration to N-body simulation. Unfortunately, N-body simulations cannot currently resolve the scale radius on all relevant scales, and are not large enough to capture the rarest, densest, and thus most phenomenologically interesting miniclusters.

Finally, just like other UBDM halos, when the effects of the gradient energy (the UBDM de Broglie wavelength) are included, miniclusters have been shown to form central axion stars~\cite{Eggemeier:2019jsu}. Axion stars\index{axion!star} in miniclusters follow approximately the same core-halo mass relation, Eq.~\eqref{eqn:core_halo}, as an ordinary halo. For the QCD axion, the resulting axion stars are on very different mass scales for the UBDM particle mass and the halo mass than the reference FDM values in Eq.~\eqref{eqn:core_halo}. 

The minicluster power spectrum, mass function, size function, and central axion stars, can all be used to constrain the QCD axion and ALPs in the post-inflation PQ symmetry breaking scenario. Some examples include:
\begin{itemize}
    \item Microlensing and femtolensing~\cite{Kolb:1995bu,Katz:2018zrn,Fairbairn:2017dmf} (see Problem~5 below).\index{gravitational lensing}
    \item Radio signals from minicluster-neutron star collisions~\cite{Tkachev:2014dpa,Iwazaki:2014wta,Bai:2017feq}.
    \item The CMB isocurvature power spectrum and large scale structure~\cite{Hardy:2016mns,Feix:2020txt}.
\end{itemize}

\begin{question}{Problem 5: Microlensing constraints on UBDM}
Show that the scaling symmetry, Eq.~\eqref{eqn:scale_symm}, is a symmetry of the SPEs. Use this relationship and the profile, Eq.~\eqref{eqn:soliton_fit}, to write down the mass-radius relation in units of solar masses $(M_\odot)$ and kiloparsecs. The \emph{Subaru Hyper Suprime Cam} (HSC)\index{Subaru Hyper Suprime Cam (HSC)}\index{HSC} microlensing survey of M31 probes PBHs in the range $10^{-11}M_\odot$ to $10^{-6}M_\odot$~\cite{Niikura:2017zjd}. The Einstein radius for gravitational microlensing, is $R_E=2[GM_\star x(1-x)d_s]^{1/2}$, where $d$ is the distance from the observer to the lens, $d_s$ is the distance from the observer to the source, and $x=d/d_s$. Compare the axion star radius to $R_E$ with $x=1/2$ and $d_s=770\text{ kpc}$, the distance to M31. What UBDM particle masses could be probed by HSC lensing due to axion stars? 

Now consider the mass-radius relation for miniclusters with initial overdensity $\delta$, Eq.~\eqref{eqn:minicluster_radius} and the minicluster mass relation Eq.~\eqref{eqn:minicluster_mass_tosc}. What range of the $(T_\text{osc},\delta)$ parameter space can be probed by microlensing?

~\

{\it{Solution on page~\pageref{ch11:prob-sol:3-5}.}}

\end{question}

%

\section{Indirect detection of UBDM}
\label{sec:astro-couplings}

\subsection{Stellar and supernova energy loss}

In this section we consider only constraints on axion-like couplings, i.e., pseudoscalar, anomalous, or shift symmetric~(see Chapter~2). Analogous bounds can of course be derived for scalar and dilaton-like couplings. Axions are pseudoscalars, and their couplings to fermions depend on the orientation of the spin, while couplings of scalar particles do not. The spin~dependence can lead to suppression of interactions, since macroscopic bodies are not in general strongly polarized. Being unsuppressed by spin effects, scalar constraints are often stronger. More details on some of the calculations are given in Chapter 5. 

First and foremost, it is \emph{extremely important} to remember that the constraints and effects we discuss in this section apply \emph{independently} of whether the axion is~(a large fraction of) the DM. The axions considered here are produced from Standard Model particles in stellar plasmas. They interact only very weakly and have a long mean-free path inside the plasma. Thus, stars are effectively transparent to axions, and the axions escape, allowing an additional cooling channel for the star.\index{stellar cooling} This changes the evolution of stars: in simple terms, it alters the progression of stars along the Hertzsprung-Russell~(HR) diagram\index{Hertzsprung-Russell diagram} of stellar luminosity versus temperature. The relationship between the mass, age, and temperature of stars is thus different than in the Standard Model. Stellar physics is generally very well understood in terms of Standard Model physics alone, and can be simulated using a code such as \textsc{mesa}~\cite{mesa}, which can be modified to include axion-induced cooling~\cite{Friedland:2012hj}. For more details see Refs~\cite{Raffelt:1990yz,1990eaun.book.....K,2008LNP...741...51R}.

Stars, including the Sun, can produce axions by the Primakoff process:\index{Primakoff effect} photons inside the star convert into axions in the ambient magnetic and electric fields of the particles (electrons and nuclear ions) in the plasma. The rate for this process is:
\begin{eqnarray}
\Gamma_{\gamma\rightarrow a} = \frac{g_{a\gamma\gamma}^2 T\kappa_s^2}{32\pi}\left[\left(1+\frac{\kappa_s^2}{4E^2}\right)\ln \left( 1+\frac{4E^2}{\kappa_s^2} \right)-1 \right]\, ,
\end{eqnarray}
where $E$ is the photon energy, $T$ is the temperature, and $\kappa_s$ is the screening length. In the Debye-H\"{u}ckel approximation we have
\begin{eqnarray}
\kappa_s^2 = \frac{4\pi\alpha}{T}\left(n_e+\sum_j Z_j^2n_j\right)\, ,
\end{eqnarray}
where $n_e$ is the free electron density, and $n_j$ is the density of the $j$-th nuclear ion of charge $Z_j$. In a neutral medium with $n_e=n_j=0$ the Primakoff rate goes to zero, since there is no background field to facilitate conversion. 

Photon energies are distributed thermally, and the temperature varies with stellar radius. Kinematically, we require $E\geq m$ to produce an axion. Typical stellar interior temperatures are in the keV range, which gives the typical energy of the emitted axions, and approximately the maximum axion mass where this cooling channel is allowed. The luminosity in axions needs to be computed for a given stellar model. Applying this to the Sun gives $L_a = 1.85\times 10^{-3}(g_{a\gamma\gamma}/10^{-10}\text{ GeV}^{-1})^2L_\odot$. It is this solar luminosity in axions that \emph{helioscope}\index{helioscope} experiments try to detect (see Chapter 5). We can derive a crude bound by demanding that the solar axion luminosity must be less than unity, since the evolution of the Sun is well described by emission dominantly in photons, i.e.\ $L_{\gamma,{\rm Sun}}=1L_\odot$ by definition. Thus:
\begin{eqnarray} 
g_{a\gamma\gamma}<2.3\times 10^{-9}\text{ GeV}^{-1}\quad \text{(luminosity of the Sun)}\, .
\label{eqn:basic_stellar_g}
\end{eqnarray}

The bound in Eq.~\eqref{eqn:basic_stellar_g} can be improved by considering the statistics of populations of stars. The best understood case is for horizontal branch (HB) stars\index{horizontal branch stars} in globular clusters. Stars in globular clusters\index{globular clusters} are all of a similar age, and differ in their masses. The distribution of the stars gives an HR diagram that can be compared to models. The observable is the ratio of HB stars to red giant branch~(RGB) stars,\index{red giant branch stars} $R$, determined by placing stars on a colour-magnitude diagram. In the Standard Model, this ratio is a function of the primordial helium abundance, $Y_{\rm He}$, stellar mass, and metallicity. Globular clusters are old systems, with ages in the range of 10~billion years. This gives a small range of available stellar masses and metallicities, which have a negligible effect on $R$. The value of $Y_{\rm He}$ can determined observationally by measurement of extragalactic H~II regions which gives $Y_{\rm He}=0.2449\pm 0.0040$~\cite{Aver:2015iza}, which is consistent with the predictions of standard BBN and the CMB measurement of the baryon abundance~\cite{Aghanim:2018eyx}. Reference~\cite{Ayala:2014pea} reports a measured average value of $R=1.38$ from 39 globular clusters, consistent with the Standard Model prediction.

With a given stellar evolution model it is possible to compute the effect of the axion-photon coupling, $g_{a\gamma\gamma}$, and the axion-electron coupling, $g_{aee}$, on $R$. At present there are two different models in the literature for the functional dependence, and each is presented in Ref.~\cite{Giannotti:2015kwo}. The specific forms are not enlightening, so we simply quote the bounds (derived in Ref.~\cite{Hoof:2018ieb}). In both cases the additional cooling channel lowers $R$ compared to the Standard Model prediction leading to a degeneracy in the combined constraints, with the maximum value of one coupling only allowed when the other is strictly zero. Setting one coupling to zero, and fixing $Y_{\rm He}=0.25$, the individual bounds are:
\begin{alignat}{2}
g_{a\gamma\gamma} &< 4.95 \; (9.56) \times 10^{-11}\,\text{GeV}^{-1} \quad &&\text{(95\% CL, HB/RGB stars)}\, ,
\label{eqn:hb_stars_g} \\
g_{aee} &< 2.95 \; (3.53) \times 10^{-13} \quad &&\text{(95\% CL, HB/RGB stars)} \, ,
\end{alignat}
where the number in brackets refers to the bound using the alternative model for $R$, which we see introduces an $\mathcal{O}(1)$ shift in the bound on $g_{a\gamma\gamma}$. For the constraints in the combined parameter space, see Refs.~\cite{Giannotti:2015kwo,Hoof:2018ieb}. 

Supernova SN1987A\index{supernova!SN1987A} provides an important bound on the axion nuclear couplings, $g_{aNN}$ and $g_d$. During the core collapse process, a proto-neutron star is formed, the gravitational field of which traps neutrinos and causes them to be emitted over a delayed period of time. This model explains the duration of the burst of two dozen observed neutrinos coincident with SN1987A. Axion emission due to nuclear bremsstrahlung:\index{bremsstrahlung}
\begin{alignat}{2}
N + N &\rightarrow N + N + a \quad &&(\text{$g_{aNN}$ coupling}) \, , \\
N + \gamma &\rightarrow N + a \quad &&(\text{$g_d$ coupling}) \, . 
\end{alignat}
would compete with neutrino emission and cool SN1987A too rapidly, shortening the neutrino burst, unless the total energy loss rate from either axion-nuclear process obeys the bound $\varepsilon_a \lesssim 10^{19}\,\text{erg g}^{-1}\text{s}^{-1} = \SI{7.2e-18}{\eV}$. 

For the first process, the cooling rate per unit mass is~\cite{2008LNP...741...51R}
\begin{eqnarray}
\varepsilon_a = \frac{1}{\rho}\left(\frac{C_N}{2f_a}\right)^2\frac{n_N}{4\pi^2}\int_0^\infty{\rm d}\omega \,\,\omega^4S_\sigma=\left(\frac{C_N}{2f_a}\right)^2\frac{T^4}{\pi^2m_N}F \, ,
\label{eqn:SN_nuclear_Brem}
\end{eqnarray}
where $\rho$ is the mass density of the supernova, $n_N$ is the nucleon number density, $\omega$ is the axion angular frequency, and $S_\sigma$ is the spin density structure function, which accounts for the fact that axions couple only to the nuclear spins. The coupling $C_N$ is the nucleon coupling weighted as $C_N^2=Y_nC_n^2+Y_pC_p^2$, where $Y_n$ and $Y_p$ are the neutron and proton abundances, estimated as $Y_p = 0.3$ and $Y_n = 1 - Y_p = 0.7$ at the relevant epoch in the supernova.

The spin density structure function is non-trivial to compute, but can be estimated in various approximations. The last equality in Eq.~\eqref{eqn:SN_nuclear_Brem} defines the dimensionless function $F$ from the integral of $S_\sigma$, which is estimated to be of order unity, and allows a simple estimate of the bound given the supernova internal temperature $T\approx 30\text{ MeV}$. A recent analysis found the more accurate bound~\cite{Chang:2018rso}
\begin{eqnarray}
\frac{C_N}{2 f_a} < \SI{1.3e-9}{\GeV^{-1}} \quad \text{(SN1987A neutrino burst)} \, ,
\end{eqnarray}
a factor of approximately four weaker than the estimate with $F=1$. The same analysis found an $\mathcal{O}(1)$ effect from the modelling of supernova temperature and density profiles. The bound from SN1987A on $C_N/f_a$ is particularly important for the QCD axion, since this couplings is always present, and so the bound can be cast as a model-independent constraint on the QCD axion mass. We use that $C_{KSVZ}^2=0.066\Rightarrow C_{KSVZ}=0.257$, leading to:
\begin{eqnarray}
m_{a,\text{QCD}}\lesssim 0.06\left(\frac{0.257}{C_N}\right)\text{\,eV} \quad \text{(SN1987A neutrino burst)}\, .
\end{eqnarray}

For the second process $\varepsilon_a$ is approximated by:
\begin{eqnarray}
\varepsilon_a = \frac{\Gamma}{V\rho} \approx \frac{\langle E_\gamma\rangle n_N n_\gamma\langle\sigma v\rangle}{\rho}\, ,
\end{eqnarray} 
where $n_{i}$ are the reactant number densities, $\langle E_\gamma\rangle$ is the average photon energy, $\rho$ is the supernova mass density, and $\langle\sigma v\rangle$ is the thermally averaged cross section. The nuclear number density and supernova mass density are known, and the other parameters are fixed in terms of the internal temperature, $T$. To estimate the bound on $g_d$, Ref.~\cite{Graham:2013gfa} approximates the cross section as $\langle\sigma v\rangle=g_d^2T^2$, leading to:
\begin{eqnarray}
g_d < 4\times 10^{-9}\text{ GeV}^{-2}\, .
\end{eqnarray}

\begin{example}{UBDM Hints: Anomalous White Dwarf Cooling}
White dwarfs~(WDs)\index{white dwarf} are stellar remnants whose electron-degenerate cores are supported by Fermi pressure against gravitational collapse. Their internal densities are relatively high as their masses are typically comparable to the mass of the Sun~($\sim 0.6\,M_\odot$), while their radii are of the order of the Earth's radius~\cite{10.1086/156841}. They cannot replenish their internal energy, and therefore continuously cool down over time.

The evolution of WDs can be altered by introducing additional cooling channels.\index{stellar cooling} These can be provided by weakly-interacting particles that efficiently carry away energy after being created in, and escaping from, the WD's core. A useful observable to infer the resulting, additional cooling rate is the so-called period increase in variable WDs. These are WDs that periodically change in brightness over time as they pulsate due to non-radial excitations, called ``gravity modes''~(see e.g.\ Ref.~\cite{Corsico:2019nmr}),\index{gravity modes} with potentially multiple pulsation periods associated with different co-existing sub-modes. For cooling WDs, the periods of their pulsations, $\Pi$, tend to increase over time with a rate $\dot{\Pi} = \mathrm{d} \Pi/\mathrm{d} t$ that can approximately be calculated via
\begin{equation}
	\frac{\dot{\Pi}}{\Pi} \approx -\frac{1}{2} \, \frac{\dot{T}}{T} + \frac{\dot{R}}{R} \approx -\frac{1}{2} \, \frac{\dot{T}}{T} \, , \label{eq:period_increase}
\end{equation}
where $T$ and $R$~are the internal temperature and radius of the WD, respectively~\cite{10.1038/303781a0}. The change in radius can usually be neglected for the observed, low-luminosity dwarfs~\cite{10.1086/163398}. Axions and ALPs induce an energy loss that is proportional to $g_{aee}^2$ since axion-electron interactions dominate in the high-density, electron-degenerate interior of the WDs~\cite{Nakagawa:1987pga,Nakagawa:1988rhp}. The resulting decrease in temperature, and therefore the additional contribution to the period increase in Eq.~(\ref{eq:period_increase}), is hence also proportional to $g_{aee}^2$.

Measurements of an anomalous period change can thus be used to estimate the associated axion-electron coupling. From the 250 known variable WDs~\cite{10.1093/mnras/stz2571}, this has so far only been done for G117-B15A \cite{Isern:1992gia,Altherr:1993zd,BischoffKim:2007ve,Isern:2008nt,Corsico:2012ki}, R548 \cite{BischoffKim:2007ve,Corsico:2012sh}, L19-2~\cite{Corsico:2016okh}, and PG~1351+489~\cite{Battich:2016htm}. The reason for this small fraction is that measuring the period change is very difficult: while the periods for the WDs listed here are of the order of a few minutes, their (inherently dimensionless) period changes, $\dot{\Pi}$, are less than about $10^{-13}$ in magnitude.

\begin{figure}
	\centering
	\includegraphics[width=0.99\linewidth]{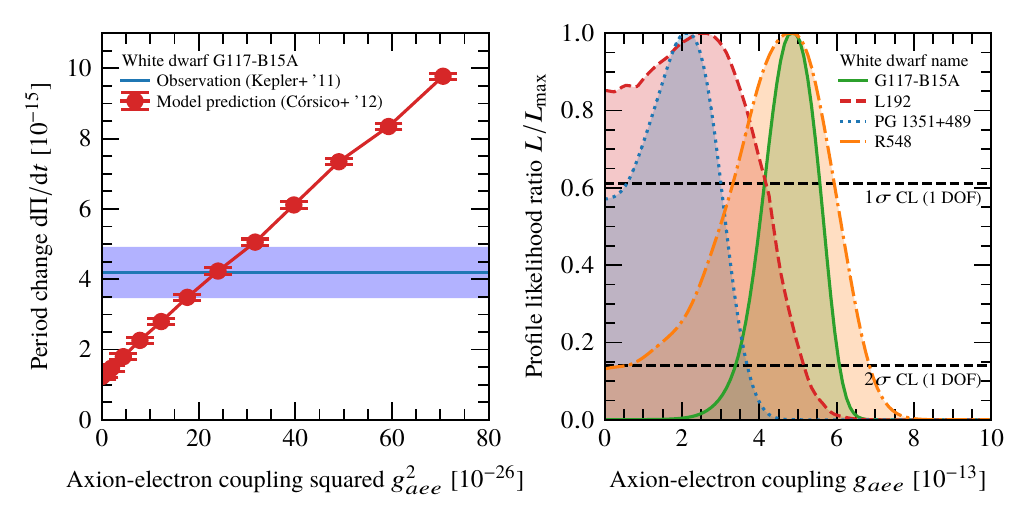}
	\caption{Cooling hints in white dwarfs. \textit{Left:}~Comparison of the predicted and measured period change as a function of axion-electron coupling squared for the WD variable G117-B15A~(data from Refs.~\cite{2012Kepler,Corsico:2012ki}). \textit{Right:}~Overview of the likelihood functions for different WD variables~(reproduced from Ref.~\cite{Hoof:2018ieb}).\label{fig:cooling_hints}}
\end{figure}

The left panel of Fig.~\ref{fig:cooling_hints} shows how the measured period increase in G117-B15A~(blue line and shading) compared to theoretical prediction from simulating WD evolution with and without axions~(red data points). To illustrate the dependence on the axion-electron coupling, we show the theoretical prediction as a function of~$g_{aee}^2$.

The right panel of Fig.~\ref{fig:cooling_hints} shows the one-dimensional profile likelihoods for the four WDs listed above. Combining these likelihoods hints at an additional cooling channel corresponding to an axion-electron coupling of a few times $10^{-13}$ at more than $3\sigma$ confidence level~\cite{Giannotti:2015kwo,Giannotti:2017hny,Hoof:2018ieb}.

In addition to difficulties of observing the period change, there are a number of uncertainties involved in the modelling of WDs and their pulsations. Multiple challenges in quantifying the statistical and systematic uncertainties in WD modelling remain, such as the modelling of the transition from the main sequence to the WD phase. More details on WD modelling can be found in Refs.~\cite{Corsico:2012ki,Corsico:2012sh,Corsico:2016okh,Battich:2016htm}. It is therefore not yet clearly established whether the cooling hints are due to systematics or indicate the presence of new physics---be it in form of a weakly-interacting particle or a completely different astrophysical cooling channel.

A recent, more general review of pulsating WDs can be found in Ref.~\cite{Corsico:2019nmr}. Apart from the period increase discussed here, the WD luminosity function can also be used to probe the evolution of WDs and be seen as a hint for ALPs~\cite{Bertolami:2014wua,Isern:2018uce}.

\end{example}

\subsection{Axion-photon conversion}\label{sec:axion-photon-astro}

In the presence of a magnetic field, axions convert into photons, and vice versa, by the Primakoff\index{Primakoff effect} and inverse Primakoff process (see Chapters~2, 4, and~5). This leads to constraints on the axion-photon coupling from any astrophysical environment penetrated by a magnetic field. In the following, we briefly mention some important instances.

Axions produced during supernova SN1987A\index{supernova!SN1987A} can escape from the supernova event. These axions are subsequently converted back into visible photons in the form of gamma rays by the magnetic field of the Milky Way. This process would have led to a gamma ray burst coincident with SN1987A, which was not observed~\cite{Chupp:1989kx}. This places constraints on the axion mass and coupling at 95\% CL~\cite{Payez:2014xsa}:
\begin{equation}
g_{a\gamma\gamma} < \SI{5e-12}{\GeV^{-1}}\quad (\text{for } m_a\lesssim \SI{e-9}{\eV})\, .
\end{equation}
The bound gets rapidly worse at higher masses due to loss of coherence of the axion field on the scale of the galactic magnetic field and the resulting reduced photon fluence. The bound has an $\mathcal{O}(1)$ dependence on the precise model of the galactic magnetic field.

Axion-photon conversion also occurs in the intergalactic medium, and leads to modulation of the X-ray spectra of active galactic nuclei (AGN) and quasars (see e.g.\ Refs~\cite{Berg:2016ese,Conlon:2017qcw}). The modulation can be modelled statistically with a stochastic model for cluster magnetic fields. The strongest bound arises from the observation of a single source, NGC1275, by the Chandra satellite, which observed no modulations and sets the 3$\sigma$ limit~\cite{Day:2018ckv,Reynolds:2019uqt}
\begin{equation}
g_{a\gamma\gamma} < \SIrange[range-phrase=\text{--},range-units=single]{6}{8e-13}{\GeV^{-1}}\quad (\text{for } m_a \lesssim \SI{e-12}{\eV})\, .
\end{equation}

Still further bounds can be derived from axion-photon conversion on cosmological scales. The conversion of CMB photons\index{cosmic microwave background radiation (CMB)}\index{CMB} by Mpc scale primordial magnetic fields leads to CMB spectral distortions (i.e.\ departure from a blackbody spectrum)~\cite{Mirizzi:2009nq,Tashiro:2013yea}. Since the cosmic background explorer (COBE) satellite determined the CMB to be the most perfect black body in the Universe~\cite{Mather:1993ij}, any departures from perfection caused by axion-photon conversion are strongly constrained. On the other hand, the origin and spectrum of large scale, primordial cosmic magnetic fields is highly uncertain (e.g.\ Ref.~\cite{Durrer:2013pga}). Thus, bounds are given relative to the amplitude of the magnetic field power spectrum averaged on cosmic length scales, $A_B = \sqrt{\langle B^2\rangle}$, as
\begin{equation}
g_{a\gamma\gamma} \lesssim \SI{e-14}{\GeV^{-1}} \left(\frac{\SI{1}{\nano\gauss}}{A_B}\right) \quad (\text{for $ m_a \lesssim \SI{e-12}{\eV}$}) \, .
\end{equation}
These bounds can be improved by up to two orders of magnitude by future CMB spectral measurements.

\begin{acknowledgement}
We are grateful to Richard Brito and Jens Niemeyer for helpful discussions, and to Jurek Bauer and David Ellis for producing Figs.~\ref{fig:scales_linear_perts} and~\ref{fig:PeakPatchEvo}. We are further indebted to Jurek Bauer and David Ellis for providing written solutions to problems~2, 3, and 5.
\end{acknowledgement}

\bibliographystyle{spphys}
\bibliography{./refs-chapter-3}
%
%
%

\newcommand*\diag{\text{diag}}
\newcommand{\mpl}{M_{pl}}
\newcommand{\GN}{G_N}
\newcommand{\dd}{\mathrm{d}}
\newcommand{\pbar}{\bar{\phi}}

\section*{Solutions to Problems}

\begin{example}{Problem 1: Background evolution of UBDM}
\label{ch11:prob-sol:3-1}

The continuity equation,\index{continuity equation} $\dot{\rho}/\rho = -3(w+1) \, \dot{a}/a$, can be integrated such that $\rho \propto a^{-3(w+1)}$. Note that this solution is also valid for the case $w = -1$ i.e.\ when $\rho$ is constant. Substituting this into the Friedmann equation,\index{Friedmann equation} we find
\begin{eqnarray}
	H^2 \equiv \left(\frac{\dot{a}}{a}\right)^2 = \frac{\rho}{3\,\mpl^2} \propto a^{-3(w+1)} \Rightarrow \dot{a} \, a^\frac{3w+1}{2} = c_1 \, ,
\end{eqnarray}
where $c_1$ is some constant. Integrating the equation above for, we obtain
\begin{eqnarray}
	a = \left\{
	\begin{array}{cl}
	c_2 \, \mathrm{e}^{c_1 t} & \text{if } w = -1\\
	\left(c_1 \, t + c_2\right)^\frac{2}{3(w+1)} & \text{else}
	\end{array}
	\right.
\end{eqnarray}
\begin{eqnarray}
	a^\frac{3(w+1)}{2} = c_1 \, t + c_2\Rightarrow a \propto t^\frac{2}{3(w+1)} \, .
\end{eqnarray}
Consequently, we find via $H = \dot{a}/a$ that
\begin{eqnarray}
	H = \left\{
		\begin{array}{cl}
		c_1 & \text{if } w = -1\\
		\frac{2 c_1}{3 (w+1)} \, \frac{1}{c_1\,t + c_2} & \text{else}
		\end{array}
		\right.
\end{eqnarray}

To change variables from conformal to physical time,\index{conformal time} note that
\begin{align}
	\pbar' &\equiv \frac{\ddot\pbar}{\dd\tau} = \frac{\dd t}{\dd\tau}  \frac{\dd\pbar}{\dd t} = a \, \dot{\pbar} \\
	\pbar'' &\equiv \frac{\dd}{\dd \tau} \pbar' = a' \dot{\pbar} + a^2 \ddot{\pbar} = a \dot{a} \dot{\pbar} + a^2 \ddot{\pbar} = a^2 (H \dot{\pbar} +  \ddot{\pbar})
\end{align}
This yields the following field equation for~$\pbar$:
\begin{eqnarray}
	0 = \ddot{\pbar} + 3 H \dot{\pbar} + m^2 \pbar = \ddot{\pbar} + \frac{2 \, \dot{\pbar}}{3(w+1) \, t} + m^2 \pbar \, . \label{eq:alp_field_equation}
\end{eqnarray}
The solutions of this equation can be expressed in terms of Bessel functions\index{Bessel function} of first~($J_\nu$) and second kind~($Y_\nu$) with $\nu = (w - 1)/2(w + 1)$ and $\delta \equiv \sqrt{1 - (2m/3H)^2}$
\begin{eqnarray}
	\pbar = \left\{
	\begin{array}{cl}
	\mathrm{e}^{-3H (1 + \delta) \, t} \left(c_3 + c_4 \, \mathrm{e}^{3H\delta \, t}\right) & \text{if } w = -1\\ t^\nu \, \left[c_3 \, J_\nu (m\,t) + c_4 Y_\nu (m\,t)\right] & \text{if } w \neq -1
	\end{array}
	\right. \, ,
\end{eqnarray}
where $c_3$ and $c_4$ are constants that depend on the initial conditions. While these may be used to derive expressions for the pressure and density, we note that the asymptotic behaviour of Eq.~(\ref{eq:alp_field_equation}) may be derived more generally. Note that, for early times, $t \rightarrow 0$,\footnote{More precisely, this is the regime where $H(t) \gg m$.} we have $H \propto 1/t \rightarrow \infty$ and the corresponding term will dominate:
\begin{eqnarray}
	\ddot{\pbar} + 3H \dot{\pbar} \simeq 0 \Rightarrow \dot{\pbar} \propto a^{-3} \Rightarrow \pbar = c_5 +  \left\{
	\begin{array}{cl}
	c_6 \mathrm{e}^{-3H \, t} & \text{if } w = -1\\ \ln(c_6 \, t) & \text{if } w = 1\\ c_6 \, t^\frac{w - 1}{w + 1} & \text{else}
	\end{array}
	\right. \, ,
\end{eqnarray}
where $c_5$ and $c_6$ are constants. In typical applications, e.g., a radiation-dominated universe\index{radiation-dominated universe} with $w = 1/3$, we have $\pbar = c_5 +  c_6 \, t^{-1/2}$. The second term is divergent for $t \rightarrow 0$ and we also see that $\pbar \simeq c_5$ for $t \gg (c_6/c_5)^2$, which are two arguments usually used for ignoring the second term and saying that the background field is simply constant at early times, $\pbar = c_5$.

On the other hand, for late times $H(t) \ll m$, Eq.~(\ref{eq:alp_field_equation}) is solved by a WKB-like\index{WKB approximation}\index{Wentzel-Kramers-Brillouin (WKB) approximation} solution,\footnote{This can be checked by substituting this in Eq.~(\ref{eq:alp_field_equation}) and noting that the non-vanishing terms are all small if $H \ll m$.} and without loss of generality we take $\pbar \propto a^{-3/2} \cos(m\,t)$ and $\dot{\pbar} \propto a^{-3/2} [H \cos(m\,t) - m \sin(m\,t)]$. Note that $H \ll m$ also implies that the oscillation periods $T \sim 1/m \ll 1/H \sim t$ such that we can assume that $H$ and $a$ do not change much over each integration
\begin{align}
	m^2\dot{\pbar}^2 = P +\rho \Rightarrow m^2\langle\dot{\pbar}^2\rangle = \langle P \rangle + \langle \rho \rangle \equiv (w_\text{eff} + 1) \langle \rho \rangle \, .
\end{align}
Using $\langle \sin(m\,t) \cos(m\,t)\rangle = 0$ and $\langle \sin^2(m\,t) \rangle = \langle \cos^2(m\,t) \rangle = 1/2$, we find that
\begin{align}
\langle\dot{\pbar}^2\rangle \propto (H/m)^2/2 + 1/2 \, ,\\
\langle\rho\rangle = m^2\langle\dot{\pbar}^2\rangle/2 + \langle V \rangle \propto H^2/4 + m^2/4 + m^2/4 \, ,\\
\Rightarrow w_\text{eff} + 1 = 2 \, \frac{(H/m)^2 + 1}{(H/m)^2 + 2} \simeq 1 \Rightarrow w_\text{eff} \simeq 0 \, .
\end{align}

For the potential $V(\phi) = \lambda \phi^4$, Eq.~(\ref{eq:alp_field_equation})  now becomes
\begin{eqnarray}
	0 = \ddot{\pbar} + \frac{2 \, \dot{\pbar}}{3(w+1) \, t} + 4\lambda \pbar^3 \, . \label{eq:other_alp_field_equation}
\end{eqnarray}
For early times, there is no change to the previous argument since the potential is irrelevant.

For later times, the shape of Eq.~(\ref{eq:other_alp_field_equation}) implies that we cannot rely on the WKB-like solutions anymore, and instead follow the general approach presented in Refs~\cite{Turner:1983he,Masso:2005zg}.\footnote{Note that, for the quartic potential $V(\pbar) = \lambda \pbar^4$, the solutions for $\pbar$ can be expressed in terms of so-called Jacobi elliptic functions\index{Jacobi elliptic functions} times an oscillating function~\cite{Masso:2005zg}, which can be used to check the general solution presented here explicitly.} We repeat this derivation for a symmetric potential~$V$, i.e.\ $V(-\pbar) = V(\pbar)$, and $V(\pbar) = \lambda \pbar^n$~($n$~even). At the maximum~$\hat{\pbar}$ the total energy density is given by the maximum potential value, $\rho(\hat{\pbar}) = V(\hat{\pbar}) \equiv \hat{V}$. Since the oscillations at late times are very rapid, the energy density of the field does not change much over one oscillation period~$T$, and $\rho \approx \hat{V}$. The definition of~$\rho$ then implies that
\begin{eqnarray}
 \dot{\pbar}^2 = \frac{2}{m^2} \, (\rho - V) = \frac{2\,\hat{V}}{m^2}\left(1 - \frac{V}{\hat{V}}\right) 
\end{eqnarray}
and we find for the oscillation period that
\begin{align}
    T &= \int \dd t = \int \frac{\dd t}{\dd \pbar} \dd \pbar = \int_{-\hat{\pbar}}^{\hat{\pbar}} \frac{1}{|\dot{\pbar}|} \dd \pbar = \frac{2 \, m}{\sqrt{2\hat{V}}} \int_{0}^{\hat{\pbar}} \frac{1}{\sqrt{1 - V/\hat{V}}} \dd \pbar \, ,
\end{align}
which in turn implies that
\begin{align}
    w_\text{eff} + 1 &= \frac{m^2 \, \langle \dot{\pbar}^2 \rangle}{\langle \rho \rangle} = \frac{m^2}{\hat{V}} \frac{1}{T} \int \dot{\pbar}^2 \dd t = \frac{m^2}{\hat{V}} \frac{1}{T} \int_{-\hat{\pbar}}^{\hat{\pbar}} \dot{\pbar}^2/|\dot{\pbar}| \dd \pbar\\
    &= \frac{2\sqrt{2} \, m}{\sqrt{\hat{V}} T} \int_{0}^{\hat{\pbar}} \sqrt{1 - V/\hat{V}} \dd \pbar\\
    & = 2 \frac{\int_{0}^{\hat{\pbar}} (1 - V/\hat{V})^{\frac{1}{2}} \dd \pbar}{\int_{0}^{\hat{\pbar}} (1 - V/\hat{V})^{-\frac{1}{2}} \dd \pbar} = \frac{2n}{n+2}
    \Rightarrow w_\text{eff} = \frac{n-2}{n+2} \, ,
\end{align}
i.e.\ $w_\text{eff} = 1/3$ for $n=4$, which corresponds to the equation-of-state parameter of radiation.

We see that when Hubble friction\index{Hubble friction} dominates, i.e., typically at sufficiently early times, any scalar particles generically behave as dark energy\index{dark energy} with $w_\text{eff} = -1$. Later on, however, the~(dominant terms of) the potential determine the behaviour of the scalar field oscillations, which need not be that of dark matter~(pressureless dust) and dark matter bounds hence do not apply.

\end{example}

\begin{example}{Problem 2: Derivation of the Schr\"{o}dinger-Poisson equations for UBDM}
\label{ch11:prob-sol:3-2}

\index{Schrodinger-Poisson equations@Schr\"{o}dinger-Poisson equations}First, let us write down the Klein-Gordon equation,\index{Klein-Gordon equation}
\begin{eqnarray}
\square \phi - \partial_{\phi} V = 0\, ,
\label{eq:KG_definition}
\end{eqnarray}
where the D'Alembertian\index{D'Alembertian} is defined as 
\begin{eqnarray}
\square=\frac{1}{\sqrt{-g}} \partial_{\mu} (\sqrt{ - g} g^{\mu \nu})\partial_{\nu}
\label{eq:box_op_definition}
\end{eqnarray}
and the potential is given by
\begin{eqnarray}
V(\phi) = \frac{m^2}{2}\phi^2 + \frac{m^2}{2} \lambda \phi^4\, .
\label{eq:potential_definition}
\end{eqnarray}

With $\Psi = \Phi$ and $a=1$, the metric is given by
\begin{eqnarray}
g = [g_{\mu\nu}] = \diag{\left[-(1+2 \Psi), 1-2 \Psi, 1-2 \Psi, 1-2 \Psi\right]}\, .
\end{eqnarray}
To first order this gives:
\begin{align*}
g^{-1} = [g^{\mu\nu}] &= \diag{\left[-(1 - 2 \Psi), 1+2 \Psi, 1+2 \Psi, 1+2 \Psi\right]}\, , \\
\sqrt{-g} &= 1 - 2 \Psi\, .
\end{align*}
Thus, the D'Alembertian\index{D'Alembertian} to first order is given by: 
\begin{eqnarray}
\square = 4 \dot{\Psi} \partial_t - (1 - 2\Psi) \partial_t^2 + (1+2\Psi) \nabla^2
\end{eqnarray}
and the Klein-Gordon equation\index{Klein-Gordon equation} (Eq. \ref{eq:KG_definition}) reads
\begin{eqnarray}
-(1-2\Psi)\ddot{\phi} + 4 \dot{\Psi}\dot{\phi} + (1+2\Psi) \nabla^2\phi - m^2 \phi - 2 m^2 \lambda \phi^3 = 0\, .
\end{eqnarray}

We can rewrite the Klein-Gordon equation by multiplying with $-(1+2\Psi)$ (since this term goes to $-1$ in the non-relativistic limit, the result remains unchanged save for an overall minus sign), which reduces the number of terms we need to consider later on:
\begin{eqnarray}
\ddot{\phi} - 4 \dot{\Psi} \dot{\phi} - (1 + 4\Psi) \nabla^2\phi + (1+2\Psi) m^2 \phi + (1+2\Psi)2m^2 \lambda \phi^3 = 0\, .
\label{eq:simiplified_KG}
\end{eqnarray}

Let us take the ansatz for $\phi$ and write:
\begin{align*}
\phi &= \frac{1}{\sqrt{2}m} \left[ \psi e^{imt} + \psi^* e^{imt}\right]\, , \\ 
\dot{\phi} &= \frac{1}{\sqrt{2}m} \left[ e^{imt}\left(\dot{\psi} + i m \psi \right) + e^{-imt}\left(\dot{\psi}^* - i m \psi^* \right)  \right]\, , \\
\ddot{\phi} &= \frac{1}{\sqrt{2}m} \left[e^{imt} \left(\ddot{\psi} + 2 i m\dot{\psi} - m^2 \psi \right) + e^{-imt} \left( ... \right)\right]\, , \\
\nabla^2 \phi &= \frac{1}{\sqrt{2}m} \left[(\nabla^2\psi) e^{imt}  + (\nabla^2 \psi^*) e^{-imt}\right]\, , \\
\phi^3 &= \frac{1}{(\sqrt{2}m)^3} \left[ e^{imt} 2 |\psi|^2\psi + e^{-imt} 2|\psi|^2 \psi^* + \psi^3 e^{3imt}  + \psi^{*3} e^{-3imt} \right] \, .
\end{align*}
Since  terms for $e^{-imt}$ are the complex conjugate, the terms in front of $e^{+imt}$ need to vanish in order for the Klein-Gordon equation to be fulfilled.
Thus, one needs to consider only the terms which go with a $e^{+imt}$ oscillation (or $e^{-imt}$, respectively). 
This gives for Eq.~\eqref{eq:simiplified_KG}:
\begin{align*}
&\frac{1}{\sqrt{2}m} [ - 4\dot{\Psi} (\dot{\psi} + i m \psi) + (\ddot{\psi} + 2 i m\dot{\psi} - m^2 \psi) - (1 + 4\Psi)(\nabla^2 \psi) \\
& \qquad + (1+2\Psi)m^2\psi + (1+2\Psi)2m^2 \lambda \frac{2}{2m^2} |\psi|^2\psi] = 0\\
\Longleftrightarrow \qquad &\frac{1}{\sqrt{2}m} [ - 4\dot{\Psi} \dot{\psi} - 4\dot{\Psi} i m \psi + \ddot{\psi} + 2 i m\dot{\psi} - \nabla^2 \psi  -  4\Psi \nabla^2 \psi \\
& \qquad + 2\Psi m^2\psi + 2\lambda |\psi|^2\psi] + 4 \Psi \lambda |\psi|^2\psi] = 0\, .
\end{align*}
Now, let us take the non-relativistic limit and consider either the limits given in the exercise or take $c\rightarrow\infty$. 
We calculated in natural units, where $c=1$. 
Remember when revoking the $c$'s that $\Psi \rightarrow \Psi/c^2$, that the time derivatives gain a factor of~$1/c$ and that -- for each $m$ within the brackets -- $m \rightarrow m c$. 
Hence, the terms in red vanish and one is left with:
\begin{align}
i \dot{\psi} - \frac{1}{2 m} \nabla^2\psi + m \Psi \psi + \frac{\lambda}{m} |\psi|^2 \psi = 0.
\end{align}
Considering the complex conjugate, this equals Eq.~\eqref{eqn:SP_1} with $\lambda_{\mathrm{GP}} = - \lambda$.

To calculate the energy density, $\rho$, to leading order we recognize that $\rho = - T^0_0$. The stress energy tensor is given by:
\begin{eqnarray}
T_{\mu \nu}=\partial_{\mu} \phi \partial_{\nu} \phi - g_{\mu \nu}\left[\frac{1}{2} g^{\alpha \beta} \partial_{\alpha} \phi \partial_{\beta} \phi+V(\phi)\right],
\label{eq:stress_energy_tensor_definition}
\end{eqnarray}
and, thus:
\begin{align*}
T^{0}_{0} &= g^{00}T_{00} \\
&= g^{00} \dot{\phi}^2 - g^{00}g_{00} \left(\frac{1}{2} g^{\alpha\beta} \partial_\alpha\phi \partial_\beta \phi + V(\phi) \right) \\
&= \frac{1}{2} g^{00} \dot{\phi}^2 - \frac{1}{2}(\nabla \phi)^2 g^{ii} - V(\phi) \\
&= - \frac{1}{2}\left[(1-2\Psi) \dot{\phi}^2 + (1 + 2\Psi) (\nabla\phi)^2 + 2V(\phi) \right]\, .
\end{align*}
The individual terms are given by:
\begin{align*}
\phi^2 &= \frac{1}{2m^2} \left[e^{2imt} \psi^2 + e^{-2imt} \psi^{*2} + 2 |\psi|^2\right]\, , \\
\dot{\phi}^2 &= \frac{1}{2m^2} [e^{2imt} ( \dot{\psi} + i m\psi)^2 + e^{-2imt} (\dot{\psi}^* - i m \psi^*)^2  + \\
& \qquad 2 (|\dot{\psi}|^2 + \underbrace{i m (\psi^*\dot{\psi} - \psi \dot{\psi}^*)}_{=2m\mathrm{Im}(\dot{\psi}\psi^*)} +  m^2 |\psi|^2)]\, , \\
(\nabla\phi)^2 &= \frac{1}{2m^2}\left[(\nabla\psi)^2 e^{2imt} + (\nabla\psi^*)^2 e^{-2imt} + 2 |\nabla\psi|^2\right]\, .\\
\end{align*}
Again, the leading order terms can be identified by either taking the limits given in the exercise or by counting powers of $c$ (remembering that the mass term in the potential has a factor of $c^2$).
Neglecting the oscillatory terms, the remaining terms are $\dot{\phi}^2 \rightarrow |\psi|^2$ and $m^2 \phi^2 \rightarrow |\psi|^2$. Thus:
\begin{align}
\rho = \frac{1}{2}\left(|\psi|^2 + |\psi|^2 \right) =|\psi|^2.
\end{align}
\end{example}

\begin{example}{Problem 3: Relaxation of UBDM}
\label{ch11:prob-sol:3-3}

Starting from 
\begin{eqnarray}
t_{\text{relax}} = 0.1 \frac{R}{v} \frac{M}{m \log \Lambda},
\label{eq:relaxation_time_definition}
\end{eqnarray}
first one notes that the host of mass $M$ with radius $R$ and the quasi-particle of mass $m$ follow from the same underlying density $\rho$. 
The host mass can be approximated by assuming it to be a sphere of constant density $\rho$, while the quasi-particle mass is effectively given by the size of the de Broglie wavelength:\index{de Broglie wavelength}
\begin{align}
M &= \frac{4 \pi}{3} \rho R^3 
\label{eq:host_mass}\\
m &\sim \rho \left(\frac{\lambda_{\text{dB}}}{2}\right)^3.
\label{eq:quasi_particle_mass}
\end{align}
The de Broglie wavelength\index{de Broglie wavelength} is given by 
\begin{eqnarray}
\lambda_{\mathrm{dB}} = \frac{h}{m v},
\label{eq:deBroglie}
\end{eqnarray}
where the velocity $v$ approximately equals the quasi-particle\index{quasi-particle} velocity. The latter is justified since the dynamics of the quasi-particle are ultimately determined by the dynamics of the underlying axion particle and, thus, one can expect $v_{\text{qp}} \sim v_a$. 

Plugging Eqns.~\eqref{eq:host_mass}, \eqref{eq:quasi_particle_mass}, and \eqref{eq:deBroglie} into \eqref{eq:relaxation_time_definition} yields:
\begin{align}
t_{\text{relax}} \sim 3.4  (m/h)^3 v^2 R^4  \frac{1}{\log{\Lambda}}.
\end{align}

Plugging in the numbers, we obtain Eq.~\eqref{eqn:t_relax_hui}:
\begin{eqnarray}
t_{\text{relax}} \sim \frac{\SI{e10}{}}{\log{\Lambda}} \left(\frac{m}{\SI{e-22}{eV}}\right)^3 \left(\frac{v}{\SI{100}{km/s}}\right)^2 \left(\frac{R}{\SI{5}{kpc}}\right)^4.
\end{eqnarray}
\end{example}

\begin{example}{Problem 4: Estimating superradiance properties of UBDM}
\label{ch11:prob-sol:3-4}

First, let us derive mass scale relevant for superradiance.\index{superradiance} In order to affect the action, the UBDM potential $m^2 \phi^2 / 2$ should be comparable gravitational term induced by the Kerr~BH,\index{black hole!Kerr}\index{Kerr black hole} $\mpl^2 R / 2$. Recalling that $8\pi \GN \equiv 1/\mpl^2$, the Ricci scalar\index{Ricci scalar} in the Kerr geometry is given by
\begin{equation}
	R^2 = 48 \, (\GN M)^2 \; \frac{\left(r^2 - a_J^2\cos^2(\theta)\right)^2 \left[\left(r^2 + a_J^2\cos^2(\theta)\right)^2 - 16r^2a_J^2\cos^2(\theta)\right]}{\left(r^2 + a_J^2\cos^2(\theta)\right)^6} \, . \label{eq:p3.4:ricci}
\end{equation}
Suppose, for simplicity, that we are at the equator~($\theta = \pi/2$) and that we just inside the ergosphere\index{ergosphere} at
\begin{equation}
	r_\text{ergo} = \GN \left( M + \sqrt{M^2 - a_J^2 \cos^2(\theta)} \right) = 2 \, \GN M \, ,
\end{equation}
which is also just the Schwarzschild radius\index{Schwarzschild radius} of the black hole.\index{black hole} We then find
\begin{equation}
    R^2 \overset{\theta=\pi/2}{=} \frac{48 \, \GN^2 M^2}{r^6} \overset{r = r_\text{ergo}}{=} \frac{3}{4} \frac{1}{ (\GN M)^4} \, .
\end{equation}

For typical field values $\phi \sim \mpl$ in the extreme environment surrounding the black hole, we find that

\begin{equation}
	\frac{m^2\phi^2}{2} \sim \frac{m^2 \mpl^2}{2} \overset{!}{\sim} \frac{\mpl^2 R}{2} \sim \frac{\mpl^2}{2} \sqrt{\frac{3}{4}} \frac{1}{ (\GN M)^2} \Rightarrow \GN M m \sim 1\, , \label{eq:p3.4:alphaeff}
\end{equation}
where we ignored the $\mathcal{O}(1)$ numerical factor.

Note that we might have used dimensional analysis~(e.g.\ that $R \sim 1/R_\text{S} \sim 1/2GM$) instead of Eq.~\ref{eq:p3.4:ricci} to arrive at a similar result.

To compute the Bosenova\index{Bosenova} condition, the self-coupling of the UBDM particles, $\lambda$, needs to be large enough to play a role. We anticipate this to happen when the corresponding term in the action, $\lambda\phi^4/4!$, becomes of the same order as the potential term, $m^2 \phi^2 / 2$. To facilitate this comparison, first note that the total energy density of the UBDM cloud can be equated to its total mass, $M_\text{cloud}$, divided by its volume, $V_\text{cloud}$, which are given by
\begin{align}
	M_\text{cloud} &= N m \, , \\
	V_\text{cloud} &= \frac{4\pi}{3} R_\text{cloud}^3 \sim \frac{4\pi}{3} r_\text{ergo}^3 = \frac{32\pi}{3} (\GN M)^3 \, ,
\end{align}
where $N$ is the number of UBDM particles. By equating $m^2 \phi^2 / 2 = M_\text{cloud}/V_\text{cloud}$, we find that
\begin{equation}
	\phi^2 = \frac{3}{16\pi} \frac{N}{(\GN M)^3 m} \, . \label{eq:p3.4:phi}
\end{equation}

From $m^2 \phi^2 / 2 \overset{!}{\sim} \lambda\phi^4/4!$ and using Eq.~\eqref{eq:p3.4:phi}, we find further that
\begin{equation}
    \phi^2 \sim \frac{12m^2}{\lambda} \Rightarrow N \sim 64\pi \frac{(\GN M m)^3}{\lambda} \sim \frac{64\pi}{\lambda} \, ,
\end{equation}
where the last step made use of Eq.~\eqref{eq:p3.4:alphaeff}. To compare to Eq.~\eqref{eqn:BHSR_Nbose}, we use that $\lambda = m^2/f_a^2$ for an axion and eliminate~$m$ via the SR condition~\eqref{eq:p3.4:alphaeff}, such that $\lambda \sim 1/f_a^2 (\GN M)^2$. Since $\mpl = \SI{2.4e18}{\GeV} = \num{2.2e-39}\,M_\odot$ we arrive at
\begin{align}
    N &\sim \frac{64\pi}{\lambda} \sim \frac{1}{\pi} \left(\frac{10\,M_\odot}{\mpl}\right)^2 \left(\frac{M}{10\,M_\odot}\right)^2 \left(\frac{f_a}{\mpl}\right)^2 \\
&\approx \num{6.6e78} \left(\frac{M}{10\,M_\odot}\right)^2 \left(\frac{f_a}{\mpl}\right)^2\, ,
\end{align}
which is very close to Eq.~\eqref{eqn:BHSR_Nbose} for the first energy level~($n=1$).

\end{example}

\begin{example}{Problem 5: Microlensing constraints on UBDM}
\label{ch11:prob-sol:3-5}

The Schr\"{o}dinger-Poisson (SP) equation\index{Schrodinger-Poisson equations@Schr\"{o}dinger-Poisson equations} is given by
\begin{eqnarray}
	i \dot{\psi} + \frac{\nabla^2}{2m}\psi + \frac{\lambda_{\mathrm{GP}}}{m}|\psi|^2\psi = 0
	\label{sp1}
\end{eqnarray}
and
\begin{eqnarray}
	\nabla^2\Phi = 4\pi G\left(|\psi|^2 - \int{\mathrm{d}^3x|\psi|^2}\right)\, .
	\label{sp2}
\end{eqnarray}
We can therefore write Eq.~\eqref{sp1} using the given scaling relation $\{t, x, \psi, \Phi, \lambda_{\mathrm{GP}} \} \rightarrow \{ \lambda^{-2}\hat{t}, \lambda^{-1}\hat{x}, \lambda^{2}\hat{\psi}, \lambda^{2}\hat{\Phi}, \lambda^{-2}\hat{\lambda}_{\mathrm{GP}} \}$.\\

\noindent
For some constant and variable, $c$ and $q$ respectively, we can write that $\frac{\partial}{\partial(cq)} = \frac{1}{c}\frac{\partial}{\partial(q)}$ and hence $\nabla \rightarrow \lambda \hat{\nabla}$. Therefore,
\begin{eqnarray}
	i \frac{\partial\lambda^2\hat{\psi}}{\partial(\lambda^{-2}\hat{t})} + \frac{\lambda^2\hat{\nabla}^2}{2m}\lambda^2\hat{\psi} + \frac{\lambda^{-2}\hat{\lambda}_{\mathrm{GP}}}{m}|\lambda^2\hat{\psi}|^2\lambda^2\hat{\psi} = 0,
\end{eqnarray}
\begin{eqnarray}
	\lambda^4 \left [ i \frac{\partial\hat{\psi}}{\partial(\hat{t})} + \frac{\hat{\nabla}^2}{2m}\hat{\psi} + \frac{\hat{\lambda}_{\mathrm{GP}}}{m}|\hat{\psi}|^2\hat{\psi} \right ]= 0,
\end{eqnarray}
\begin{eqnarray}
	 i \frac{\partial\hat{\psi}}{\partial(\hat{t})} + \frac{\hat{\nabla}^2}{2m}\hat{\psi} + \frac{\hat{\lambda}_{\mathrm{GP}}}{m}|\hat{\psi}|^2\hat{\psi} = 0.
\end{eqnarray}
Therefore, Eq.~\eqref{sp1} is invariant under the rescaling. This can similarly be shown for the second equation using the fact that for the solitons\index{soliton} $|\psi|^2 \gg 1$.

For $\lambda = 1$, the soliton profile is given by
\begin{eqnarray}
	\frac{\rho_{\mathrm{sol}}(r)}{m^2M^2_{pl}} = \chi^2(mr) = \frac{1}{(1+\alpha^2m^2r^2)^8}\, .
\end{eqnarray}
The mass of the soliton is then given by 
\begin{eqnarray}
	M_{\mathrm{sol}} = 4\pi\int^{r_s}_0\rho_{\mathrm{sol}}(r)r^2dr\, .
\end{eqnarray}
The rescaling relates to $\chi \rightarrow \lambda^2\hat{\chi}$. Then, since $\rho_{\mathrm{sol}}(r) = |\chi|^2$, 
\begin{eqnarray}
	\begin{split}
	M_{\mathrm{sol}} & = 4\pi\int^{r_c}_0   | \chi |^2r^2dr \\
	       & = 4\pi\int^{r_c}_0 \lambda^4|\hat{\chi}|^2\frac{1}{\lambda^3}\hat{r}^2d\hat{r} \\
	       & = \lambda \hat{M}_{\mathrm{sol}}\, .
	\end{split}
\end{eqnarray}
Similarly, $\rho_{\mathrm{sol}}(r) = \lambda^4 \hat{\rho}_{\mathrm{sol}}(\hat{r})$. We can therefore rescale the profile to 
\begin{eqnarray}
	\frac{\lambda^4\hat{\rho}_{\mathrm{sol}}(r)}{m^2M^2_{pl}} =  \frac{1}{(1+\alpha^2\lambda^{-2}m^2\hat{r}^2)^8}\, .
\end{eqnarray}
We can define a scale radius $r_c = \frac{\lambda}{\alpha m}$ allowing us to write the density profile as 

\begin{eqnarray}
	\rho_{\mathrm{sol}}(r) = \frac{M^2_{pl}}{r_c^4\alpha^4m^2}\frac{1}{{(1+(\hat{r}/r_c)^2)^8}}\, .
\end{eqnarray}
Integrating and making the change of variables, $u = \hat{r}/r_c$, we find that
\begin{eqnarray}
	\hat{M}_{\mathrm{sol}} =\frac{4\pi \mpl^2 }{r_c\alpha^4m^2} \int^{1}_0\frac{u^2}{(1+u^2)^8}du\, .
\end{eqnarray}
The integral is now a constant ($\sim 0.246$)\footnote{This integral can be solved analytically. The result is insensitive to whether we define the mass to be $M(r<r_c)$ or $M(r<\infty)$.}. Fixing the units and dropping the hats, we then find
\begin{eqnarray}
	M_{\mathrm{sol}} \sim 4 \times 10^8 \left( \frac{m}{10^{-22} \text{eV}} \right)^{-2} \left( \frac{r_c}{\text{kpc}} \right)^{-1} M_{\odot}\, .
\end{eqnarray}

The Einstein radius\index{Einstein radius} for such a lens is given by,
\begin{eqnarray}
	R_E = 2\times10^{-7} \left( \frac{M_*}{M_{\odot}} \right)^{1/2} \text{kpc}\, .
\end{eqnarray}
Rearranging our mass-radius relation we see
\begin{eqnarray}
	r_c \sim 4 \times 10^8 \left( \frac{m}{10^{-22} \text{eV}} \right)^{-2} \left( \frac{M_{\mathrm{sol}}}{M_{\odot}} \right)^{-1} \text{kpc}\, .
\end{eqnarray}
For the object to lens like a point-mass, we require that $r_c<R_E$, therefore

\begin{eqnarray}
	4 \times 10^8 \left( \frac{m}{10^{-22} \text{eV}} \right)^{-2} \left( \frac{M_{\mathrm{sol}}}{M_{\odot}} \right)^{-1} <  2\times10^{-7} \left( \frac{M_{\mathrm{sol}}}{M_{\odot}} \right)^{1/2} \text{kpc}\, ,
\end{eqnarray}
which can be rearranged to
\begin{eqnarray}
	m < 5 \times 10^{-15} \left ( \frac{M_{\mathrm{sol}}}{M_{\odot}} \right)^{-3/4} \text{eV}\, .
\end{eqnarray}
This upper limit is maximized by being sensitive to masses as small as possible. Setting $M_{\mathrm{sol}}$ to the smallest mass detectable by HSC\index{Subaru Hyper Suprime Cam (HSC)}\index{HSC}
\begin{eqnarray}
	m < 1.4 \times 10^{-10} \text{eV}\, .
\end{eqnarray}

To calculate the range of the $(T_{\rm osc},\delta)$~prameter space, we neglect the activation function in Eq.~\eqref{eqn:minicluster_mass_tosc} (since $S(x) \sim \mathcal{O}(1)$). Requiring again that $R_{\mathrm{MC}} < R_{\mathrm{E}}$ and substituting our MC mass into the equation for the Einstein radius we find the region range of  ($T_{\mathrm{osc}}$, $\delta$) parameter space can be probed by microlensing\index{gravitational lensing} to be 
\begin{eqnarray}
	\frac{1}{\delta (1+\delta)^{1/3}} \left( \frac{T_{\mathrm{osc}}}{2 \text{GeV}} \right)^{1/2} \lesssim 7\times10^{-4}\, .
\end{eqnarray}

\end{example}

\bibliographystyle{spphys}
\bibliography{refs-chapter-11}

\begin{thebibliography}{1}
\providecommand{\url}[1]{{#1}}
\providecommand{\urlprefix}{URL }
\expandafter\ifx\csname urlstyle\endcsname\relax
  \providecommand{\doi}[1]{DOI \discretionary{}{}{}#1}\else
  \providecommand{\doi}{DOI \discretionary{}{}{}\begingroup
  \urlstyle{rm}\Url}\fi

\bibitem{Turner:1983he}
M.S. Turner, Phys. Rev. D \textbf{28}, 1243 (1983)

\bibitem{Masso:2005zg}
E.~Masso, F.~Rota, G.~Zsembinszki, Phys. Rev. D \textbf{72}, 084007 (2005)

\end{thebibliography}


\begin{thebibliography}{100}
\providecommand{\url}[1]{{#1}}
\providecommand{\urlprefix}{URL }
\expandafter\ifx\csname urlstyle\endcsname\relax
  \providecommand{\doi}[1]{DOI \discretionary{}{}{}#1}\else
  \providecommand{\doi}{DOI \discretionary{}{}{}\begingroup
  \urlstyle{rm}\Url}\fi

\bibitem{Marsh:2015xka}
D.J.E. {Marsh}, Phys. Rep. \textbf{643}, 1 (2016)

\bibitem{2013JCAP...10..020A}
M.~{Archidiacono}, S.~{Hannestad}, A.~{Mirizzi}, G.~{Raffelt}, Y.Y.Y. {Wong},
  J. Cosmol. Astropart. Phys. \textbf{10}, 020 (2013)

\bibitem{Aghanim:2018eyx}
N.~Aghanim, Y.~Akrami, M.~Ashdown, J.~Aumont, C.~Baccigalupi, M.~Ballardini,
  A.~Banday, R.~Barreiro, N.~Bartolo, S.~Basak, et~al., Astron. Astrophys.
  \textbf{641}, A6 (2020)

\bibitem{1976JPhA....9.1387K}
T.W.B. {Kibble}, J. Phys. A Math. Theor. \textbf{9}(8), 1387 (1976)

\bibitem{1985Natur.317..505Z}
W.H. {Zurek}, Nature \textbf{317}(6037), 505 (1985)

\bibitem{Hiramatsu:2012gg}
T.~Hiramatsu, M.~Kawasaki, K.~Saikawa, T.~Sekiguchi, Phys. Rev. D \textbf{85},
  105020 (2012).
\newblock [Erratum: Phys. Rev. D 86, 089902 (2012)]

\bibitem{Klaer:2017ond}
V.B. Klaer, G.D. Moore, J. Cosmol. Astropart. Phys. \textbf{1711}(11), 049
  (2017)

\bibitem{Gorghetto:2018myk}
M.~Gorghetto, E.~Hardy, G.~Villadoro, J. High Energy Phys. \textbf{07}, 151
  (2018)

\bibitem{Armengaud:2019uso}
E.~Armengaud, et~al., J. Cosmol. Astropart. Phys. \textbf{1906}(06), 047 (2019)

\bibitem{1990eaun.book.....K}
E.W. {Kolb}, M.S. {Turner}, \emph{{The early universe}} (Addison-Wesley, 1990)

\bibitem{Dodelson:2003ft}
S.~Dodelson, \emph{{Modern Cosmology}} (Academic Press, Amsterdam, 2003)

\bibitem{mukhanov}
V.~Mukhanov, \emph{{Physical Foundations of Cosmology}} (Cambridge University
  Press, 2005)

\bibitem{Copeland:2006wr}
E.J. Copeland, M.~Sami, S.~Tsujikawa, Int. J. Mod. Phys. D \textbf{15}, 1753
  (2006)

\bibitem{Clifton:2011jh}
T.~Clifton, P.G. Ferreira, A.~Padilla, C.~Skordis, Phys. Rep. \textbf{513}, 1
  (2012)

\bibitem{2019ApJS..240...23A}
D.S. Aguado, R.~Ahumada, A.~Almeida, S.F. Anderson, B.H. Andrews, B.~Anguiano,
  E.A. Ort{\'\i}z, A.~Arag{\'o}n-Salamanca, M.~Argudo-Fern{\'a}ndez, M.~Aubert,
  et~al., Astrophys. J., Suppl. Ser. \textbf{240}(2), 23 (2019)

\bibitem{Primack:2001ib}
J.R. Primack, SLAC Beam Line \textbf{31N3}, 50 (2001)

\bibitem{bertschinger1995}
C.P. Ma, E.~Bertschinger, Astrophys. J. \textbf{455}, 7 (1995)

\bibitem{camb}
A.~{Lewis}, A.~{Challinor}, A.~{Lasenby}, Astrophys. J. \textbf{538}, 473
  (2000)

\bibitem{class}
J.~Lesgourgues, arXiv:1104.2932  (2011)

\bibitem{Hlozek:2014lca}
R.~{Hlo\v{z}ek}, D.~{Grin}, D.J.E. {Marsh}, P.G. {Ferreira}, Phys. Rev. D
  \textbf{91}, 103512 (2015)

\bibitem{Lewis:2006fu}
A.~Lewis, A.~Challinor, Phys. Rep. \textbf{429}, 1 (2006)

\bibitem{Hlozek:2017zzf}
R.~Hlo\v{z}ek, D.J.E. Marsh, D.~Grin, Mon. Not. Roy. Astron. Soc.
  \textbf{476}(3), 3063 (2018)

\bibitem{2014PhRvD..89h3536L}
B.~{Li}, T.~{Rindler-Daller}, P.R. {Shapiro}, Phys. Rev. D \textbf{89}(8),
  083536 (2014)

\bibitem{2009PhLB..680....1H}
J.C. {Hwang}, H.~{Noh}, Phys. Lett. B \textbf{680}, 1 (2009)

\bibitem{Bauer:2020zsj}
J.B. Bauer, D.J. Marsh, R.~Hlo{\v{z}}ek, H.~Padmanabhan, A.~Lagu{\"e},
  arXiv:2003.09655  (2020)

\bibitem{hu2000}
W.~Hu, R.~Barkana, A.~Gruzinov, Phys. Rev. Lett. \textbf{85}, 1158 (2000)

\bibitem{amendola2005}
L.~Amendola, R.~Barbieri, Phys. Lett. B \textbf{642}, 192 (2006)

\bibitem{2010PhRvD..82j3528M}
D.J.E. {Marsh}, P.G. {Ferreira}, Phys. Rev. D \textbf{82}(10), 103528 (2010)

\bibitem{Poulin:2018dzj}
V.~Poulin, T.L. Smith, D.~Grin, T.~Karwal, M.~Kamionkowski, Phys. Rev. D
  \textbf{98}(8), 083525 (2018)

\bibitem{SKA_RedBook}
D.J. Bacon, R.A. Battye, P.~Bull, S.~Camera, P.G. Ferreira, I.~Harrison,
  D.~Parkinson, A.~Pourtsidou, M.G. Santos, L.~Wolz, et~al., Publ. Astron. Soc.
  Aust. \textbf{37} (2020)

\bibitem{HIRAX}
L.B. Newburgh, et~al., Proc. SPIE Int. Soc. Opt. Eng. \textbf{9906}, 99065X
  (2016)

\bibitem{Khmelnitsky:2013lxt}
A.~Khmelnitsky, V.~Rubakov, J. Cosmol. Astropart. Phys. \textbf{1402}, 019
  (2014)

\bibitem{Porayko:2018sfa}
N.K. Porayko, et~al., Phys. Rev. D \textbf{98}(10), 102002 (2018)

\bibitem{Edwards:2018ccc}
F.~Edwards, E.~Kendall, S.~Hotchkiss, R.~Easther, J. Cosmol. Astropart. Phys.
  \textbf{1810}(10), 027 (2018)

\bibitem{Springel:2005mi}
V.~Springel, Mon. Not. Roy. Astron. Soc. \textbf{364}, 1105 (2005)

\bibitem{Widrow&Kaiser1993}
L.M. Widrow, N.~Kaiser, Astrophys. J. \textbf{416}(2), L71 (1993)

\bibitem{Uhlemann:2014npa}
C.~Uhlemann, M.~Kopp, T.~Haugg, Phys. Rev. D \textbf{90}(2), 023517 (2014)

\bibitem{Mocz:2018ium}
P.~Mocz, L.~Lancaster, A.~Fialkov, F.~Becerra, P.H. Chavanis, Phys. Rev. D
  \textbf{97}(8), 083519 (2018)

\bibitem{wyatt_trajectories}
R.E. {Wyatt}, \emph{{Quantum Dynamics with Trajectories}} (Springer, 2005)

\bibitem{ballentine_book}
L.E. {Ballentine}, \emph{{Quantum Mechanics}} (World Scientific, 1998)

\bibitem{Levkov:2018kau}
D.G. Levkov, A.G. Panin, I.I. Tkachev, Phys. Rev. Lett. \textbf{121}(15),
  151301 (2018)

\bibitem{Schive:2014dra}
H.Y. Schive, T.~Chiueh, T.~Broadhurst, Nature Phys. \textbf{10}, 496 (2014)

\bibitem{Mocz:2019emo}
P.~Mocz, et~al., Phys. Rev. Lett. \textbf{123}(14), 141301 (2019)

\bibitem{Schive:2015kza}
H.Y. Schive, T.~Chiueh, T.~Broadhurst, K.W. Huang, Astrophys. J.
  \textbf{818}(1), 89 (2016)

\bibitem{Schultz:2014eia}
C.~Schultz, J.~O\~{n}orbe, K.N. Abazajian, J.S. Bullock, Mon. Not. Roy. Astron.
  Soc. \textbf{442}(2), 1597 (2014)

\bibitem{Bouwens:2014fua}
R.J. Bouwens, et~al., Astrophys. J. \textbf{803}(1), 34 (2015)

\bibitem{Windhorst:2005as}
R.A. Windhorst, S.H. Cohen, R.A. Jansen, C.~Conselice, H.J. Yan, New Astron.
  Rev. \textbf{50}, 113 (2006)

\bibitem{Leung:2018evj}
E.~Leung, T.~Broadhurst, J.~Lim, J.M. Diego, T.~Chiueh, H.Y. Schive,
  R.~Windhorst, Astrophys. J. \textbf{862}(2), 156 (2018)

\bibitem{Bozek:2014uqa}
B.~Bozek, D.J.E. Marsh, J.~Silk, R.F.G. Wyse, Mon. Not. Roy. Astron. Soc.
  \textbf{450}(1), 209 (2015)

\bibitem{Corasaniti:2016epp}
P.S. Corasaniti, S.~Agarwal, D.J.E. Marsh, S.~Das, Phys. Rev. D \textbf{95}(8),
  083512 (2017)

\bibitem{Viel:2013apy}
M.~Viel, G.D. Becker, J.S. Bolton, M.G. Haehnelt, Phys. Rev. D \textbf{88},
  043502 (2013)

\bibitem{Gnedin:2001wg}
N.Y. Gnedin, A.J.S. Hamilton, Mon. Not. Roy. Astron. Soc. \textbf{334}, 107
  (2002)

\bibitem{Tegmark:2002cy}
M.~Tegmark, M.~Zaldarriaga, Phys. Rev. D \textbf{66}, 103508 (2002)

\bibitem{Hui:2016ltb}
L.~Hui, J.P. Ostriker, S.~Tremaine, E.~Witten, Phys. Rev. D \textbf{95}(4),
  043541 (2017)

\bibitem{Chabanier:2019eai}
S.~Chabanier, M.~Millea, N.~Palanque-Delabrouille, Mon. Not. Roy. Astron. Soc.
  \textbf{489}(2), 2247 (2019)

\bibitem{Armengaud:2017nkf}
E.~Armengaud, N.~Palanque-Delabrouille, C.~Y\`eche, D.J.E. Marsh, J.~Baur, Mon.
  Not. Roy. Astron. Soc. \textbf{471}(4), 4606 (2017)

\bibitem{Irsic:2017yje}
V.~Ir\v{s}i\v{c}, M.~Viel, M.G. Haehnelt, J.S. Bolton, G.D. Becker, Phys. Rev.
  Lett. \textbf{119}(3), 031302 (2017)

\bibitem{Zhang:2017dpp}
U.H. Zhang, T.~Chiueh, Phys. Rev. D \textbf{96}(6), 063522 (2017)

\bibitem{Leong:2018opi}
K.H. Leong, H.Y. Schive, U.H. Zhang, T.~Chiueh, Mon. Not. Roy. Astron. Soc.
  \textbf{484}(3), 4273 (2019)

\bibitem{2019ApJ...871...28B}
B.~{Bar-Or}, J.B. {Fouvry}, S.~{Tremaine}, Astrophys. J. \textbf{871}(1), 28
  (2019)

\bibitem{El-Zant:2019ios}
A.A. El-Zant, J.~Freundlich, F.~Combes, A.~Halle,   (2019)

\bibitem{Lancaster:2019mde}
L.~Lancaster, C.~Giovanetti, P.~Mocz, Y.~Kahn, M.~Lisanti, D.N. Spergel,
  (2019)

\bibitem{Marsh:2018zyw}
D.J.E. Marsh, J.C. Niemeyer, Phys. Rev. Lett. \textbf{123}(5), 051103 (2019)

\bibitem{2019MNRAS.485.2861C}
B.V. {Church}, P.~{Mocz}, J.P. {Ostriker}, Mon. Not. Roy. Astron. Soc.
  \textbf{485}(2), 2861 (2019)

\bibitem{BinneyTremaine2008}
J.~{Binney}, S.~{Tremaine}, \emph{{Galactic Dynamics: Second Edition}}
  ({Princeton University Press}, 2008)

\bibitem{Li:2016utv}
T.S. Li, et~al., Astrophys. J. \textbf{838}(1), 8 (2017)

\bibitem{2016ApJ...824L..14C}
D.~{Crnojevi{\'c}}, D.J. {Sand}, D.~{Zaritsky}, K.~{Spekkens}, B.~{Willman},
  J.R. {Hargis}, Astrophys. J. Lett. \textbf{824}(1), L14 (2016)

\bibitem{Schive:2019rrw}
H.Y. Schive, T.~Chiueh, T.~Broadhurst,   (2019)

\bibitem{Rogers:2020ltq}
K.K. Rogers, H.V. Peiris, Phys. Rev. Lett. \textbf{126}, 071302 (2021)

\bibitem{Marsh:2015wka}
D.J.E. Marsh, A.R. Pop, Mon. Not. Roy. Astron. Soc. \textbf{451}(3), 2479
  (2015)

\bibitem{2004sgig.book.....C}
S.M. {Carroll}, \emph{{Spacetime and geometry. An introduction to general
  relativity}} (Addison Wesley, 2004)

\bibitem{2011PhRvD..83d4026A}
A.~{Arvanitaki}, S.~{Dubovsky}, Phys. Rev. D \textbf{83}(4), 044026 (2011)

\bibitem{2015LNP...906.....B}
R.~{Brito}, V.~{Cardoso}, P.~{Pani} (eds.).
\newblock \emph{{Superradiance}}, \emph{Lect. Notes Phys.}, vol. 906 (2015)

\bibitem{Abbott:2016blz}
B.P. Abbott, et~al., Phys. Rev. Lett. \textbf{116}(6), 061102 (2016)

\bibitem{Stott:2018opm}
M.J. Stott, D.J.E. Marsh, Phys. Rev. D \textbf{98}(8), 083006 (2018)

\bibitem{Dolan:2018dqv}
S.R. Dolan, Phys. Rev. D \textbf{98}(10), 104006 (2018)

\bibitem{Brito:2013wya}
R.~Brito, V.~Cardoso, P.~Pani, Phys. Rev. D \textbf{88}(2), 023514 (2013)

\bibitem{East:2017ovw}
W.E. East, F.~Pretorius, Phys. Rev. Lett. \textbf{119}(4), 041101 (2017)

\bibitem{Yoshino:2012kn}
H.~Yoshino, H.~Kodama, Prog. Theor. Phys. \textbf{128}, 153 (2012)

\bibitem{Arvanitaki:2014wva}
A.~Arvanitaki, M.~Baryakhtar, X.~Huang, Phys. Rev. D \textbf{91}(8), 084011
  (2015)

\bibitem{Rosa:2017ury}
J.G. Rosa, T.W. Kephart, Phys. Rev. Lett. \textbf{120}(23), 231102 (2018)

\bibitem{Ikeda:2018nhb}
T.~Ikeda, R.~Brito, V.~Cardoso, Phys. Rev. Lett. \textbf{122}(8), 081101 (2019)

\bibitem{Day:2019bbh}
F.V. Day, J.I. McDonald, J. Cosmol. Astropart. Phys. \textbf{10}, 051 (2019)

\bibitem{Arvanitaki:2016qwi}
A.~Arvanitaki, M.~Baryakhtar, S.~Dimopoulos, S.~Dubovsky, R.~Lasenby, Phys.
  Rev. D \textbf{95}(4), 043001 (2017)

\bibitem{Hlozek:2016lzm}
R.~Hlo\v{z}ek, D.J.E. Marsh, D.~Grin, R.~Allison, J.~Dunkley, E.~Calabrese,
  Phys. Rev. D \textbf{95}(12), 123511 (2017)

\bibitem{Kobayashi:2017jcf}
T.~Kobayashi, R.~Murgia, A.~De~Simone, V.~Ir\v{s}i\v{c}, M.~Viel, Phys. Rev. D
  \textbf{96}(12), 123514 (2017)

\bibitem{Aghamousa:2016zmz}
A.~Aghamousa, J.~Aguilar, S.~Ahlen, S.~Alam, L.E. Allen, C.A. Prieto, J.~Annis,
  S.~Bailey, C.~Balland, O.~Ballester, et~al., arXiv:1611.00036  (2016)

\bibitem{Grin:2019mub}
D.~Grin, M.A. Amin, V.~Gluscevic, R.~Hlǒzek, D.J. Marsh, V.~Poulin,
  C.~Prescod-Weinstein, T.L. Smith, arXiv:1904.09003  (2019)

\bibitem{1969PhRv..187.1767R}
R.~{Ruffini}, S.~{Bonazzola}, Phys. Rev. \textbf{187}, 1767 (1969)

\bibitem{1991PhRvL..66.1659S}
E.~{Seidel}, W.M. {Suen}, Phys. Rev. Lett. \textbf{66}, 1659 (1991)

\bibitem{Clough:2015sqa}
K.~Clough, P.~Figueras, H.~Finkel, M.~Kunesch, E.A. Lim, S.~Tunyasuvunakool,
  Class. Quant. Grav. \textbf{32}(24), 245011 (2015)

\bibitem{Helfer:2016ljl}
T.~Helfer, D.J.E. Marsh, K.~Clough, M.~Fairbairn, E.A. Lim, R.~Becerril, J.
  Cosmol. Astropart. Phys. \textbf{1703}(03), 055 (2017)

\bibitem{Levkov:2016rkk}
D.G. Levkov, A.G. Panin, I.I. Tkachev, Phys. Rev. Lett. \textbf{118}(1), 011301
  (2017)

\bibitem{Michel:2018nzt}
F.~Michel, I.G. Moss, Phys. Lett. B \textbf{785}, 9 (2018)

\bibitem{1959PhRv..116.1322A}
R.~{Arnowitt}, S.~{Deser}, C.W. {Misner}, Phys. Rev. \textbf{116}(5), 1322
  (1959)

\bibitem{Schive:2014hza}
H.Y. Schive, M.H. Liao, T.P. Woo, S.K. Wong, T.~Chiueh, T.~Broadhurst, W.Y.P.
  Hwang, Phys. Rev. Lett. \textbf{113}(26), 261302 (2014)

\bibitem{Du:2016aik}
X.~Du, C.~Behrens, J.C. Niemeyer, B.~Schwabe, Phys. Rev. D \textbf{95}(4),
  043519 (2017)

\bibitem{Mina:2020eik}
M.~Mina, D.F. Mota, H.A. Winther, arXiv:2007.04119  (2020)

\bibitem{Nori:2020jzx}
M.~Nori, M.~Baldi, arXiv:2007.01316  (2020)

\bibitem{Veltmaat:2018dfz}
J.~Veltmaat, J.C. Niemeyer, B.~Schwabe, Phys. Rev. D \textbf{98}(4), 043509
  (2018)

\bibitem{JacksonKimball:2017qgk}
D.F. Jackson~Kimball, D.~Budker, J.~Eby, M.~Pospelov, S.~Pustelny, T.~Scholtes,
  Y.V. Stadnik, A.~Weis, A.~Wickenbrock, Phys. Rev. D \textbf{97}(4), 043002
  (2018)

\bibitem{Derevianko:2013oaa}
A.~Derevianko, M.~Pospelov, Nature Phys. \textbf{10}, 933 (2014)

\bibitem{Giudice:2016zpa}
G.F. Giudice, M.~McCullough, A.~Urbano, J. Cosmol. Astropart. Phys.
  \textbf{1610}(10), 001 (2016)

\bibitem{Dietrich:2018jov}
T.~Dietrich, F.~Day, K.~Clough, M.~Coughlin, J.~Niemeyer, Mon. Not. Roy.
  Astron. Soc. \textbf{483}(1), 908 (2019)

\bibitem{1988PhLB..205..228H}
C.J. {Hogan}, M.J. {Rees}, Phys. Lett. B \textbf{205}, 228 (1988)

\bibitem{Vaquero:2018tib}
A.~Vaquero, J.~Redondo, J.~Stadler, J. Cosmol. Astropart. Phys. \textbf{04},
  012 (2019)

\bibitem{Ellis:2020gtq}
D.~Ellis, D.J. Marsh, C.~Behrens, arXiv:2006.08637  (2020)

\bibitem{Green:2003un}
A.M. Green, S.~Hofmann, D.J. Schwarz, Mon. Not. Roy. Astron. Soc. \textbf{353},
  L23 (2004)

\bibitem{Wantz:2009it}
O.~Wantz, E.P.S. Shellard, Phys. Rev. D \textbf{82}, 123508 (2010)

\bibitem{Borsanyi:2016ksw}
S.~Borsanyi, et~al., Nature \textbf{539}(7627), 69 (2016)

\bibitem{Kolb:1994fi}
E.W. Kolb, I.I. Tkachev, Phys. Rev. D \textbf{50}, 769 (1994)

\bibitem{sigl}
G.~{Sigl}, \emph{{Astroparticle Physics: Theory and Phenomenology}}
  ({Springer}, 2017)

\bibitem{peacock}
J.~{Peacock}, \emph{{Cosmological Physics}} ({Cambridge University Press},
  1999)

\bibitem{Zurek:2006sy}
K.M. Zurek, C.J. Hogan, T.R. Quinn, Phys. Rev. D \textbf{75}, 043511 (2007)

\bibitem{Gosenca:2017ybi}
M.~Gosenca, J.~Adamek, C.T. Byrnes, S.~Hotchkiss, Phys. Rev. D \textbf{96}(12),
  123519 (2017)

\bibitem{Eggemeier:2019khm}
B.~Eggemeier, J.~Redondo, K.~Dolag, J.C. Niemeyer, A.~Vaquero, Phys. Rev. Lett.
  \textbf{125}(4), 041301 (2020)

\bibitem{Enander:2017ogx}
J.~Enander, A.~Pargner, T.~Schwetz, J. Cosmol. Astropart. Phys. \textbf{12},
  038 (2017)

\bibitem{Fairbairn:2017sil}
M.~Fairbairn, D.J.E. Marsh, J.~Quevillon, S.~Rozier, Phys. Rev. D
  \textbf{97}(8), 083502 (2018)

\bibitem{Eggemeier:2019jsu}
B.~Eggemeier, J.C. Niemeyer, Phys. Rev. D \textbf{100}(6), 063528 (2019)

\bibitem{Kolb:1995bu}
E.W. Kolb, I.I. Tkachev, Astrophys. J. Lett. \textbf{460}, L25 (1996)

\bibitem{Katz:2018zrn}
A.~Katz, J.~Kopp, S.~Sibiryakov, W.~Xue, J. Cosmol. Astropart. Phys.
  \textbf{12}, 005 (2018)

\bibitem{Fairbairn:2017dmf}
M.~Fairbairn, D.J.E. Marsh, J.~Quevillon, Phys. Rev. Lett. \textbf{119}(2),
  021101 (2017)

\bibitem{Tkachev:2014dpa}
I.~Tkachev, JETP Lett. \textbf{101}(1), 1 (2015)

\bibitem{Iwazaki:2014wta}
A.~Iwazaki, arXiv:1412.7825  (2014)

\bibitem{Bai:2017feq}
Y.~Bai, Y.~Hamada, Phys. Lett. B \textbf{781}, 187 (2018)

\bibitem{Hardy:2016mns}
E.~Hardy, J. High Energy Phys. \textbf{02}, 046 (2017)

\bibitem{Feix:2020txt}
M.~Feix, S.~Hagstotz, A.~Pargner, R.~Reischke, B.M. Schaefer, T.~Schwetz,
  arXiv:2004.02926  (2020)

\bibitem{Niikura:2017zjd}
H.~Niikura, et~al., Nat. Astron. \textbf{3}(6), 524 (2019)

\bibitem{mesa}
B.~{Paxton}, L.~{Bildsten}, A.~{Dotter}, F.~{Herwig}, P.~{Lesaffre},
  F.~{Timmes}, Astrophys. J., Suppl. Ser. \textbf{192}(1), 3 (2011)

\bibitem{Friedland:2012hj}
A.~Friedland, M.~Giannotti, M.~Wise, Phys. Rev. Lett. \textbf{110}(6), 061101
  (2013)

\bibitem{Raffelt:1990yz}
G.G. Raffelt, Phys. Rep. \textbf{198}, 1 (1990)

\bibitem{2008LNP...741...51R}
G.G. Raffelt, in \emph{Axions} (Springer, 2008), p.~51

\bibitem{Aver:2015iza}
E.~Aver, K.A. Olive, E.D. Skillman, J. Cosmol. Astropart. Phys.
  \textbf{1507}(07), 011 (2015)

\bibitem{Ayala:2014pea}
A.~Ayala, I.~Dom\'inguez, M.~Giannotti, A.~Mirizzi, O.~Straniero, Phys. Rev.
  Lett. \textbf{113}(19), 191302 (2014)

\bibitem{Giannotti:2015kwo}
M.~Giannotti, I.~Irastorza, J.~Redondo, A.~Ringwald, J. Cosmol. Astropart.
  Phys. \textbf{1605}(05), 057 (2016)

\bibitem{Hoof:2018ieb}
S.~Hoof, F.~Kahlhoefer, P.~Scott, C.~Weniger, M.~White, J. High Energy Phys.
  \textbf{03}, 191 (2019)

\bibitem{Chang:2018rso}
J.H. Chang, R.~Essig, S.D. McDermott, J. High Energy Phys. \textbf{09}, 051
  (2018)

\bibitem{Graham:2013gfa}
P.W. Graham, S.~Rajendran, Phys. Rev. D \textbf{88}, 035023 (2013)

\bibitem{10.1086/156841}
H.L. {Shipman}, Astrophys. J. \textbf{228}, 240 (1979)

\bibitem{Corsico:2019nmr}
A.H. Corsico, L.G. Althaus, M.M. Miller~Bertolami, S.~Kepler, Astron.
  Astrophys. Rev. \textbf{27}(1), 7 (2019)

\bibitem{10.1038/303781a0}
D.E. {Winget}, C.J. {Hansen}, H.M. {van Horn}, Nature \textbf{303}(5920), 781
  (1983)

\bibitem{10.1086/163398}
S.D. {Kawaler}, D.E. {Winget}, C.J. {Hansen}, Astrophys. J. \textbf{295}, 547
  (1985)

\bibitem{Nakagawa:1987pga}
M.~Nakagawa, Y.~Kohyama, N.~Itoh, Astrophys. J. \textbf{322}, 291 (1987)

\bibitem{Nakagawa:1988rhp}
M.~Nakagawa, T.~Adachi, Y.~Kohyama, N.~Itoh, Astrophys. J. \textbf{326}, 241
  (1988)

\bibitem{10.1093/mnras/stz2571}
A.D. {Romero}, L.A. {Amaral}, T.~{Klippel}, D.~{Sanmartim}, L.~{Fraga},
  G.~{Ourique}, I.~{Pelisoli}, G.R. {Lauffer}, S.O. {Kepler}, D.~{Koester},
  Mon. Not. R. Astron. Soc. \textbf{490}(2), 1803 (2019)

\bibitem{Isern:1992gia}
J.~Isern, M.~Hernanz, E.~Garcia-Berro, Astrophys. J. \textbf{392}, L23 (1992)

\bibitem{Altherr:1993zd}
T.~Altherr, E.~Petitgirard, T.~del Rio~Gaztelurrutia, Astropart. Phys.
  \textbf{2}, 175 (1994)

\bibitem{BischoffKim:2007ve}
A.~Bischoff-Kim, M.H. Montgomery, D.E. Winget, Astrophys. J. \textbf{675}, 1512
  (2008)

\bibitem{Isern:2008nt}
J.~Isern, E.~Garcia-Berro, S.~Torres, S.~Catalan, Astrophys. J. \textbf{682},
  L109 (2008)

\bibitem{Corsico:2012ki}
A.H. Corsico, L.G. Althaus, M.M.M. Bertolami, A.D. Romero, E.~Garcia-Berro,
  J.~Isern, S.O. Kepler, Mon. Not. Roy. Astron. Soc. \textbf{424}, 2792 (2012)

\bibitem{Corsico:2012sh}
A.H. Corsico, L.G. Althaus, A.D. Romero, A.S. Mukadam, E.~Garcia-Berro,
  J.~Isern, S.O. Kepler, M.A. Corti, J. Cosmol. Astropart. Phys. \textbf{1212},
  010 (2012)

\bibitem{Corsico:2016okh}
A.H. Corsico, A.D. Romero, L.G. Althaus, E.~Garcia-Berro, J.~Isern, S.O.
  Kepler, M.M. Miller~Bertolami, D.J. Sullivan, P.~Chote, J. Cosmol. Astropart.
  Phys. \textbf{1607}(07), 036 (2016)

\bibitem{Battich:2016htm}
T.~Battich, A.H. Corsico, L.G. Althaus, M.M. Miller~Bertolami, M.M.M.
  Bertolami, J. Cosmol. Astropart. Phys. \textbf{1608}(08), 062 (2016)

\bibitem{2012Kepler}
S.~Kepler, Publ. Astron. Soc. Pac.: Conf. Ser. \textbf{462}, 322 (2012)

\bibitem{Giannotti:2017hny}
M.~Giannotti, I.G. Irastorza, J.~Redondo, A.~Ringwald, K.~Saikawa, J. Cosmol.
  Astropart. Phys. \textbf{10}, 010 (2017)

\bibitem{Bertolami:2014wua}
M.M. Miller~Bertolami, B.E. Melendez, L.G. Althaus, J.~Isern, J. Cosmol.
  Astropart. Phys. \textbf{1410}(10), 069 (2014)

\bibitem{Isern:2018uce}
J.~Isern, E.~Garcia-Berro, S.~Torres, R.~Cojocaru, S.~Catalan, Mon. Not. Roy.
  Astron. Soc. \textbf{478}(2), 2569 (2018)

\bibitem{Chupp:1989kx}
E.~Chupp, W.~Vestrand, C.~Reppin, Phys. Rev. Lett. \textbf{62}, 505 (1989)

\bibitem{Payez:2014xsa}
A.~Payez, C.~Evoli, T.~Fischer, M.~Giannotti, A.~Mirizzi, A.~Ringwald, J.
  Cosmol. Astropart. Phys. \textbf{02}, 006 (2015)

\bibitem{Berg:2016ese}
M.~Berg, J.P. Conlon, F.~Day, N.~Jennings, S.~Krippendorf, A.J. Powell,
  M.~Rummel, Astrophys. J. \textbf{847}(2), 101 (2017)

\bibitem{Conlon:2017qcw}
J.P. Conlon, F.~Day, N.~Jennings, S.~Krippendorf, M.~Rummel, J. Cosmol.
  Astropart. Phys. \textbf{07}, 005 (2017)

\bibitem{Day:2018ckv}
F.~Day, S.~Krippendorf, Galaxies \textbf{6}(2), 45 (2018)

\bibitem{Reynolds:2019uqt}
C.S. Reynolds, M.D. Marsh, H.R. Russell, A.C. Fabian, R.~Smith, F.~Tombesi,
  S.~Veilleux, Astrophys. J. \textbf{890}(1), 59 (2020)

\bibitem{Mirizzi:2009nq}
A.~Mirizzi, J.~Redondo, G.~Sigl, J. Cosmol. Astropart. Phys. \textbf{08}, 001
  (2009)

\bibitem{Tashiro:2013yea}
H.~Tashiro, J.~Silk, D.J.E. Marsh, Phys. Rev. D \textbf{88}(12), 125024 (2013)

\bibitem{Mather:1993ij}
J.C. Mather, et~al., Astrophys. J. \textbf{420}, 439 (1994)

\bibitem{Durrer:2013pga}
R.~Durrer, A.~Neronov, Astron. Astrophys. Rev. \textbf{21}, 62 (2013)

\end{thebibliography}


\end{document}